\documentclass[%
 reprint,
superscriptaddress,
 amsmath,amssymb,
]{revtex4-2}

\usepackage{graphicx}
\usepackage{dcolumn}
\usepackage{bm}
\usepackage{bbm}
\usepackage[caption=false]{subfig}
\usepackage{float}
\usepackage{hyperref}
\usepackage{algorithm}
\usepackage{algpseudocode}
\usepackage{color}
\usepackage{multirow}
\graphicspath{{images/}}

\usepackage{amsthm}

\begin{document}

\preprint{APS/123-QED}

\title{Optimizing Continuous-Variable Quantum Key Distribution with Phase-Shift Keying Modulation and Postselection}
\author{Florian Kanitschar}
 \email{florian.kanitschar@outlook.com}
\affiliation{AIT  Austrian  Institute  of  Technology,  Center  for  Digital  Safety\&Security,  Giefinggasse  4,  1210  Vienna, Austria}
 \affiliation{TU Wien, Faculty of Physics, Wiedner Hauptstraße 8, 1040 Vienna, Austria}
 
 \author{Christoph Pacher}
  \affiliation{AIT  Austrian  Institute  of  Technology,  Center  for  Digital  Safety\&Security,  Giefinggasse  4,  1210  Vienna, Austria}
\affiliation{fragmentiX Storage Solutions GmbH, IST Park, Plöcking 1, 3400 Klosterneuburg, Austria}

\date{\today}

\begin{abstract}
A numerical security proof technique is used to analyse the security of continuous-variable quantum key distribution (CV-QKD) protocols with phase-shift keying modulation against collective attacks in the asymptotic limit. We argue why it is sufficient to consider protocols with a maximum number of eight signal states and analyse different postselection strategies for protocols with four (QPSK) and eight (8PSK) signal states for untrusted ideal  and trusted nonideal detectors. We introduce a \emph{cross-shaped} postselection strategy, and show that both cross-shaped and radial and angular postselection clearly outperform a radial postselection scheme (and no postselection) for QPSK protocols. For all strategies studied, we provide analytical results for the operators defining the respective regions in phase space. We outline several use-cases of postselection, which can easily be introduced in the data processing of both new and existing CV-QKD systems:
Motivated by the high computational effort for error-correction, we studied the case when a large fraction of the raw key is eliminated by postselection and observed that this can be achieved while increasing the secure key rate. Postselection can also be used to partially compensate the disadvantage of QPSK protocols over 8PSK protocols for high transmission distances, while being experimentally less demanding. 
Finally, we highlight that postselection can be used to reduce the key rate gap between scenarios with trusted and untrusted detectors while relying on less assumptions on Eve's power.
\end{abstract}

\maketitle



\section{\label{sec:Intro}Introduction}
Quantum key distribution (QKD) enables two remote parties to expand a short symmetric preshared secret key into a long   secret key without any assumptions on the computational power of a potential adversary. Perhaps the most famous cryptographic protocol based on quantum mechanics, BB84, was published by Bennett and Brassard \cite{Bennett_Brassard_1984} in 1984 and is based on (discretely) polarized photons. Discrete-variable (DV) QKD protocols like BB84 rely on single-photon detectors that are rather expensive components. In contrast, continuous-variable (CV) QKD protocols are based on the field quadratures of light that are measured by homodyne or heterodyne detection using much cheaper photodiodes. Continuous-variable QKD goes back to Ralph in 1999 \cite{Ralph_1999} and is based on discrete modulation.  Postselection, i.e. selection of certain signals for further processing based on measurement results, can increase the secure key rate for continuous-variable QKD-protocols by reducing the information available to an adversary \cite{Silberhorn_2002}. Postselection areas for multiletter phase-shift keying (PSK) protocols in linear channels are discussed in \cite{Sych_2010}. References \cite{Pirandola_2020, Diamanti_2015, Scarani_2009} give comprehensive reviews of the entire field of quantum key distribution, implementations, and security proofs.

The process of finding a lower bound on the achievable secure key rate is called security proof. Analytical attempts to prove the security of a certain QKD protocol are usually very technical, introduce looseness in the lower bounds and are often difficult to generalize. In contrast, numerical attempts are typically more flexible concerning changes in the protocol structure, but have only finite precision, and it cannot be expected that the optimization tasks involved achieve the optimum with arbitrary accuracy. In particular, for CV protocols, we have to approximate physical quantities living in infinite-dimensional Hilbert spaces by finite-dimensional representatives to make the key-rate-finding problem computationally feasible. Finally, as the most interesting discrete modulated (DM) CV-QKD protocols involve four or more states, the numerical tasks are high-dimensional, hence computation time is crucial.

Several security proofs for DM CV-QKD protocols in the asymptotic limit are known. Analytical approaches are restricted to certain scenarios such as linear quantum channels \cite{Heid_2006, Sych_2010}, or a fixed number of signal states \cite{Zhao_2009}, \cite{Bradler_2018}. A general attempt \cite{Leverrier_2021} gives an analytical lower bound on the secure key rate of DM CV-QKD protocols with arbitrary modulation, but is loose for a low number of signal states and does not consider postselection. In contrast, numerical approaches \cite{Coles_2016, Winick_2018, Lin_2019, Ghorai_2019} are more flexible, but suffer from high computational complexity and assume that it is sufficient to solve the key rate finding problem occurring by truncating the infinite Fock space. This so-called photon-number cutoff assumption was removed in \cite{Upadhyaya_2021}. 

While these security proofs are valid in the limit of infinitely long keys, establishing general security in the finite-size regime is an open problem. Recently, \cite{Pirandola_2021} proved the finite-size security of a QKD protocol with discrete encoding under the restriction of collective \emph{Gaussian} attacks, while \cite{Matsuura_2021} introduced a general security proof of a special two-state DM CV-QKD protocol.

\subsection{Contribution}
In the present work, we adapt a numerical framework  to calculate secure key rates of DM CV-QKD protocols under the assumption of collective attacks in the asymptotic limit \cite{Coles_2016,Winick_2018} and  use it to analyze and optimize different phase-shift keying protocols.  We consider both the untrusted and trusted detector scenarios. We reduc the computation time and increase the accuracy of the calculated key rates by replacing numerical approximations for certain operators with analytical expressions.

We point out why it is not useful to consider more than eight PSK signal states for phase-shift keying modulation protocols using state-of-the-art postselection strategies. Backed by considerations about the bit error probability for heterodyne measurement outcomes, we introduce a simple postselection strategy for QPSK protocols (cross-shaped postselection) and show that it clearly outperforms state-of-the-art radial \cite{Lin_2019} postselection in terms of the achievable secure key rate and performs comparably with radial and angular postselection. We show how postselection can be used to reduce the amount of \emph{raw} key (i.e., data that must be error-corrected) significantly, tackling a well-known bottleneck in practical implementations,  at the cost of only a much smaller decrease in the amount of secure key.
We highlight how a smart choice of the postselection strategy can also reduce the gap in the key rate between QPSK and 8PSK modulation. 

\subsection{Organization}
The remainder of this work is structured as follows. In Section~\ref{sec:proof_approach} we give a brief summary of the numerical security proof approach we use, including the changes necessary to consider trusted detectors.
In Section~\ref{sec:protocols} we introduce a general phase-shift keying protocol and motivate our particular choices for the number of signal states and the postselection strategies. This is followed by Section~ \ref{sec:Implementation} where we comment on details of the implementation. In the main part of the paper, Section~\ref{sec:Results}, we present and discuss our results for both untrusted and trusted detectors. We show that postselection is an important improvement to discretely modulated CV-QKD. Section~\ref{sec:Discussion} concludes the paper. Detailed explanations, calculations and additional examinations are provided in the appendices. 

\subsection{Choice of Units}
Throughout the paper, we use natural units, that is, the quadrature operators are defined by $\hat{q} := \frac{1}{\sqrt{2}}(\hat{a}^{\dagger}+ \hat{a})\\
     \hat{p} := \frac{i}{\sqrt{2}}(\hat{a}^{\dagger}- \hat{a}),$
where $\hat{a}$ and $\hat{a}^{\dagger}$ are the bosonic ladder operators and the commutation relation between the $q$- and $p$-quadrature operators has the form $[\hat{q}, \hat{p}] = i$.\\


\section{\label{sec:proof_approach}The security proof approach}
First, we give a brief overview of the numerical security proof approach against collective attacks in the asymptotic limit, following \cite{Winick_2018} and \cite{Lin_2019}. Interested readers will find a more detailed discussion in Appendix \ref{APP:Two-step-process}. The physical intuition of the key rate finding problem is the following: we search for the optimal attack, that is, the density matrix of Alice's and Bob's joint quantum state that (1) minimizes the achievable secure key rate while (2) matching expected values of experimentally accessible observables (e.g. quadrature amplitudes, photon numbers). 
The well-known Devetak-Winter formula \cite{Devetak_Winter_2006} gives the secret key rate in the asymptotic limit. In the case of reverse reconciliation, it can be reformulated \cite{Winick_2018} in terms of the quantum relative entropy $D(\rho||\sigma) := \textrm{Tr}\left[\rho \left(\log_2(\rho)-\log_2(\sigma)\right)\right]$, a quantity measuring the distinguishability of two states $\rho$ and $\sigma$, the sifting probability $p_{\textrm{pass}}$, and the amount of information leakage per signal in the error-correction phase $\delta_{EC}$ as follows
\begin{equation}
R^{\infty} = \min_{\rho_{AB} \in \mathcal{S}} D\left(\mathcal{G}(\rho_{AB}) || \mathcal{Z}(\mathcal{G}(\rho_{AB}))\right) - p_{\textrm{pass}} \delta_{EC}.    
\end{equation}
Here, $\mathcal{G}$ is a completely positive, trace-nonincreasing map describing classical postprocessing steps, and $\mathcal{Z}$ is a pinching quantum channel required to 'read out' the results of the key map. We note that $p_{\textrm{pass}}$ is contained implicitly in the first term of the target function. A more detailed explanation is given in \cite{Lin_2019}. The set $\mathcal{S}$ is the feasible set of the minimization, which is a subset of the set of density operators $\mathcal{D}(\mathcal{H}_{AB})$, where $\mathcal{H}_{AB} = \mathcal{H}_A \otimes \mathcal{H}_B$, and is defined by a set of linear constraints, 
\begin{equation}
\mathcal{S} := \left\{ \rho_{AB} \in \mathcal{D}(\mathcal{H}_{AB}) ~|~ \forall i \in I: \textrm{Tr}\left[ \Gamma_i \rho_{AB} \right] = \gamma_i  \right\},
\end{equation}
with Hermitian operators $\Gamma_i$, real numbers $\gamma_i$ and some finite set $I$. In what follows, we denote the objective function of this minimization by $f$.

\begin{figure*}
    \centering
\includegraphics[width=0.8\textwidth]{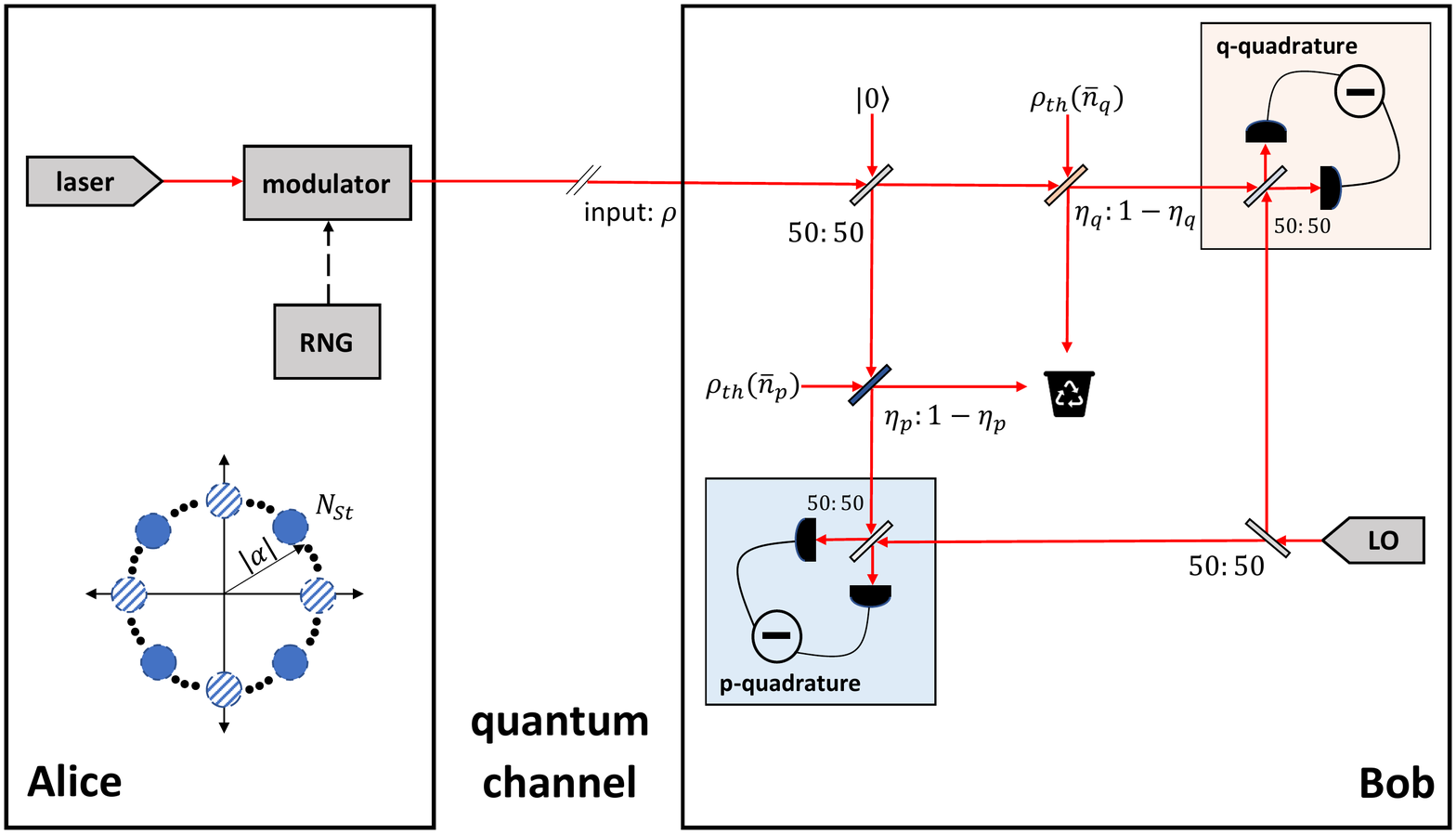}
\makeatletter\long\def\@ifdim#1#2#3{#2}\makeatother
\caption{Sketch of the realistic (nonideal) QKD experiment, where on Bob's side we depict the physical model of an imperfect, trusted, noisy heterodyne detector (for ideal detector, remove the blue and light-red beam splitter). We denote the local oscillator by LO and the random number generator in Alice's lab with RNG. }\label{fig:sketch_trusted}
\end{figure*} 

Using Lindblad's theorem \cite{Lindblad_1974} and the linearity of $\mathcal{G}$ and $\mathcal{Z}$, it can be shown that we face a convex minimization problem, with linear and semidefinite constraints, hence a semidefinite program (SDP). The present problem requires optimization over a subset of the set of all density operators, which is an infinite-dimensional vector space. In order to make the problem computationally feasible, we approximate it by a finite-dimensional vector space. While Bob's infinite-dimensional state space is spanned by the basis of Fock states $\{|n\rangle ~ : ~ n \in \mathbb{N}\}$, following the photon-number cutoff assumption in \cite{Lin_2019}, we approximate the infinite-dimensional state space by $\mathcal{H}_B^{N_c} :=\textrm{span}\{|n\rangle ~ : ~ 0 \leq n \leq N_c\}$, where $N_c \in \mathbb{N}$ is the cutoff number.
In contrast to most of the minimization problems, for the present problem it is not sufficient to come close to the minimum as this would give us only an upper bound on the secure key rate. Therefore, the approach in \cite{Winick_2018} tackles the key rate finding problem in a two-step process. First, a numerical solving algorithm is applied to find an eavesdropping attack that is close to optimal. Then, in a second step, this upper bound of the secure key rate is converted into a lower bound, taking numerical imprecision into account.

To start the key generation process, Alice prepares $|\psi_x\rangle$ drawn from a set of $N_{\textrm{St}}$ states with probability $p_x$ and sends them to Bob, using the quantum channel. Thanks to the source-replacement scheme \cite{Curty_2004, Ferenzci_2012}, equivalently the corresponding formulation in the entanglement-based scheme can be considered, where Alice prepares the bipartite state $|\Psi\rangle_{AA'} = \sum_{x} \sqrt{p_x} |x\rangle_A |\psi_x\rangle_{A'}$. $A$ denotes the register that is kept by Alice and the state labeled with $A'$ is sent to Bob. The quantum channel is modeled as a completely positive trace-preserving map $\mathcal{E}_{A'\rightarrow B}$. Hence, the bipartite state shared by Alice and Bob reads $\rho_{AB} = \left(\mathbbm{1}_A \otimes \mathcal{E}_{A'\rightarrow B} \right)\left(|\Psi\rangle\langle \Psi|_{AA'} \right)$.

Bob performs heterodyne measurements, hence determines the first and second moments of $\hat{q}$ and $\hat{p}$. These observations can be used to calculate the mean photon number $\hat{n} = \frac{1}{2} \left(\hat{q}^2 + \hat{p}^2 -1 \right)$ and $\hat{d} = \hat{q}^2 - \hat{p}^2$ to constrain the density matrix $\rho_{AB}$. Furthermore, we know that Eve has no access to Alice's system and, hence, she cannot modify the states held by Alice. That can be expressed mathematically as $\textrm{Tr}_B\left[ \rho_{AB} \right] = \sum_{x,y = 0}^{N_{\textrm{St}}-1} \sqrt{p_x p_y} \langle \psi_y|\psi_x\rangle~ |x\rangle \langle y|_A$, which is a matrix-valued constraint, where $p_x$ is the probability that the state $|\Psi_x\rangle$ is prepared, for $x \in \{0,...,N_{\textrm{St}}-1\}$.  Therefore, we have the following SDP: \cite{Lin_2019},
\begin{equation} \label{eq:SDP}
\begin{aligned}
   \textrm{minimize } &D(\mathcal{G}(\rho_{AB}) || \mathcal{G}(\mathcal{Z}(\rho_{AB}))  )\\
   \textrm{subject to: } &\\
    &\textrm{Tr}\left[ \rho_{AB} \left( |x\rangle\langle x|_A \otimes \hat{q} \right) \right] = p_x \langle \hat{q} \rangle_x\\
    &\textrm{Tr}\left[ \rho_{AB} \left( |x\rangle\langle x|_A \otimes \hat{p} \right) \right] = p_x \langle \hat{p} \rangle_x\\
    &\textrm{Tr}\left[ \rho_{AB} \left( |x\rangle\langle x|_A \otimes \hat{n} \right) \right] = p_x \langle \hat{n} \rangle_x\\
    &\textrm{Tr}\left[ \rho_{AB} \left( |x\rangle\langle x|_A \otimes \hat{d} \right) \right] = p_x \langle \hat{d} \rangle_x\\
    &\textrm{Tr}_B\left[ \rho_{AB} \right] = \sum_{i,j=0}^{N_{\textrm{St}}-1} \sqrt{p_i p_j} \langle \psi_j | \psi_i\rangle ~|i\rangle\langle j|_A\\
    & \rho_{AB} \geq 0,
\end{aligned}
\end{equation}
where $x \in \{0,..., N_{\textrm{St}}-1\}$.

\subsection{Trusted, nonideal detector approach}

Up to now, we have assumed that Bob performs measurements with an ideal, noiseless, heterodyne detector having unit detection efficiency. However, detectors used in QKD devices are not ideal, but are noisy and have an efficiency smaller than unity. These deviations from the ideal model can be included in the security analysis differently. 

Firstly, in the most conservative scenario,  we assume that Bob always uses an ideal detector and all nonideal detector properties occur on the transmission line (effectively increasing the transmission loss and excess noise)\footnote{This leads to a secure protocol even if we assume that the attacker Eve could manipulate the heterodyne detector in Bob's device, i.e. reduce its noise (or know partly or fully the noise signal) and/or increase its efficiency. Eve would first manipulate the detector and then try to hide an attack on the transmission line by reproducing the effects of the original detector.}. This is easily done in theory as it does not need any change to the security proof. However, it leads to much lower key rates and maximal achievable distances compared with the next scenario.

Secondly, as the detector is located in Bob's lab, it may less conservatively be assumed that Eve cannot improve the detector's efficiency and cannot reduce or control its electronic noise. 
In \cite{Lin_2020}, the numerical security proof method is extended to this second, so-called ``trusted'', detector scenario. We give a brief summary of the adaptations introduced there: A nonideal heterodyne detector is modeled by two homodyne detectors and a beam splitter, where each of the homodyne detectors has non-unit detector efficiency $\eta_q, \eta_p$ and suffers from electronic noise $\nu_q, \nu_p$. Similar to the excess noise, the electronic noise is measured in shot noise units. The quantum optical model by Lodewyck \cite{Lodewyck_2007} includes these quantities in the following way (see Figure~\ref{fig:sketch_trusted}). After the input signal is split into two parts at a $50:50$ beam splitter (where it is mixed with the vacuum state) each part of the signal passes another beam splitter with transmissions $\eta_q$ and $\eta_p$ (hence, reflectances of $1-\eta_q$ and $1-\eta_p$), respectively. There, we mix each part of the split signal with a thermal state with mean photon number $\overline{n}_q$ and $\overline{n}_p$. If we choose the mean photon numbers to be $\overline{n}_i = \frac{\nu_i}{2(1-\eta_i)}$, $i \in \{q,p\}$, we relate each thermal state to the observed amount of electronic noise introduced by the corresponding detector. Finally, two ideal homodyne detectors measure the signals.


\section{\label{sec:protocols}Phase-shift Keying Protocols}
We consider prepare-and-measure (P\&M) phase-shift keying protocols, which are generalizations of 'Protocol 2' in \cite{Lin_2019} which is a QPSK protocol with radial (and angular) postselection.

Therefore, we consider two distant parties, the sender Alice and the receiver Bob, who want to establish a symmetric key. They are connected by an authenticated classical channel and a quantum channel. Eve, an adversary, can listen to the classical communication and manipulate and store signals that are exchanged via the quantum channel. 
In what follows, we denote the number of signal states by $N_{\textrm{St}}$, and the raw key block size by $N \in \mathbb{N}$.

\begin{enumerate}
    \item[1) ] Alice prepares for every $n \leq N$ one out of $N_{\textrm{St}}$ coherent states $\left| \Psi_n \right\rangle = \left| |\alpha| e^{i \frac{(2k+1)\pi}{N_{\textrm{St}}} } \right\rangle$ corresponding to the symbol $x_n = k$,  where $|\alpha| > 0$ (arbitrary but fixed) is the coherent state amplitude and $k \in \{0,...,N_{\textrm{St}}-1\}$, according to some probability distribution (see Figure~\ref{fig:phase_space_both}). In the present work, this is chosen to be the uniform distribution. 
     This phase is called \textbf{state preparation}. After preparing one of these states, Alice sends it to Bob using the quantum channel.
    
    \item[2) ] Once the state is transmitted to Bob, he performs a heterodyne measurement and obtains some complex number $y_n$. This is called the \textbf{measurement phase}.
    
 \begin{figure}
\includegraphics[width=0.48\textwidth]{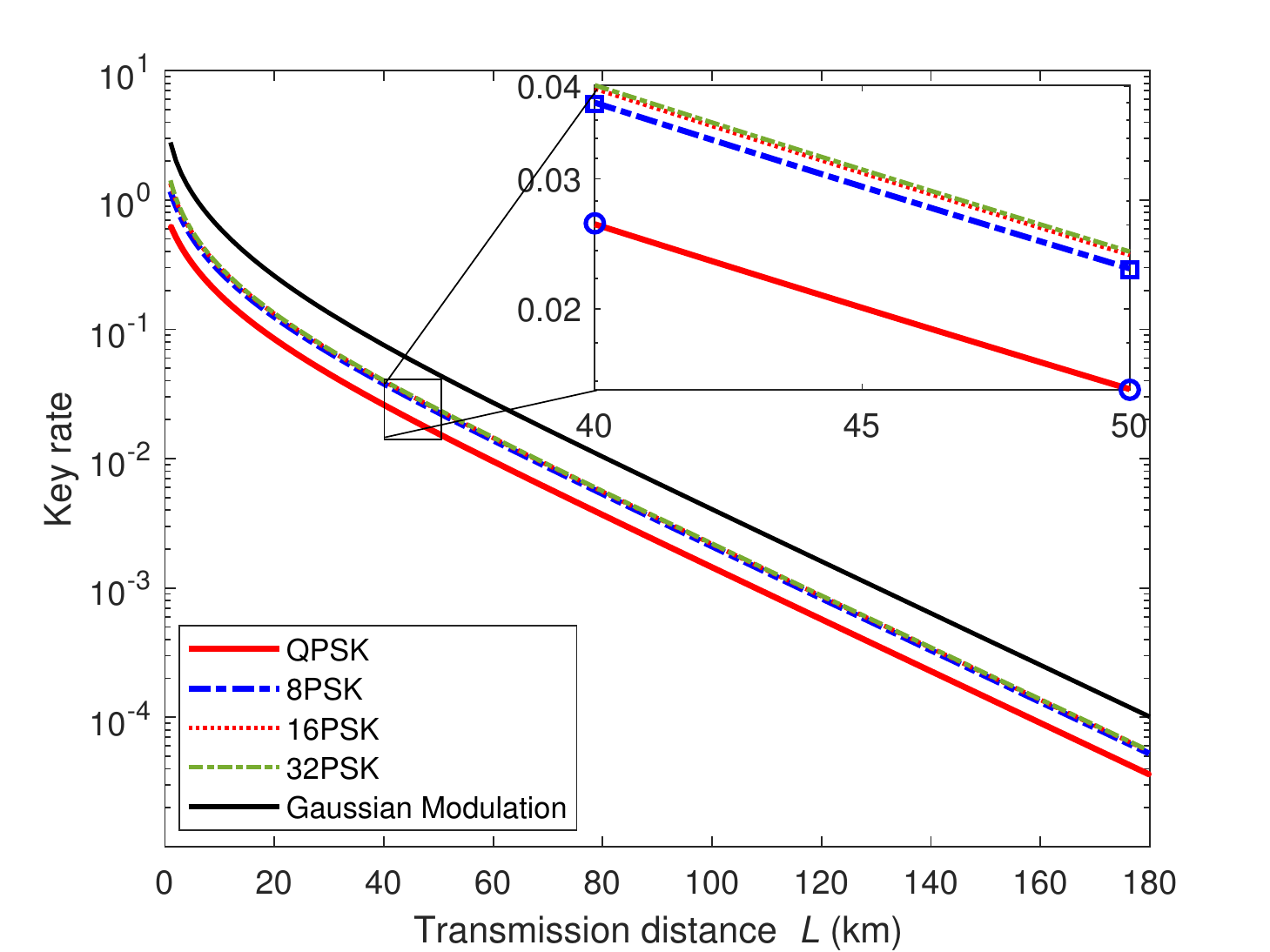}
\makeatletter\long\def\@ifdim#1#2#3{#2}\makeatother
\caption{\label{fig:N_states_incl_Gauss} Secure key rates for phase-shift keying protocols with $4, 8, 16$ and $32$ signal states in loss-only channels. For comparison, the achievable secure key rate for Gaussian modulation is shown. }
\end{figure}

    \item[3) ] Alice and Bob agree via the classical channel to choose some random subset $\mathcal{I}_{\textrm{Test}} \subset \{n \in\mathbb{N}~:~ n \leq N\}$ and reveal the corresponding sent symbols $x_l$ and measurement results $y_l$ for $l\in\mathcal{I}_{\textrm{Test}}$ to perform \textbf{parameter estimation}, that is, they determine the expected values of the first and second moments in Equation~(\ref{eq:SDP}). The remaining rounds $\mathcal{I}_{\textrm{Key}} := \{n \in\mathbb{N}~:~ n \leq N\} \setminus \mathcal{I}_{\textrm{test}}$ will be used for key generation. For simplicity, assume that $\mathcal{I}_{\textrm{Key}}$ contains the first $m := |\mathcal{I}_{\textrm{Key}}|$ rounds that can be used for key generation (this can be assumed without loss of generality, as we always find some bijective map that reorders the set). After this step, Alice holds a key string $\mathbf{X} := (x_1, ..., x_m)$.  
    
    \item[4) ] In the simplest case, Bob applies a \textbf{key map} to obtain his key string $\textbf{Z} = (z_j)_{j \in \mathcal{I}_{\textrm{Key}} }$. He maps each of his measurement outcomes $y_l$ for rounds $l \in \mathcal{I}_{\textrm{Key}}$  to an element in a finite set $\{0,..., N_{\textrm{St}}-1\}$. .

    \item[4*) ] In a more sophisticated version of the protocol, Bob is allowed to additionally \textbf{perform postselection}, that is, he can discard certain rounds according to some rule. He applies a modified \textbf{key map} to obtain his key string $\textbf{Z} = (z_j)_{j \in \mathcal{I}_{\textrm{Key}} }$. He adds an additional symbol $\perp$ to his finite set $\{0,..., N_{\textrm{St}}-1, \perp \}$, and maps discarded rounds to $\perp$ while kept rounds are treated as described in 4).

\item[5)] After having sent and transmitted a block of symbols, Alice and Bob perform classical \textbf{error correction}. For example, as considered in the present paper, Bob may transmit information about his key string and Alice corrects her key string according to his information, which is called reverse reconciliation (since the information flow is the opposite direction to the quantum signals sent). The error correction routine used cannot be expected to work perfectly, but with an efficiency of $0\leq \beta \leq 1$. After finishing the error correction, Alice and Bob use almost universal hash functions to upper-bound the probability that the error correction phase has failed. They omit the key if the hash values do not coincide and restart the key generation process  
\item[6)] Finally, Alice and Bob apply \textbf{privacy amplification} algorithms to reduce Eve's knowledge about their common information by omitting parts of their shared key, for example, by using seeded randomness extractor algorithms.    
\end{enumerate}

\subsection{Practical choices for the number of signal states}
\label{sec:number_signal_states}
Recent works on protocols with phase-shift keying modulation \cite{Ghorai_2019, Lin_2019, Ghalaii_2020, Hirano_2017, Bradler_2018} have focused on protocols with two, three or four signal states. 
Since there is no obvious upper limit on the number of signal states, we generalized an analytical security proof \cite{Heid_2006} for loss-only channels to an arbitrary number of signal states (for details, see Appendix \ref{sec:Th_Calc}) to investigate the effect of varying $N_{\textrm{St}}$. 

For the key map in step 4) of the protocol, we divide the phase space into $N_{\textrm{St}}$ equally sized wedges and associate each wedge with an element in the set $\{0,..., N_{\textrm{St}}-1\}$.

We examined the effect of different numbers of signal states and depict our findings in Figure~\ref{fig:N_states_incl_Gauss}. There, we plot the secure key rate as function of the distance for $4,~8,~16$ and $32$ signal states. For each distance, the coherent state amplitude $\alpha$ is optimized. Although our approach is not restricted to powers of $2$, similarly to classical communication, only $2^{N_{\textrm{St}}}$ states can be mapped directly to $N_{\textrm{St}}$ bits; for other values the generation of binary keys is significantly more complicated. Furthermore, it turned out to be sufficient and practical to find the maximal number of signal states $N_{\textrm{St}}$ that still increase the secure key rate noticeable. Our results show that using eight PSK signal states increases the secure key rate considerably compared with the QPSK protocol for loss-only channels, while increasing the number of signal states further has no significant impact on the secure key rate. Additionally, for comparison, we added the secure key rate for protocols with Gaussian modulation \cite{Laudenbach_2018, Grosshans_Grangier_2002a, Grosshans_2005, Braunstein_2008}. The gap in the secure key rate between PSK and Gaussian modulation can be explained by the additional amplitude modulation for Gaussian protocols. Hence, higher key rates than with discretely modulated CV protocols can be obtained with a combination of phase and amplitude modulation, for example, by adding signals on a second circle with a radius greater than $|\alpha|$.

\begin{figure}
\includegraphics[width=0.48\textwidth]{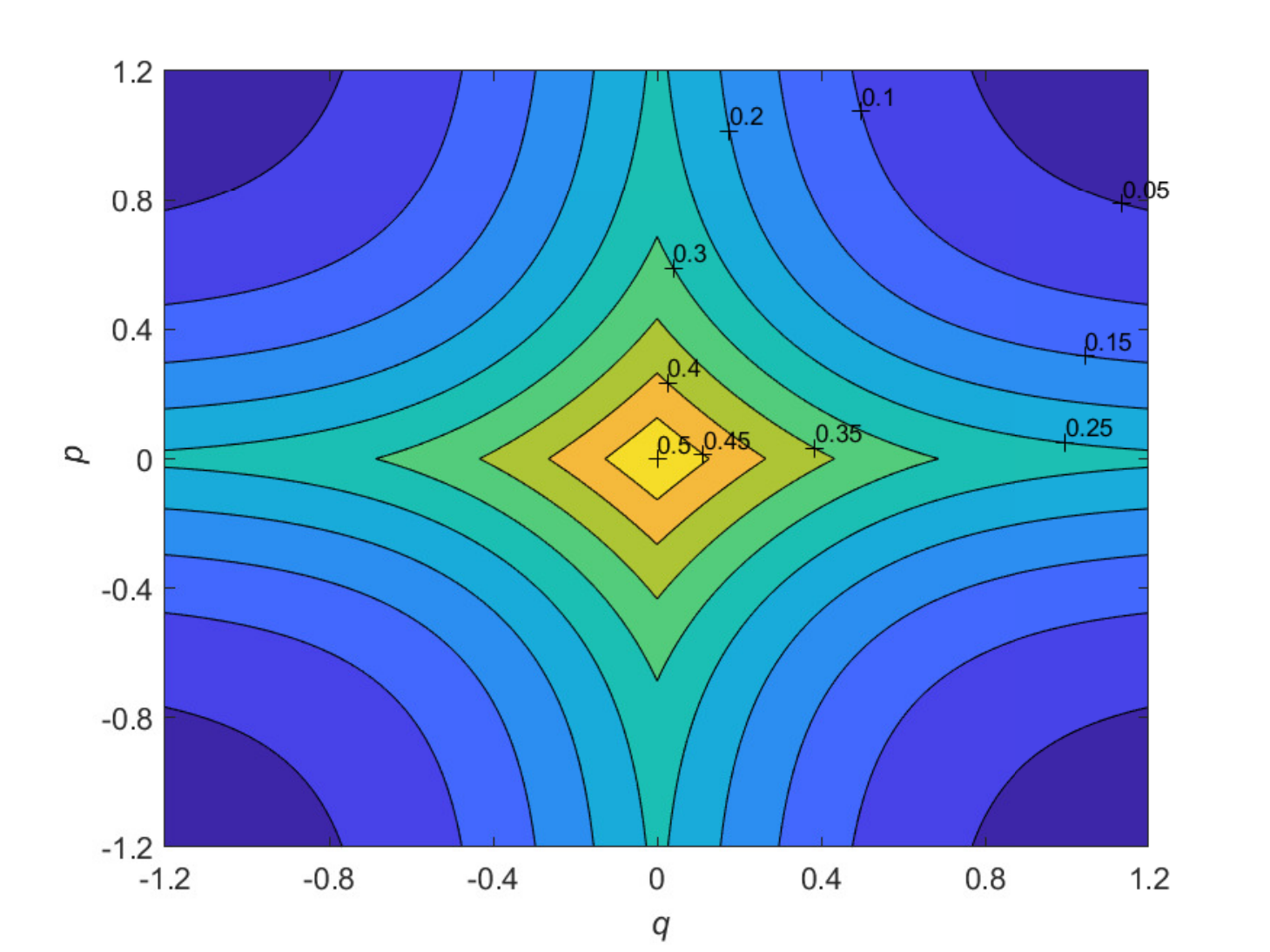}
\makeatletter\long\def\@ifdim#1#2#3{#2}\makeatother
\caption{\label{fig:BER_contourplot} Contour-plot of the expected bit error rate for heterodyne measurement without noise for the QPSK protocol for coherent state amplitude $|\alpha| = 0.8$.}
\end{figure}

\subsection{Phase shift keying with four or eight signal states and postselection}
Since doubling the number of signal states quadruples the dimension of the problem (see Footnote \footnote{In fact, doubling the number of signal states increases the system-dimension even by more than a factor of four. This is, because a higher number of signal states leads to a higher optimal coherent state amplitude $|\alpha|$, which again requires a higher photon-number cutoff.} for a more detailed discussion), backed by our finding in the previous subsection, we focus on protocols with four and eight signal states (arranged on one circle with arbitrary but fixed radius $|\alpha|$) with our numerical method, as we do not expect a significant improvement of the secure key rate for noisy channels as well. Constellations with additional amplitude modulation are left for future work. 

In what follows, we motivate and discuss postselection for QPSK and 8PSK protocols and describe the key maps of the resulting protocols.

\subsubsection{Motivation for the postselection strategies}\label{sec:Motivation_PS}
It is well known that the bit error rate (BER) is a major source for key rate reduction in DM CV-QKD at higher distances. In Figure~\ref{fig:BER_contourplot}, we plot the contour lines of the bit error probability for a four-state protocol without noise. Since adding noise increases the probability of bit errors further, in reality, we expect even higher bit error rates near the axes. Based on this observation, it seems reasonable to discard certain measurement results before proceeding with the key-generation process on the remaining ones. This is known as postselection. Optimally, a postselection strategy which omits signals in regions bounded by lines similar to the contour lines in Figure~\ref{fig:BER_contourplot} would be chosen. It would be subject to optimization to determine the exact parameters of the boundary. Unfortunately, for computational and numerical reasons, it is not possible to calculate secure key rates for postselection areas with this shape directly. Therefore, inspired by the bit error probability contour plot, we consider simplified postselection schemes which remove measurement results close to the axes. Technically, this is realized by adding an additional symbol ($\perp$) to the key map which is assigned whenever a signal lies within the discarded area. A more detailed explanation of the BER, its calculation and a discussion of numerical and computational reasons for working with simplified postselection strategies can be found in Appendix \ref{APDX:Discussion_PS}.

\begin{figure}
\subfloat[QPSK constellation. \label{fig:phase_space} ]{
    \includegraphics[width=0.35\textwidth]{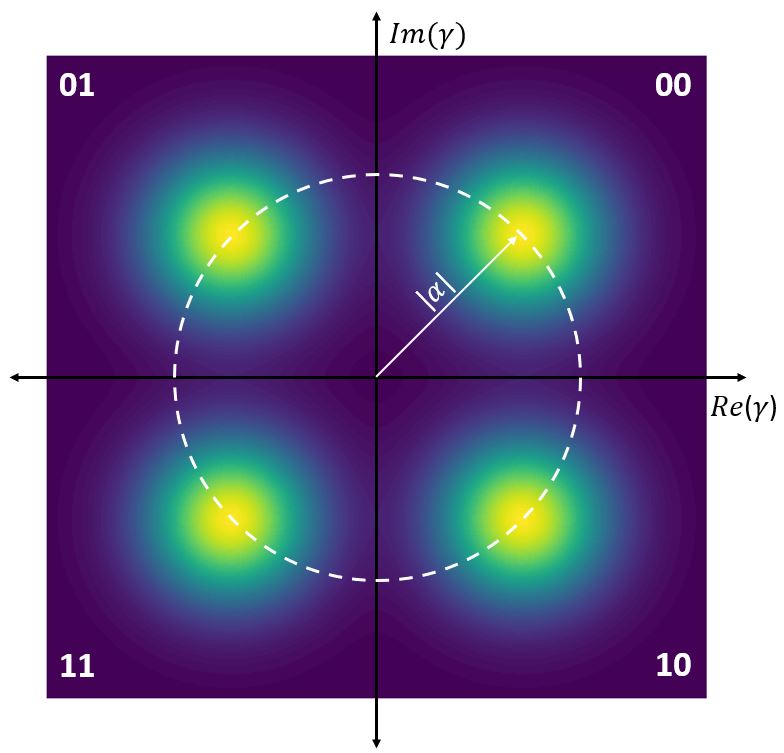}}\\
\subfloat[8PSK constellation. \label{fig:phase_space_8PSK}]{
    \includegraphics[width=0.35\textwidth]{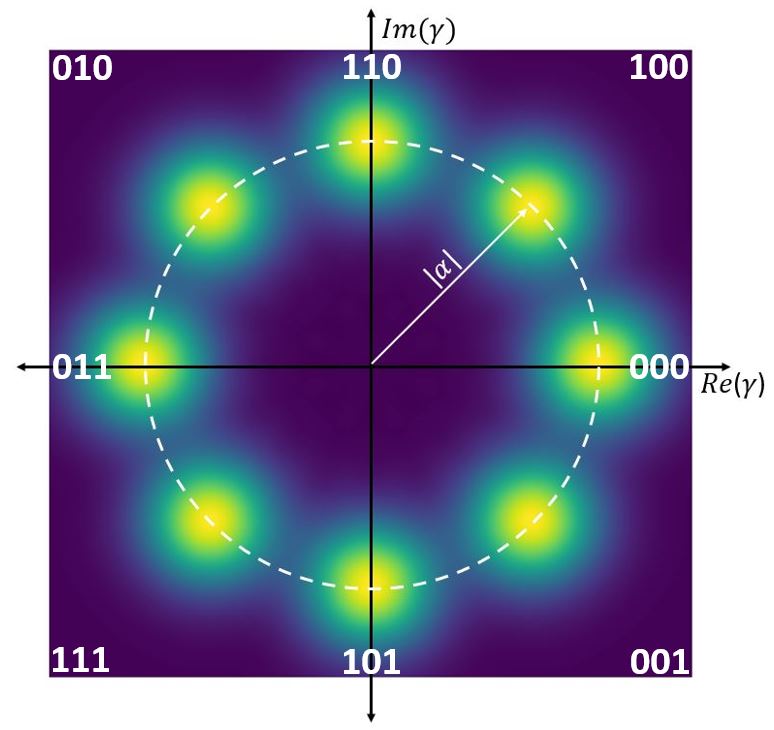}}
\makeatletter\long\def\@ifdim#1#2#3{#2}\makeatother
\caption{Key maps for QKSP protocols with (a) radial and angular postselection, (b) cross-shaped postselection and (c) radial postselection for the 8PSK protocol. Bob's measurement outcomes $\gamma \in \mathbb{C}$ that are in one of the blue--shaded areas are dropped (i.e., are assigned the symbol $\perp$). The remaining outcomes are assigned bit values that are associated with the corresponding areas. $\Delta_r$ is the radial and $\Delta_a$ is the angular postselection parameter, and $\Delta_c$ denotes the parameter for the cross-shaped postselection strategy.}
\end{figure}

\begin{figure}
\subfloat[\label{fig:radial_angular_PS} QPSK radial and angular postselection]{
\includegraphics[width=0.35\textwidth]{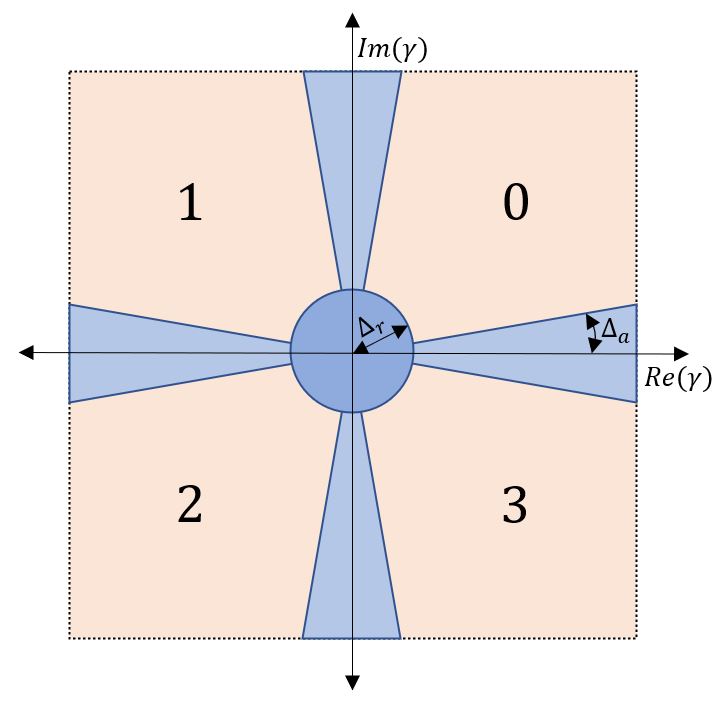}}\\
\subfloat[\label{fig:cross_PS} QPSK cross-shaped postselection]{
\includegraphics[width=0.35\textwidth]{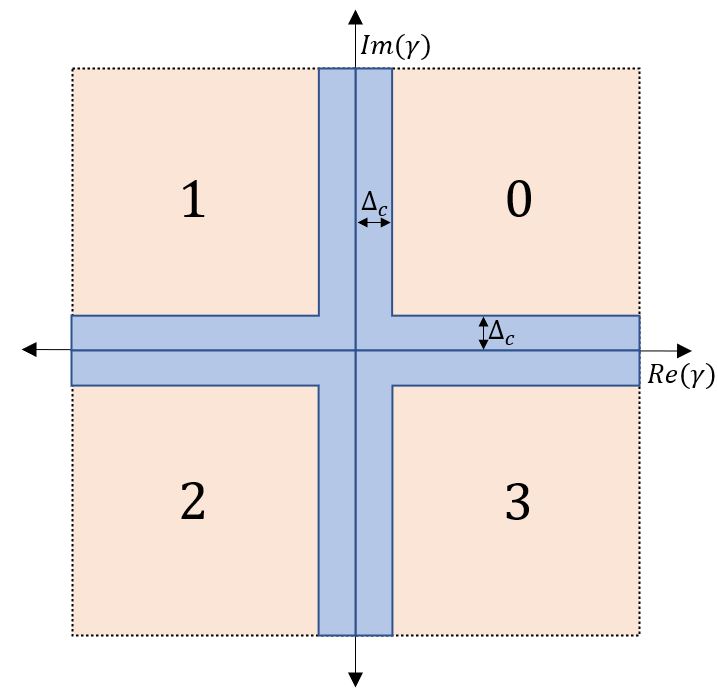}}\\
\subfloat[\label{fig:key_map8PSK} 8PSK radial postselection]{
\includegraphics[width=0.35\textwidth]{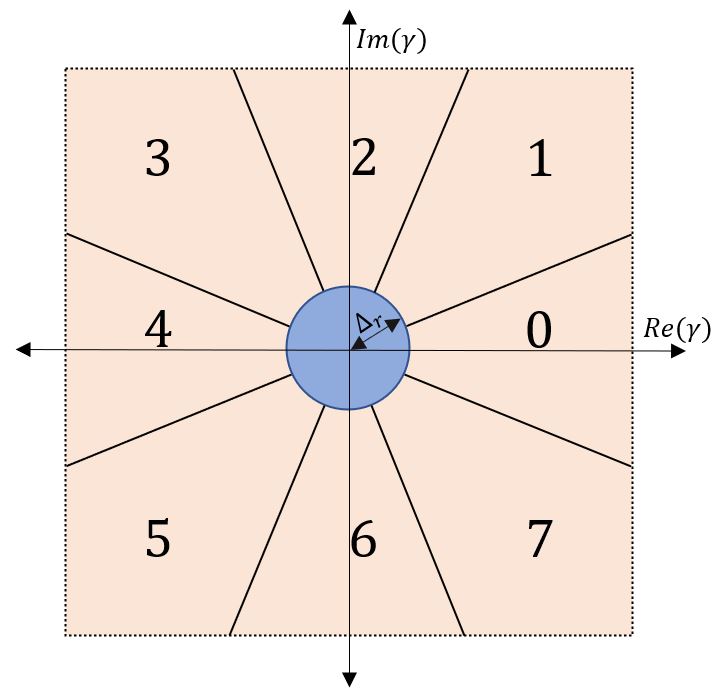}}
\makeatletter\long\def\@ifdim#1#2#3{#2}\makeatother
\caption{Key maps for QPSK protocols with a) radial and angular postselection and b) cross-shaped postselection and c) for radial postselection for the 8PSK protocol. Bob's measurement outcomes $\gamma \in\mathbb{C}$ that are in one of the blue-shaded areas are dropped, i.e., are assigned to the symbol $\perp$. The remaining outcomes are assigned to the bit-values that are associated with the corresponding areas. $\Delta_r$ is the radial- and $\Delta_a$ is the angular postselection parameter and $\Delta_c$ denotes the parameter for the cross-shaped postselection strategy. \label{fig:PS_sketches} }
\end{figure}

\subsubsection{Description of protocols with postselection}
For protocols with postselection, we proceed with step 4*) of the protocol description above. Based on our considerations in the previous subsections, in the rest of the paper we focus on protocols with four or eight signal states (see Figure~\ref{fig:phase_space_both} for a sketch in phase space). As outlined above, choosing postselection regions bounded exactly by a contour line of the bit error probability is not feasible. Instead, we consider postselection strategies that discard signals close to the (symmetry) axes of the protocol. We consider the following three different postselection strategies:
\begin{itemize}
\item[a) ] \textbf{QPSK radial and angular Postselection (raPS):} Fix some $0 \leq \Delta_r \in \mathbb{R}$ and $0 \leq \Delta_a \in \mathbb{R}$ and determine Bob's key string according to Figure~\ref{fig:radial_angular_PS}. Note that radial postselection (rPS) is the special case where we set $\Delta_a = 0$.

\item[b) ] \textbf{QPSK cross-shaped Postselection (cPS):} Fix some $0 \leq \Delta_c \in \mathbb{R}$ and determine Bob's key string according to Figure~\ref{fig:cross_PS}.
        
\item[c) ] \textbf{8PSK radial Postselection (8rPS):} Fix some $0 \leq \Delta_r \in \mathbb{R}$. Bob obtains his key string as shown in Figure~\ref{fig:key_map8PSK}.
\end{itemize}
The corresponding areas associated with the logical symbols are illustrated in Figure~\ref{fig:PS_sketches}. We note that initially we also examined angular postselection for 8PSK, but we did not observe an increase in the secure key rate.
A more formal description of the postselection areas can be found in Appendix \ref{sec:postprocessing}.
Setting $\Delta_r = \Delta_a = 0$ in a) or $\Delta_c = 0$ in b) results in the same protocol \textbf{QPSK  without postselection (noPS)}. Likewise, setting in c) $\Delta_r =0$ results in the protocol \textbf{8PSK without postselection (8noPS)}. Furthermore, we note that, in contrast to \cite{Lin_2019}, we rotated the signal states for QPSK protocols by $\pi/4$ in the $p$,$q$-plane such that they are not located on the axes but on the diagonals. This is in accordance with the constellation of the signal states in classical QPSK schemes and makes it easier to consider cross-shaped postselection. As outlined in the protocol description, postselection can be introduced by modifying the key map. Mathematically, the key map is included in the postprocessing map $\mathcal{G}$, which is part of the objective function of the present optimization problem (see Equation (\ref{eq:SDP})). In the upcoming section, we give analytical expressions for the so-called region operators, which represent the key map within the frame of the present security proof method. For details about how exactly the region operators are used to calculate secure key rates, we refer to Appendix~\ref{APDX:Details_Implementation}.

\subsection{Region operators for the untrusted detector}\label{sec:RO_trusted}
We start with the untrusted detector scenario for QPSK protocols. As Bob performs heterodyne measurement of the incident quantum states, the positive operator-valued measure (POVM) of an ideal, untrusted homodyne detector has the form $\{E_{\gamma} = \frac{1}{\pi} |\gamma\rangle \langle \gamma | ~:\gamma \in \mathbb{C} \}$ \cite{Sanders_2004}.
Then, the measurement operators, called region operators, corresponding to the symbol $z=k$, $k \in \{0,1,2,3\}$, are defined by
\begin{equation}\label{eq:Rra}
    R^{\textrm{ra}}_z := \int_{A^{\textrm{ra}}_z} E_{\gamma} \,d^2\gamma= \frac{1}{\pi} \int_{A^{\textrm{ra}}_z} |\gamma\rangle\langle \gamma| \,d^2\gamma
\end{equation}
\begin{equation}\label{eq:Rc}
    R^{\textrm{c}}_z := \int_{A^{\textrm{c}}_z} E_{\gamma} \,d^2\gamma= \frac{1}{\pi} \int_{A^{\textrm{c}}_z} |\gamma\rangle\langle \gamma| \,d^2\gamma.
\end{equation}
    
As we later approximate the infinite-dimensional problem by a problem living in a finite-dimensional Fock space, we express the region operators in the number basis,
\begin{align}\label{eq:R_ra_inf_}
R_z^{\textrm{ra}} &= \sum_{n=0}^{\infty} \sum_{m=0}^{\infty} \langle n |R_z^{\textrm{ra}} |m\rangle |n\rangle\langle m|
\end{align}
\begin{align}\label{eq:R_c_inf_}
R_z^{\textrm{c}} &= \sum_{n=0}^{\infty} \sum_{m=0}^{\infty} \langle n |R_z^{\textrm{c}} |m\rangle |n\rangle\langle m|.
\end{align}
Then, once it comes to numerical treatment, we may replace the upper limit in the sums by the cutoff number $N_c$ (see Section~\ref{sec:proof_approach}) . It remains to find expressions for the matrix elements $\langle n |R_z^{\textrm{ra}} |m\rangle$ in Equation (\ref{eq:R_ra_inf_}) and $\langle n |R_z^{\textrm{c}} |m\rangle$ in Equation (\ref{eq:R_c_inf_}). We show in Appendix \ref{apdx:untrusted} that they can be calculated analytically and have the form 

\begin{widetext}

\begin{align}
\langle n |R_z^{\textrm{ra}} |m\rangle &= \left\{
\begin{array}{ll}
\frac{\Gamma\left(n+1, \Delta_r^2 \right) }{\pi (n!)} \left(\frac{\pi}{4} - \Delta_a\right), & n = m, \\
\frac{\Gamma\left(\frac{m+n}{2}+1, \Delta_r^2 \right) }{\pi(m-n)\sqrt{n!}\sqrt{m!}} e^{-i(m-n)\left(z+\frac{1}{2} \right)\frac{\pi}{2}} \sin\left[\left(\frac{\pi}{4}-\Delta_a \right)(m-n) \right], & \, n \neq m, \\
\end{array}
\right.
\end{align}
\begin{align}
\langle n |R_z^{\textrm{c}} |m\rangle &= \left\{
\begin{array}{ll}
\frac{1}{4\pi (n!)} \sum\limits_{j=0}^{n} \binom{n}{j} \Gamma\left(j+\frac{1}{2}, \Delta_c^2 \right) \Gamma\left(n-j+\frac{1}{2}, \Delta_c^2\right), & n = m, \\
\frac{1}{4\pi \sqrt{n!} \sqrt{m!}}  \sum\limits_{j=0}^{n}\sum\limits_{k=0}^{m} \binom{n}{j}\binom{m}{k} \Gamma\left(\frac{j+k+1}{2}, \Delta_c^2 \right) \Gamma\left(\frac{n+m-j-k+1}{2}, \Delta_c^2 \right)  D_{j,k,m,n}^{(z)},
 & \, n \neq m, \\
\end{array}
\right.    
\end{align}

where 
\begin{equation}
    D_{j,k,m,n}^{(z)} = i^{n-m+k-j} \cdot \left\{
\begin{array}{ll}
1, & z = 0, \\
(-1)^{k-j}, & z = 1, \\
(-1)^{n-m}, & z = 2, \\
(-1)^{n-m+k-j}, & z = 3.
\end{array}
\right. 
\end{equation}

By a similar calculation with adapted angular integration, we obtain for the present 8PSK protocol
\begin{align}
&\langle n |R_z^{8ra} |m\rangle = \left\{
\begin{array}{ll}
\frac{\Gamma\left(n+1, \Delta_r^2 \right) }{ \pi (n!)} \left(\frac{\pi}{8} - \Delta_a\right), & n = m, \\
\frac{\Gamma\left(\frac{m+n}{2}+1, \Delta_r^2 \right) }{\pi (m-n)\sqrt{n!}\sqrt{m!}} e^{-i(m-n)z\frac{\pi}{4}}  \sin\left[\left(\frac{\pi}{8}-\Delta_a \right)(m-n) \right], & \, n \neq m, \\
\end{array}
\right. 
\end{align}
where now $z \in\{0,...,7\}$.
\end{widetext}

\subsection{Region operators for the trusted detector}\label{sec:RO_untrusted}
In this section we give analytical expressions for the region operators in the trusted detector scenario. Since we mainly discuss only trusted detectors for QPSK protocols, we derive the operators only for protocols with four signal states. However, the radial and angular case can be generalized easily to $N_{\textrm{St}}$ states.

Similarly, for $G_y$, the POVM corresponding to the trusted detector (see Equation\ (\ref{eq:POVM_Gy})), we need to express the region operators
\begin{equation}\label{eq:RraTr}
    R^{\textrm{ra, tr}}_z := \int_{A^{\textrm{ra}}_z} G_{\gamma} \,d^2\gamma,
\end{equation}
\begin{equation}\label{eq:RcTr}
    R^{\textrm{c, tr}}_z := \int_{A^{\textrm{c}}_z} G_{\gamma} \,d^2\gamma,
\end{equation}
for the trusted noise scenario in the number basis,
\begin{align}
    R_z^{\textrm{ra, tr}} &= \sum_{n=0}^{\infty}\sum_{m=0}^{\infty} \langle n|R_z^{\textrm{ra, tr}}|m\rangle |n\rangle \langle m| \\
    R_z^{\textrm{c, tr}} &= \sum_{n=0}^{\infty}\sum_{m=0}^{\infty}\langle n|R_z^{\textrm{c, tr}}|m\rangle |n\rangle \langle m|.
\end{align}
 We show in Appendix \ref{apdx:trusted} that the coefficients for the trusted noise scenario have the form

\begin{widetext}
\begin{align}
\langle n |R_z^{\textrm{ra, tr}} |m\rangle &= \left\{
\begin{array}{ll}
C_{n,n}\left(\frac{\pi}{4}-\Delta_a \right) \sum\limits_{j=0}^{n} \binom{n}{n-j} \frac{\Gamma(j+1, a\Delta_r^2)}{ a^{j+1} b^j  j!}, & n = m, \\
\frac{C_{n,m}}{(m-n)a^{\frac{m-n}{2}}} e^{-i(m-n)\left(z+\frac{1}{2}\right)\frac{\pi}{2}} \sin\left[(m-n)\left(\frac{\pi}{4}-\Delta_a \right) \right] \sum\limits_{j=0}^{n} \binom{m}{n-j} \frac{\Gamma\left(j+1+\frac{m-n}{2},a\Delta_r^2\right)}{ a^{j+1} b^j j!},  & \, n < m, \\
\overline{\langle m |R_z^{\textrm{ra, tr}} |n\rangle}, &\, n>m,
\end{array}
\right.
\end{align}
\begin{align}
\begin{aligned}
\langle n|& R_z^{\textrm{c, tr}}|m\rangle =  \\&\left\{
\begin{array}{ll}
 C_{n,n} \sum\limits_{j=0}^{n} \binom{n}{n-j} \frac{1}{a^{j+1}b^j j!} \sum\limits_{k=0}^{j} \binom{j}{k} \Gamma\left(k+\frac{1}{2}, a\Delta_c^2 \right)\Gamma\left(j-k+\frac{1}{2}, a\Delta_c^2 \right), & n = m, \\
\frac{C_{n,m}}{4 a^{\frac{m-n}{2}}} \sum\limits_{j=0}^{n} \binom{m}{n-j} \frac{1}{a^{j+1} b^j j!} \sum\limits_{k=0}^{m-n} \binom{m-n}{k} D_{k,m,n}^{(z)}\sum\limits_{l=0}^{j}\binom{j}{l}  \Gamma\left(l+\frac{k+1}{2}, a\Delta_c^2 \right) \Gamma\left(j-l+\frac{m-n-k+1}{2}, a\Delta_c^2 \right), & \, n < m, \\
\overline{\langle m |R_z^{\textrm{c, tr}} |n\rangle}, &\, n>m,
\end{array}
\right.    
\end{aligned}
\end{align}

where $C_{n,m} := \frac{1}{\pi \eta_d{\frac{m-n}{2}+1}} \sqrt{\frac{n!}{m!}} \frac{\overline{n}_d^n}{(1+\overline{n}_d)^{m+1}}$, $a := \frac{1}{\eta_d (1+\overline{n}_d)}$, $b:= \eta_d \overline{n}_d(1+\overline{n}_d)$ and
\begin{equation}
    D_{k,m,n}^{(z)} = i^{m-n-k} \cdot \left\{
\begin{array}{ll}
(-1)^{m-n-k}, & z = 0, \\
(-1)^{m-n},& z = 1, \\
(-1)^{k}, & z = 2, \\
1, & z = 3.
\end{array}
\right. .
\end{equation}
\end{widetext}
We note that \cite{Lin_2020} derives an expression for nonrotated signal states (i.e., signal states lying on the axis) for the case with only radial postselection, relying on Taylor series expansion. In contrast, our result for the radial and angular-case is more general, as it additionally includes angular postselection. Furthermore, we give a direct expression that does not require Taylor series coefficients. Again, both results have been validated with numerical solutions of the relevant integrals with \textsc{MATLAB}\texttrademark, version R2020a.

Furthermore, for the sake of completeness, we give analytical expressions for the Fock-basis representation of the first- and second-moment observables $\hat{F}_Q, \hat{F}_P, \hat{S}_Q, \hat{S}_P$ in Appendix \ref{sec:observables}.

Not relying on numerical integration is not only significantly faster but also more accurate and eliminates integration errors, which have not been considered in the security analysis so far.

The main advantage of cross-shaped postselection over radial and angular postselection is its simplicity since it can be described by merely one parameter ($\Delta_c$) while the radial and angular strategy requires two parameters ($\Delta_r, \Delta_a$). Therefore, the parameter space for optimizations for radial and angular postselection is twice as large as the parameter space for cross-shaped postselection. Furthermore, depending on just one parameter, makes it easier to grasp the influence of postselection on the raw key (see Section~\ref{Sec:p_pass_examinations}).


\section{\label{sec:Implementation}Details and optimization of the implementation}
In this section, we comment briefly on the details and parameters of our implementation.
The numerical method used is explained in Section~\ref{sec:proof_approach} (and a more detailed explanation can be found in Appendix \ref{APP:Two-step-process}). In Appendix \ref{APDX:Details_Implementation} we derive analytical expressions for operators used to model the problem. We note that the usage of the analytical expressions instead of numerical solutions of the integrals is highly recommended, as (1) the time saving even for small systems is formidable, (2) the numerical precision is higher, and (3) the influence of errors due to numerical integration on the key rate has not yet been considered, representing a gap between the security proof and its numerical implementation. The coding is carried out in \textsc{MATLAB}\texttrademark, version R2020a, and we used CVX \cite{cvx1, cvx2} to model the linear SDPs that appear in step 1 and step 2 of the method and employed the MOSEK solver (Version 9.1.9) \cite{mosek}  as well as SDPT3 (Version 4.0) \cite{SDPT3_a, SDPT3_b}  to conduct the SDP optimization tasks. It turned out that the line search at the end of every (modified) Frank-Wolfe step,
\begin{equation*}
   \textrm{minimize}_{t \in [0,1]}~ f(\rho_i + t \Delta \rho),
\end{equation*}
can be solved efficiently by bisection.

We found an initial value required to start the Frank-Wolfe algorithm in two different ways. The first method utilizes the SDP solver, where we formulate the problem exactly as in Equation~(\ref{eq:SDP}) but replace the target function by $f(\rho) = 1$. Then, the SDP solver returns a density matrix that satisfies all constraints, hence lying in the feasible set $\mathcal{S}$. The second method uses a model for a two-mode Gaussian channel (see \cite{Weedbrook_2012}) with excess noise $\xi$ and transmittance $\eta$ to calculate a density matrix on Bob's side, given Alice's density matrix. The first method is faster, in particular for systems with cutoff numbers $N_c = 10$ and larger. On the other hand, the second method is numerically more stable and yields density matrices with only positive eigenvalues, even for very 'exotic' parameter regimes (e.g., very low $\xi$, very high $L$), where solver imprecisions cause slightly negative eigenvalues for the first method.

Unless mentioned otherwise, we used the cutoff number $N_c = 12$ for QPSK protocols and $N_c = 14$ for 8PSK protocols, which turned out to be an ideal compromise between accuracy and computation speed for most of the reasonable parameter inputs. A more detailed discussion of that choice can be found in Appendix \ref{APDX:choice_cutoff}. Therefore, all infinite-dimensional operators and quantities are replaced by their finite-dimensional representations in Fock spaces of size $N_c$. For example, the upper limit of the sum appearing in the Fock representation of the region operators is replaced by $N_c$. The maximal number of Frank-Wolfe steps for QPSK protocols is chosen between $N_{\textrm{FW}} = 30$ and $N_{\textrm{FW}} = 150$, and for 8PSK protocols between $N_{\textrm{FW}}=30$ and $N_{\textrm{FW}} = 200$, depending on the system parameters applied. In general, $N_{\textrm{FW}}$ is chosen as small as possible under the condition that the bound obtained for the key rate does not improve significantly for higher values of $N_{\textrm{FW}}$. We used $\epsilon_{\textrm{FW}} = 10^{-7}$ for the threshold of the Frank-Wolfe algorithm and $\tilde{\epsilon}  = 10^{-11}$ for the perturbation.

In the present work, we performed all calculations with the following model for the transmittance, $\eta = 10^{-0.02 L}$. That is a transmittance of -0.2 dB or about $95.5\%$ per kilometer, which is realistic for practical implementations. Recall, that both the excess noise $\xi$ and the electronic noise $\nu_{el}$ are measured in shot noise units. Unless mentioned otherwise, we work with a reconciliation efficiency of $\beta = 0.95$.

Secure key rates in the present paper are obtained by optimizing over $|\alpha|$ and $\Delta_r$, $\Delta_c$ or $\Delta_r$ and $\Delta_a$ respectively (depending on postselection strategy). In general, $|\alpha|$ is varied in steps of $0.05$ in the interval $|\alpha| \in [0.4, 1.2]$ for QPSK and in the interval $|\alpha| \in [0.7, 2.0]$ for 8PSK protocols. For given transmission distance, the interval can be narrowed down significantly (see discussion in Section~\ref{sec:opt_coh_amp}). The postselection parameters are varied in steps of $0.025$ in the intervals $\Delta_c \in [0,0.45]$, $\Delta_r \in [0,0.70]$ and $\Delta_a \in [0,0.35]$, unless mentioned otherwise. 

In Appendix \ref{sec:Th_Calc}, we validate our implementation by comparing it with analytical results for loss-only channels. We observe that steps 1 and 2 (hence, the upper and lower bounds obtained on the secure key rate) are separated only by negligible gaps. Therefore, in the rest of the work we omit the curves for step 1 and plot only step 2, which is the relevant (lower) bound, for the sake of clarity.

\begin{figure}
\includegraphics[width=0.48\textwidth]{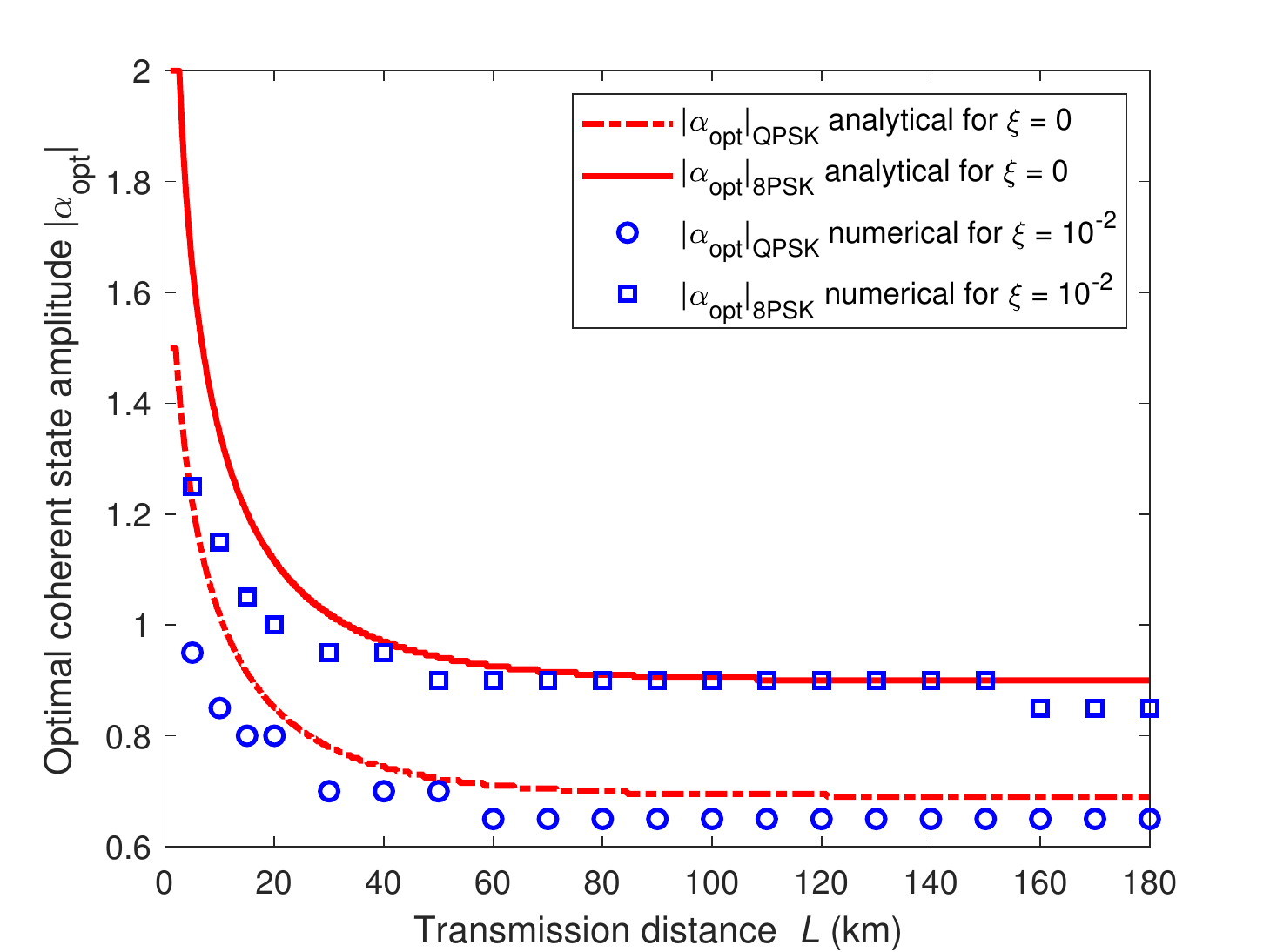}
\makeatletter\long\def\@ifdim#1#2#3{#2}\makeatother
\caption{\label{fig:alpha_th_vs_num} Optimal choice of the coherent state amplitude $|\alpha|$ for $\xi > 0$ obtained by coarse-grained search compared with predicted optimal choice of loss-only channel for QPSK and 8PSK protocol. As our results for $\xi = 0.01$ and $\xi > 0.01$ do not differ significantly, we plot only the data points for $\xi = 0.01$ to improve clarity. In what follows, we use these optimal values for $|\alpha|$.}
\end{figure}

\section{Results\label{sec:Results}}
Recall that throughout the paper we work with a reconciliation efficiency of $\beta = 0.95$, unless mentioned otherwise, and measure the excess noise $\xi$ in shot noise units.
\subsection{Optimal coherent state amplitude}\label{sec:opt_coh_amp}
Before we investigate the influence of postselection, we enquire about the influence of $|\alpha|$ on the secure key rate, as the optimal choice of the postselection parameter might heavily depend on the chosen $|\alpha|$. In Figure~\ref{fig:alpha_th_vs_num} we investigate the optimal choice of the coherent state amplitude $|\alpha|$ for both QPSK and 8PSK protocols and an excess noise level of $\xi = 0.01$ and compare the results with the analytical prediction for $\xi = 0$ (see Appendix~\ref{sec:Th_Calc} for details). We note that, according to our observations, the optimal coherent state amplitudes for $\xi = 0.02$ do not differ significantly from those for $\xi = 0.01$. We examined transmission distances up to $180$km for QPSK and up to $250$km for 8PSK. Since in the latter case both the values of the analytical prediction and the results of our numerical investigation remain constant for transmission distances higher than $80$km and $160$km respectively, we omit the part of the plot exceeding $180$km. As expected, the optimal coherent state amplitude for noisy channels is slightly lower than that for a loss-only channel. We observe that the optimal choice for the coherent state amplitude decreases with increasing transmission distance, hence with increasing losses. This is in accordance with our expectations, as for high channel losses, Eve can theoretically receive a much stronger signal than Bob (Eve is assumed to extract Alice's signal right after leaving her lab). Hence, the amplitude has to be small for high transmission distances to keep Eve's advantage as small as possible.

The QPSK-values for $20,~50,~80$ and $100$km match the values reported by \cite{Lin_2019} for a similar protocol with rotated signal states. For example, they report optimal values of about $0.78$ for $20$km and $0.66$ for $80$km while we obtained $0.80$ and $0.65$, respectively. Within the frame of the accuracy of our coarse-grained search with steps of $\Delta_{|\alpha|} = 0.05$ this coincides with our results. Furthermore, we observe that there are only minor differences between the optimal values found for different values of excess noise. Therefore, we can later limit the search for $\left|\alpha_{opt}\right|$ to a restricted interval around the optimal coherent state amplitude, obtained from the noiseless case.


\subsection{\label{sec:Main}Postselection strategies in the untrusted detector scenario}
In this section we present numerical results and findings for the so-called untrusted detector scenario in which Alice and Bob attribute any detector imperfection to the channel which is under Eve's control. 
Without loss of generality we assume in this section ideal detectors with efficiency $\eta_d=1$ and zero electronic noise. However, untrusted, \emph{nonideal detectors} can easily be modeled by multiplying the channel transmission by the detector efficiency $\eta_d$, and adding the detector noise to the channel noise. Therefore, for example, curves in key rate versus transmission distance plots will be shifted to the left for nonideal detectors compared with ideal detectors. 

\subsubsection{Secure key rates for QPSK and 8PSK}\label{sec:PS_Strategies}
In this section, we examine optimal postselection strategies for QPSK and 8PSK protocols and compare their performance for transmission distances up to $200$km. Our numerical examinations showed that the optimal coherent state amplitudes obtained without performing postselection (see Section~\ref{sec:opt_coh_amp}) remain optimal or very close to optimal with nonzero postselection parameters. Therefore, all data points in this section represent key rates optimized over the coherent state amplitude $\alpha$ and the postselection parameter(s) corresponding to the chosen postselection strategy. The optimizations over the postselection parameters are carried out as fine-grained searches in steps of $0.025$.

\begin{figure}
\includegraphics[width=0.48\textwidth]{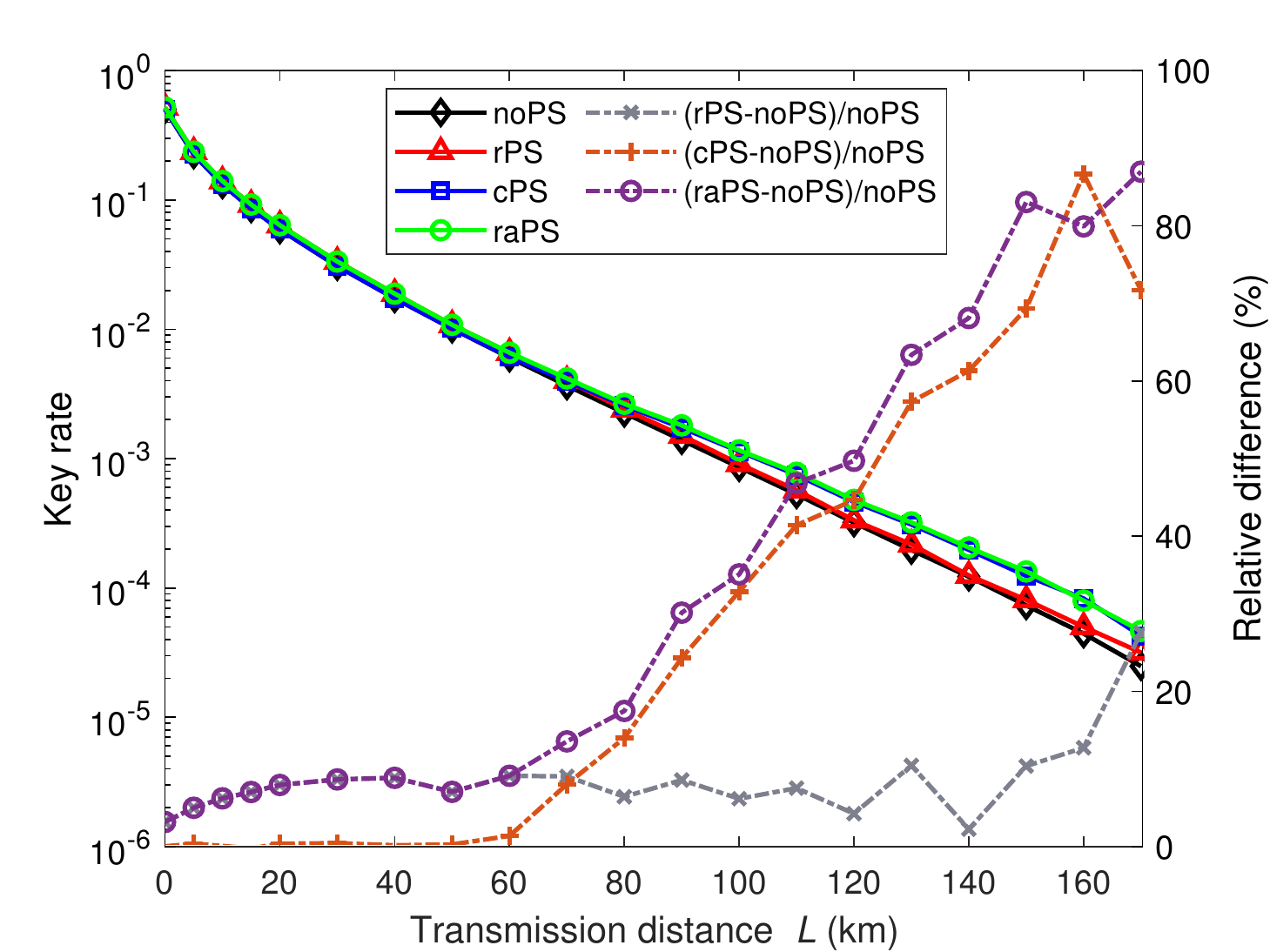}
\makeatletter\long\def\@ifdim#1#2#3{#2}\makeatother
\caption{\label{fig:comp_ps_strategies} Secure key rates for the untrusted detector scenario without postselection (reference curve), as well as for radial, cross-shaped, and radial and angular postselection for $\xi = 0.01$. The noPS curve is equivalent to the results in \cite{Lin_2019}. 
Furthermore, we plotted relative differences (right $y$ axis) between the secure key rates obtained with different postselection strategies and secure key rates obtained without performing postselection.}
\end{figure}

We first examine four-state protocols. In Figure~\ref{fig:comp_ps_strategies} we plot the calculated (lower bounds on the) secure key rate for three different postselection strategies (rPS, cPS, raPS). Recall that radial postselection (rPS) is the special case of radial and angular postselection (raPS) where $\Delta_a = 0$. Therefore, the key rates obtained by radial postselection are always lower or equal to the key rates obtained by radial and angular postselection. However, since radial postselection is the best-known postselection strategy for QPSK protocols, we plot the curves for radial postselection separately to enable better comparison and to highlight the outperformance of radial and angular postselection and cross-shaped postselection over radial postselection. On the secondary (right) $y$ axis, we plot the relative improvements of the examined postselection strategies compared with the secure key rates obtained without performing postselection (noPS). Note that the (black) noPS curve is equivalent to the secure key rates without postselection reported in \cite{Lin_2019} for the same four-state protocol with both signal states and key map rotated by $\pi/4$. It can be observed that performing radial postselection improves the secure key rate only slightly by about $10\%$, while radial and angular postselection performs similar to radial postselection up to distances of $60$km. For longer transmission distances, radial and angular postselection shows a clear outperformance which increases with increasing distance, leading to a relative improvement of $88\%$ at $170$km for the radial and angular strategy compared with the no postselection scenario. For distances less than $60$km the optimal angular postselection parameter $\Delta_a$ is zero, which explains why there is no improvement compared with radial postselection for short distances. Finally, the cross-shaped postselection strategy does not improve the secure key rates for distances up to $50$km (as the optimal cross-shaped postselection parameter $\Delta_c = 0$) and performs comparably to the radial and angular scheme for longer transmission distances. 

\begin{figure}
\subfloat[$\beta = 0.90$\label{fig:8PSK_QPSK_noPS_beta_090}]{
    \includegraphics[width=0.48\textwidth]{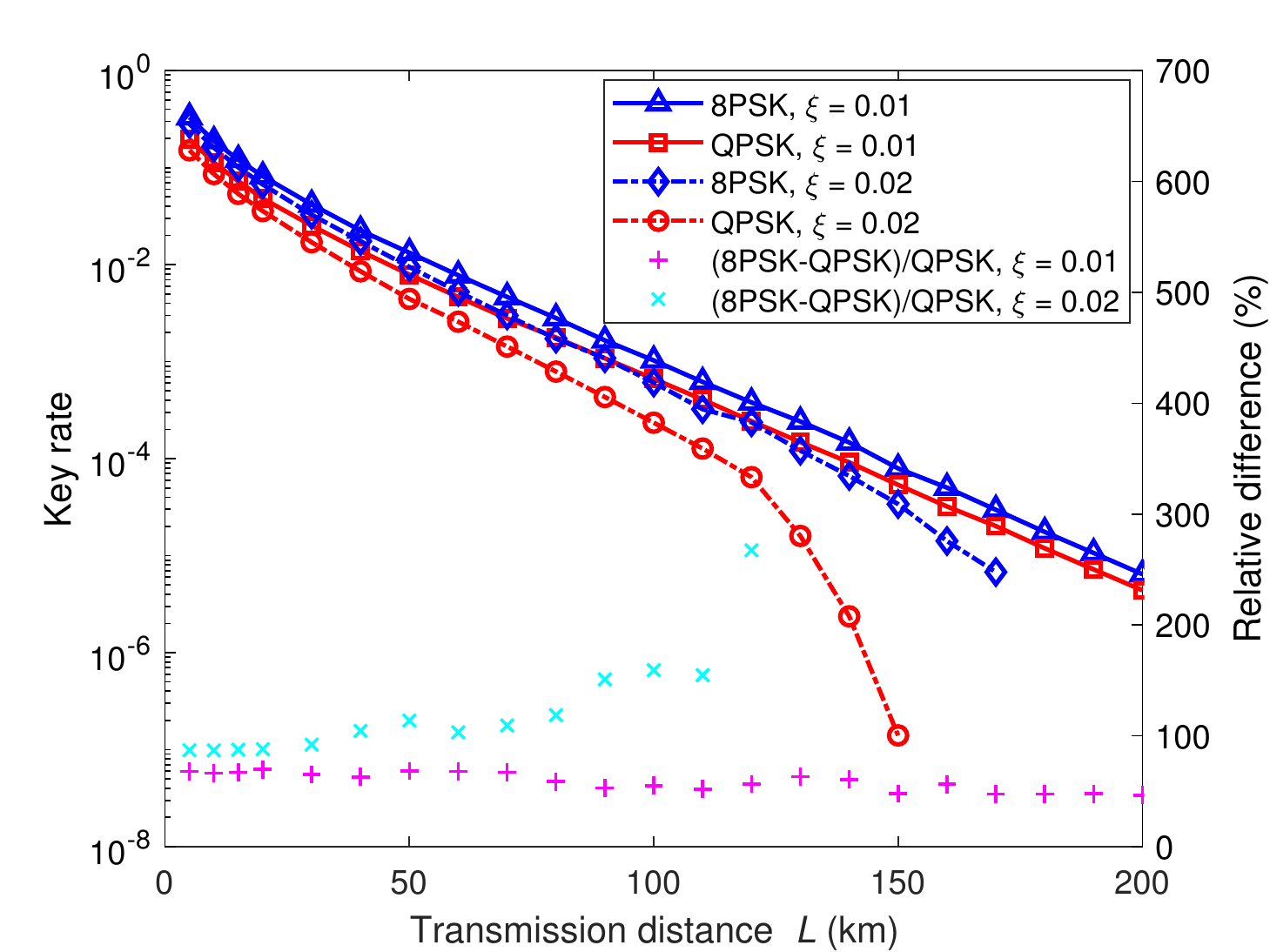}}\\
\subfloat[$\beta = 0.95$\label{fig:8PSK_QPSK_noPS_beta_095}]{
    \includegraphics[width=0.48\textwidth]{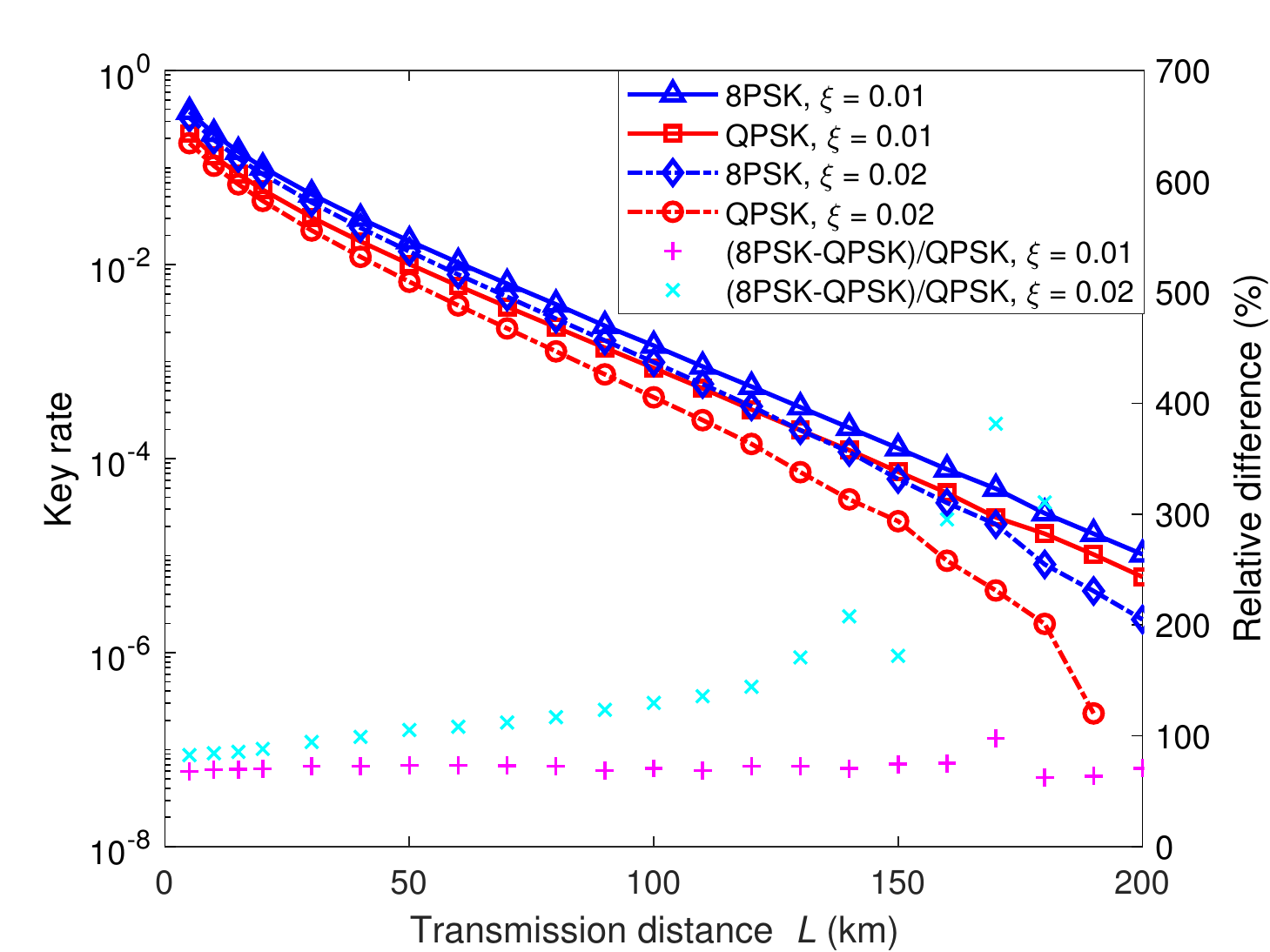}}
    \makeatletter\long\def\@ifdim#1#2#3{#2}\makeatother
\caption{\label{fig:8PSK_QPSK_no_PS} Comparison of the secure key rates for the 8PSK and QPSK protocol without postselection for $\xi = 0.01$ and $\xi = 0.02$ and two different values for the reconciliation efficiency $\beta \in \{0.90, 0.95\}$. 
The (red) QPSK curves in (b) are equivalent to the results in \cite{Lin_2019}.
}
\end{figure}

Next we study how the picture changes when we increase the number of states from four to eight.
In Figure~\ref{fig:8PSK_QPSK_no_PS}, we compare the secure key rates obtained for the 8PSK protocol with those for the QPSK protocol for two different values of excess noise ($\xi = 0.01$ and $\xi = 0.02$) and without performing postselection. In Figure~\ref{fig:8PSK_QPSK_noPS_beta_090} we chose the reconciliation efficiency $\beta$ to be $0.90$, while in Figure~\ref{fig:8PSK_QPSK_noPS_beta_095} $\beta$ is $0.95$. The results for  noisy channels confirm our observations for loss-only channels in Figure~ \ref{fig:N_states_incl_Gauss} as the secure key rates for the 8PSK protocol in all scenarios are clearly higher than those for the QPSK protocol. The relative improvement for $\xi = 0.01$ and $\beta = 0.95$ is between $60\%$ and $95\%$ (depending on the transmission distance) while for $\xi = 0.01$ and $\beta = 0.90$ the relative differences are between $45\%$ and $70\%$. The advantage of 8PSK increases even more for higher values of excess noise, as the secure key rates for QPSK begin to drop steeply at $160$km (for $\beta = 0.95$) and at $130$km (for $\beta = 0.90$) while the secure key rates for the 8PSK protocol remain stable. Additionally, this results in longer achievable maximal transmission distances with the 8PSK protocol than with the QPSK protocol for those cases where the QPSK key rates drop.

Similar to the four-state protocol, the secure key rates for the 8PSK protocol can be improved further by applying additional postselection. For 8PSK protocols our considerations focus on the radial scheme. This is because we selectively examined the influence of additional angular postselection and did not observe a significant impact for eight signal states. Since the computational effort of our numerical method is already very high for the eight-state protocol, we therefore did not investigate angular postselection parameters $\Delta_a > 0$ for all data points. 

\begin{figure}
\includegraphics[width=0.48\textwidth]{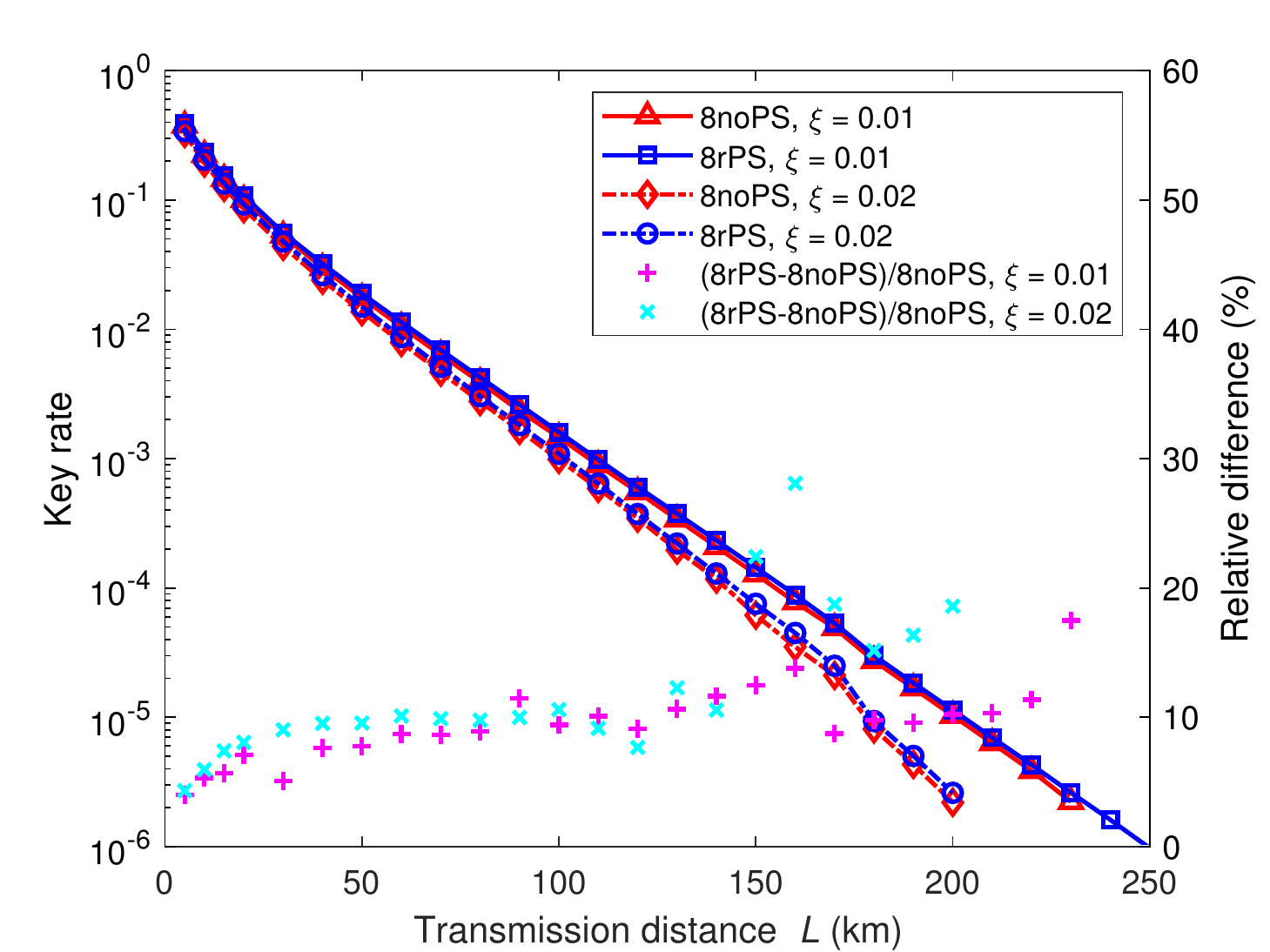}
\makeatletter\long\def\@ifdim#1#2#3{#2}\makeatother
\caption{\label{fig:8PSK_noPS_rPS_examination} Comparison of the secure key rates obtained for the 8PSK protocol with radial postselection (8rPS) and without postselection (8noPS) for $\xi = 0.01$ and $\beta = 0.95$.}
\end{figure}

In Figure~\ref{fig:8PSK_noPS_rPS_examination}, we plot the secure key rates obtained with radial postselection (8rPS) and without postselection (8noPS) for two different values of excess noise ($\xi = 0.01$ and $\xi = 0.02$) for fixed reconciliation efficiency of $\beta = 0.95$. The secondary $y$ axis represents the relative difference between the secure key rates with and without postselection for fixed excess noise. We observe a moderate improvement in the medium to long single-digit percent range for short transmission distances and up to $28\%$ for medium to long transmission distances (and $\xi = 0.01$). Furthermore, it is remarkable that the protocol with 8rPS is able to generate nonzero secure key rates for transmission distances up to at least $250$km. 

\begin{figure}
\includegraphics[width=0.48\textwidth]{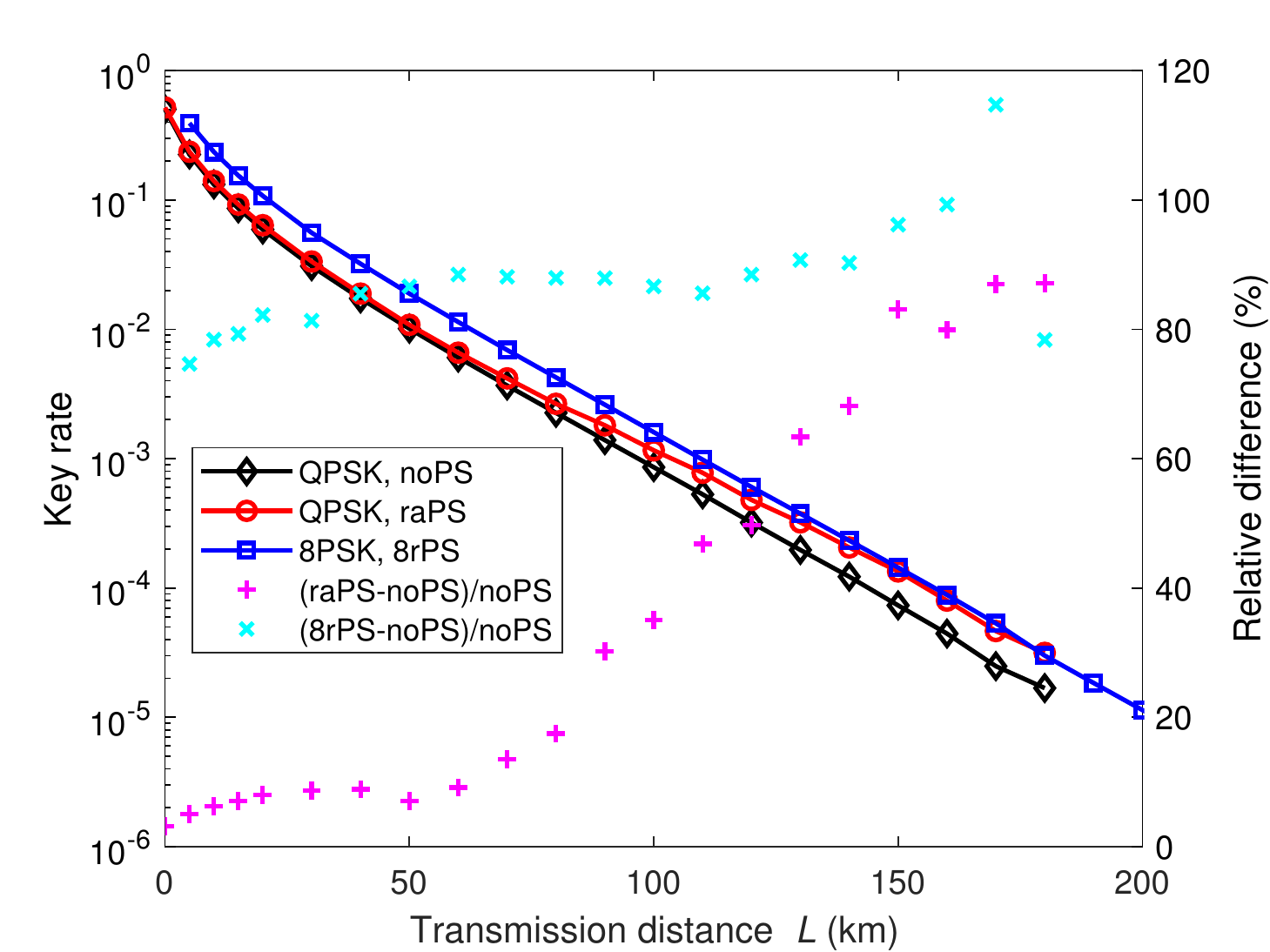}
\makeatletter\long\def\@ifdim#1#2#3{#2}\makeatother
\caption{\label{fig:8PSK_QPSK_best_strategies} Comparison of the best postselection strategies for both modulation schemes for $\xi = 0.01$ and $\beta = 0.95$. Note that the (black) QPSK key rate curve without postselection is equivalent to the results in \cite{Lin_2019}.}
\end{figure}

Finally, we summarize the results of our investigations in Figure~\ref{fig:8PSK_QPSK_best_strategies}, where we plot the best postselection strategy for four-state protocols, which is radial and angular postselection (raPS), and the best postselection strategy for eight-state protocols, which is radial postselection (8rPS). For reference, we also plot the curve representing the achievable secure key rate for the four-state protocol obtained without performing postselection. Note that, again, this curve corresponds to the results reported in \cite{Lin_2019} for a rotated version of the four-state protocol examined. Therefore, the relative differences plotted in Figure~\ref{fig:8PSK_QPSK_best_strategies} display the improvements achieved in the present work. As described earlier, performing radial and angular postselection for the four-state protocol increases the secure key rate considerably, compared with not performing any postselection, where the advantage increases with increasing transmission distances, peaking at an outperformance of $88\%$ for $170$km and $180$km. Recall that the cross-shaped postselection strategy performs comparably, in particular for medium to long transmission distances. However, for clarity, we plotted only the results for radial and angular postselection. The secure key rates for eight signal states with radial postselection are between $80\%$ and $100\%$ higher than those for four signal states without performing postselection. We observe that the relative advantage of 8PSK with radial postselection over QPSK without postselection remains approximately stable over the examined range for the transmission distance $L$, while the advantage of QPSK with radial and angular postselection over QPSK without postselection increases with $L$. For distances greater than $140$km QPSK with radial and angular postselection (as well as QPSK with cross-shaped postselection) performs comparably to 8PSK with radial postselection. Thus, for high transmission distances, the advantage of a higher number of signal states can be compensated by a suitable postselection strategy.
This is relevant both from a theoretical and an applied point of view: On the one hand, calculating the secure key rates for a higher number of signal states is computationally costly. On the other hand, preparing coherent states which differ in phase by a smaller angle with high accuracy is experimentally more challenging. Both tasks can be circumvented by introducing radial and angular (or cross-shaped) postselection to a four-state protocol
which requires only minor software adaptations and can be implemented easily.

\subsubsection{Dependency of the secure key rate on the probability of passing the postselection step}\label{Sec:p_pass_examinations}
In step 5) of the protocol, Alice and Bob perform error correction to reconcile their raw keys and obtain keys which are identical. This task, in general, is complicated and computationally expensive and therefore often a bottleneck in many practical implementations.

\begin{figure}
\subfloat[\label{fig:ppass_vs_keyrate_50km} $L=50$km]{
\includegraphics[width=0.48\textwidth]{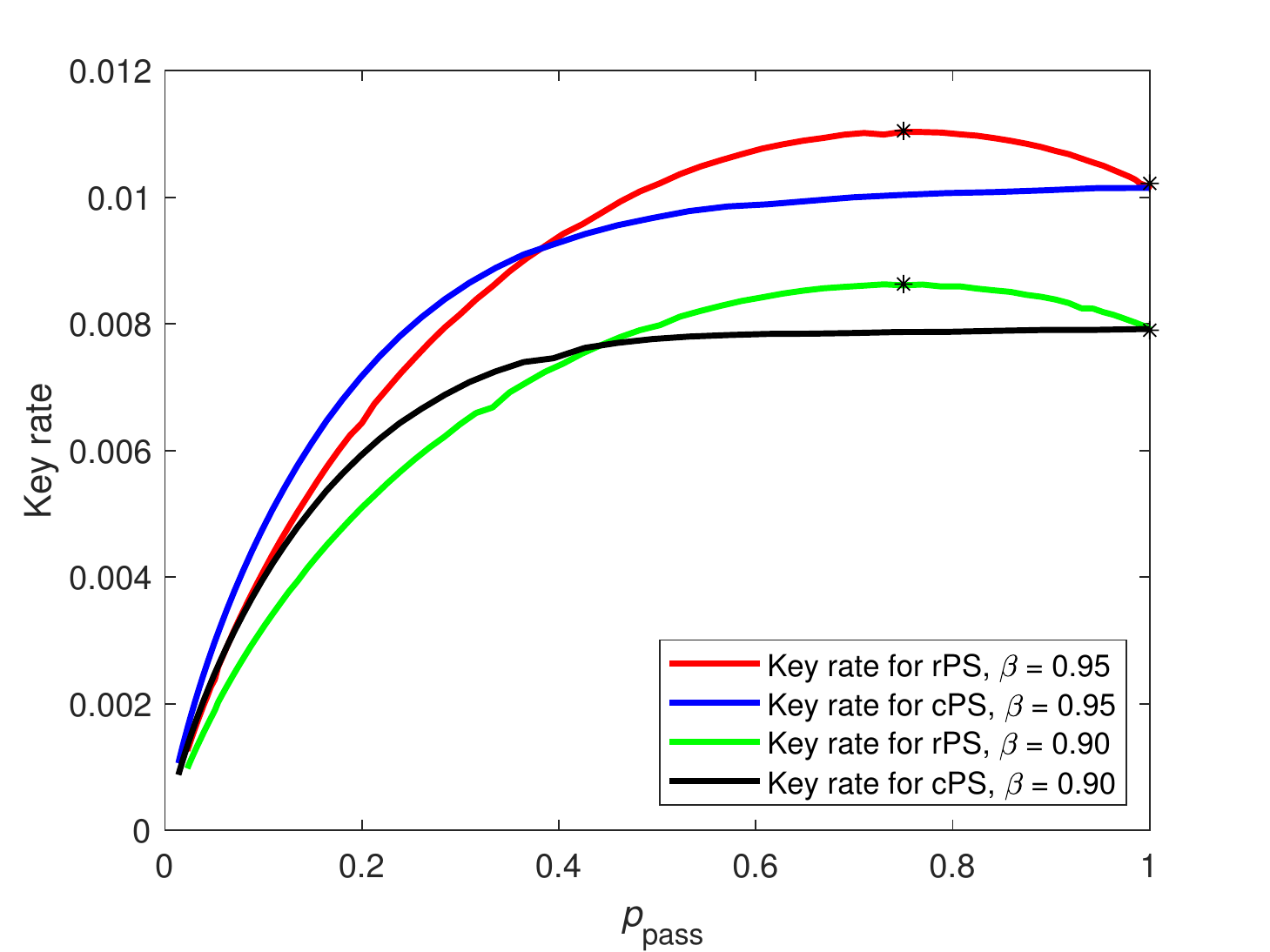}}\\
\subfloat[\label{fig:ppass_vs_keyrate_100km} $L= 100km$]{
\includegraphics[width=0.48\textwidth]{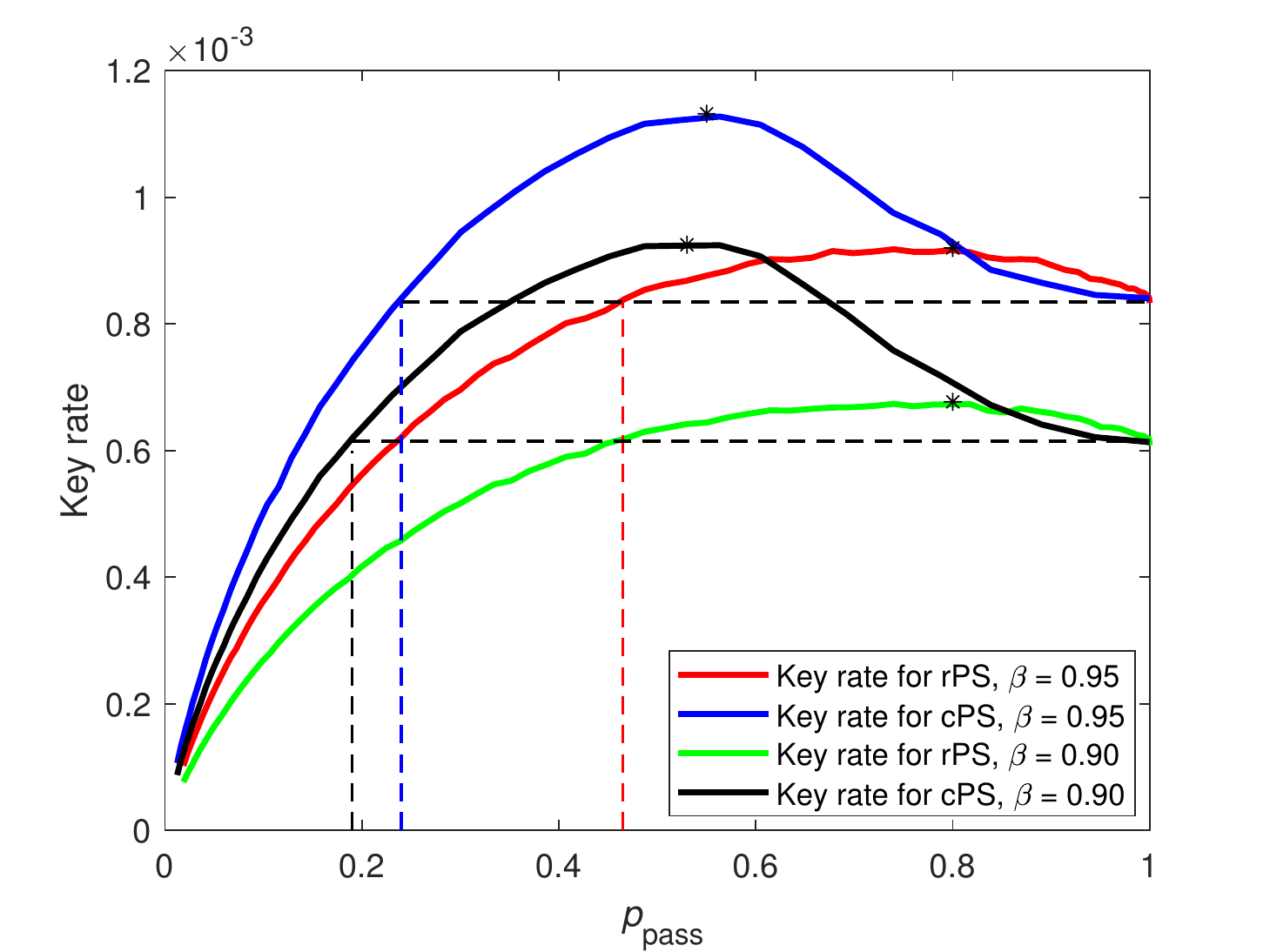}}
\makeatletter\long\def\@ifdim#1#2#3{#2}\makeatother
\caption{\label{fig:p_pass_vs_keyrate_rPS_cPS} Secure key rates versus probability of passing the postselection phase for radial and cross-shaped postselection schemes for QPSK with $|\alpha| = 0.7$. The excess noise is set to $\xi = 0.01$ and we plot curves for $\beta = 0.95$ and $\beta = 0.90$. Note that $p_{\textrm{pass}}=1$ indicates the case without postselection.}
\end{figure}

To address this issue, we examine the influence of postselection on the fraction of the raw key which passes the postselection phase (i.e., the fraction of the raw key which has to be error-corrected). We begin with the QPSK protocol, where we fix the excess noise at $\xi = 0.01$ and the coherent state amplitude $|\alpha| = 0.70$ and examine transmission distances of $L=50$km and $L=100$km. Note that, according to Figure~\ref{fig:alpha_th_vs_num}, $|\alpha| = 0.70$ is the optimal value for $50$km and very close to the optimal choice for $100$km. We compare radial postselection and cross-shaped postselection since both strategies depend merely on one postselection parameter (in contrast to radial and angular postselection, which depends on $\Delta_r$ and $\Delta_a$). We varied the postselection parameters in the intervals $\Delta_r \in [0,2]$ and $\Delta_c \in [0, 1.125]$ in steps of size $0.025$ and plotted the secure key rate achieved against the probability of passing the postselection phase $p_{\textrm{pass}}$. Note that $1-p_{\textrm{pass}}$ corresponds to the fraction of the raw key which is removed by the postselection procedure. In Figure~\ref{fig:p_pass_vs_keyrate_rPS_cPS} we plot our results for two different values of the reconciliation efficiency $\beta \in \{0.90,  0.95\}$.

For $L=50$km (see Figure~\ref{fig:ppass_vs_keyrate_50km}), we observe that the maximal secure key rate is achieved with radial postselection (which confirms our earlier results that for short distances rPS yields slightly higher key rates than cPS) at $p_{\textrm{pass}} = 0.75$, while the cross-shaped postselection strategy increases monotonically, reaching its maximum at $p_{\textrm{pass}} = 1$, that is, for the case without postselection (which, again, confirms our earlier results in  Section~\ref{sec:PS_Strategies} that the optimal choice at $50$km is $\Delta_c = 0$). For $p_{\textrm{pass}} \gtrsim 40\%$ radial postselection yields slightly higher secure key rates than the cross-shaped scheme .

We discover the opposite for $L=100$km (see Figure~\ref{fig:ppass_vs_keyrate_100km}), where the cross-shaped postselection strategy yields higher secure key rates than the radial scheme, which, again, is in accordance with our earlier results in  Section~\ref{sec:PS_Strategies}. The cross-shaped strategy obtains its maximum at $p_{\textrm{pass}} \approx 0.55$, while the (much lower) maximum of the radial postselection scheme is obtained at $p_{\textrm{pass}} = 0.80$. Note, that using the cross-shaped postselection strategy increases the secure key rates by about $35\%$ although the raw key is reduced by almost $50\%$ (i.e., the part of the raw key which has to be error-corrected is halved compared with the protocol without postselection)! This shows the clear advantage of the cross-shaped over the radial postselection scheme, where the raw key is only reduced by $20\%$, while the secure key rate increases by merely $10\%$. 

Assume we aim to obtain the same secure key rate as without performing postselection (corresponding to $p_{\textrm{pass}} = 1)$ while removing as much raw key as possible. For $L=100$km, performing radial postselection can remove about $53\%$ of the raw key without decreasing the secure key rate while with cross-shaped postseletion $77\%$ of the raw key can be removed such that merely $23\%$ of the raw key need to be error corrected (see dashed lines in Figure~\ref{fig:ppass_vs_keyrate_100km}).

With the aim of reducing the data that has to be error-corrected drastically, this idea can be taken even further. To reduce the raw key by, for example, $80\%$ (or even more) the cross-shaped postselection scheme yields higher key rates than the radial scheme for both transmission distances examined, $L=50$km and $L=100$km. For $100$km and cross-shaped postselection, the secure key rates obtained are even almost equal to those obtained without postselection, while the key rates for the radial scheme are clearly lower. Since, according to earlier examinations, the cross-shaped strategy remains superior for higher transmission distances, we expect similar results for all $L\geq 50$km. This shows the clear advantage of cross-shaped postselection, in particular for medium to long transmission distances, and the potential application to reduce the data that has to be error-corrected. 

\begin{figure}
\subfloat[QPSK, $|\alpha| = 0.70$.\label{fig:p_pass_examination_QPSK_rps_95}]{
    \includegraphics[width=0.48\textwidth]{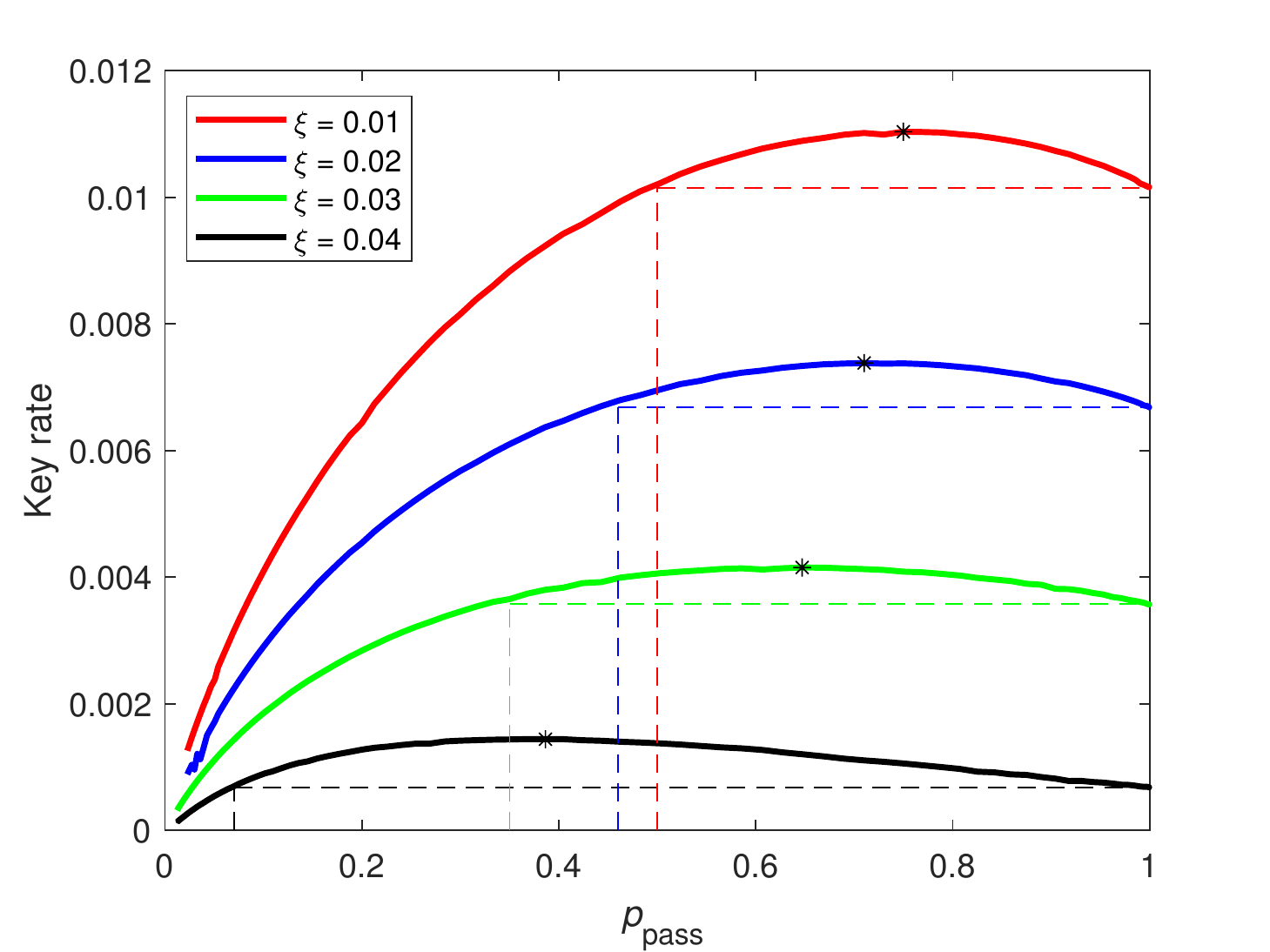}}\\
\subfloat[8PSK, $|\alpha| = 0.90$. \label{fig:p_pass_examination_8PSK_rps_95}]{
    \includegraphics[width=0.48\textwidth]{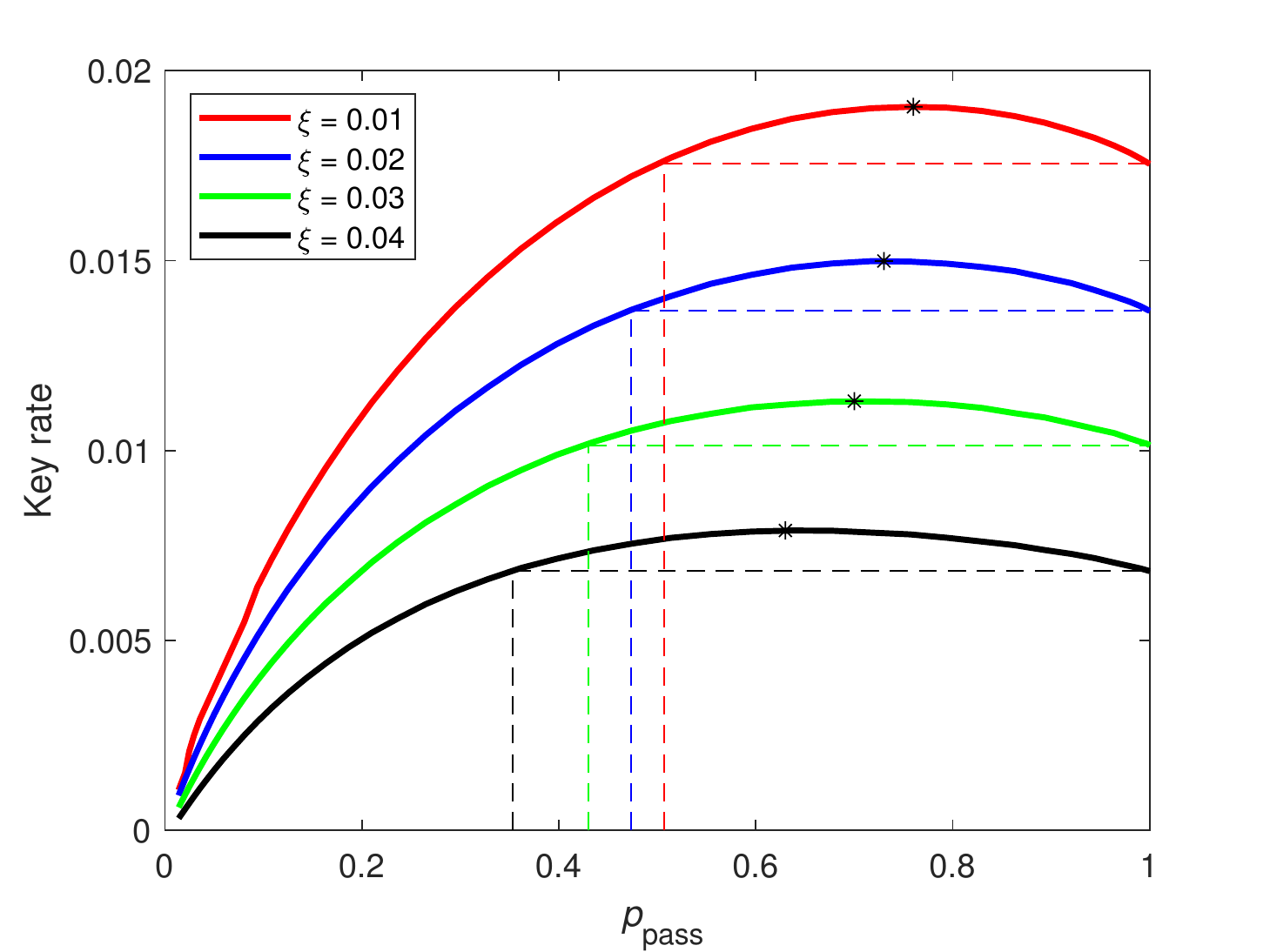}}\\
\makeatletter\long\def\@ifdim#1#2#3{#2}\makeatother    
\caption{Secure key rate versus the probability of passing the postselection phase $p_{pass}$ for radial postselection and four different values of excess noise and a fixed transmission distance of $L=50$km. The underlying data are calculated by varying the postselection parameter $\Delta_r$ in the interval $[0,2.15]$ with a step size of $0.025$. The reconciliation efficiency is fixed to $\beta = 0.95$. \label{fig:p_pass_examination_rPS}}
\end{figure}

For the 8PSK protocol and radial postselection, we investigated the secure key rates for $L=50$km and $|\alpha| = 0.90$ (which is the optimal choice for $50$km) for four different values of excess noise $\xi \in \{0.01, 0.02, 0.03, 0.04\}$ and two different values for the reconciliation efficiency $\beta \in \{0.90, 0.95\}$.
In Figure~\ref{fig:p_pass_examination_rPS} we plot the achievable secure key rates for $\beta = 0.95$ against the probability of passing the postselection for QPSK modulation (Figure~\ref{fig:p_pass_examination_QPSK_rps_95}) and for 8PSK modulation (Figure~\ref{fig:p_pass_examination_8PSK_rps_95}). Qualitatively similar curves are obtained for $\beta = 0.90$ (not shown). 

\begin{table}[]
\begin{tabular}[t]{|c||c|c||c|c|}
\hline 
\multicolumn{5}{|c|}{max key rate at ($p_{\textrm{pass}}$) }\\
\hline &  \multicolumn{2}{c|}{$\beta = 0.90$} & \multicolumn{2}{c|}{$\beta = 0.95$} \\ \hline
$\xi$ & QPSK             & 8PSK            & QPSK                & 8PSK                 \\ \hline \hline

0.01  & 0.0086 (0.73)    & 0.014 (0.75)    & 0.0110 (0.75)   & 0.019 (0.76) \\ \hline
0.02  & 0.0049 (0.69)    & 0.010 (0.71)    & 0.0074 (0.71)   & 0.015 (0.73) \\ \hline
0.03  & 0.0018 (0.49)    & 0.007 (0.67)    & 0.0042 (0.64)   & 0.011 (0.70) \\ \hline
0.04  & -                & 0.003 (0.51)    & 0.0014 (0.37)   & 0.079 (0.63) \\ \hline 

\end{tabular}
\makeatletter\long\def\@ifdim#1#2#3{#2}\makeatother
\caption{Maximal achievable secure key rate for QPSK and 8PSK protocols at $L=50$km with radial postselection for four different values of excess noise and two values for the reconciliation efficiency. The values of $p_{\textrm{pass}}$ at which the maximal secure key rate is obtained are given in parenthesis.}
\label{table:max_KR}

\end{table}
\begin{table}
\begin{tabular}[t]{|c||c|c||c|c|}
\hline 
\multicolumn{5}{|c|}{$p_{\textrm{pass}}$ }\\
\hline
      & \multicolumn{2}{c|}{$\beta = 0.90$ } & \multicolumn{2}{c|}{$\beta = 0.95$ } \\ \hline
$\xi$  & QPSK     & 8PSK       & QPSK     & 8PSK     \\ \hline \hline

0.01   & 0.48    & 0.50    & 0.51         & 0.51               \\ \hline
0.02   & 0.41    & 0.44    & 0.44         & 0.47               \\ \hline
0.03   & 0.17    & 0.36    & 0.32         & 0.43               \\ \hline
0.04   & -       & 0.21    & 0.07         & 0.34                                                   \\ \hline 

\end{tabular}
\makeatletter\long\def\@ifdim#1#2#3{#2}\makeatother
\caption{The values of $p_{\textrm{pass}}$ at which secure key rate when performing radial postselection is equal to the secure key rate obtained without performing postselection. We consider two different values of $\beta$ and four different values of excess noise. The transmission distance $L$ is fixed at $50$km}
\label{table:keyRate_vs_ppass}
\end{table}

\begin{table}
\begin{tabular}[t]{|c||c|c||c|c|}
\hline
\multicolumn{5}{|c|}{change of secure key rate}  \\
\hline
      & \multicolumn{2}{c|}{$\beta = 0.90$ } & \multicolumn{2}{c|}{ $\beta = 0.95$ } \\ \hline
$\xi$ & QPSK        & 8PSK    & QPSK        & 8PSK       \\ \hline\hline

0.01  & $-19\%$   & $-20\%$    & $-20\%$       & $-21\%$        \\ \hline
0.02  & $-13\%$   & $-16\%$    & $-15\%$       & $-19\%$        \\ \hline
0.03  & $+25\%$   & $-8\%$     & $-3\%$        & $-14\%$        \\ \hline
0.04  & $-$       & $+19\%$    & $+109\%$      & $-8\%$         \\ \hline
\end{tabular}
\makeatletter\long\def\@ifdim#1#2#3{#2}\makeatother
\caption{Relative change in the secure key rate when omitting $70$\% of the raw key compared with the secure key rate obtained without performing postselection for two different values of $\beta$, four different values of excess noise and $L=50$km.}
\label{table:keyRate_omit}
\end{table}

We observe that the secure key rates obtained with 8PSK modulation are in all scenarios clearly higher than those for QPSK modulation, confirming our earlier results. For both modulation schemes, the maximal secure key rate is obtained at lower $p_{\textrm{pass}}$ for increasing excess noise, indicating that the advantage of postselection increases with increasing noise. 
The curves in Figure \ref{fig:p_pass_examination_rPS} again  motivate various strategies to increase the achievable secure key rate maximally and/or reduce the secure key rate significantly, similarly to our QPSK discussion. We summarise the key rates, associated with various scenarios for both modulation schemes, both reconciliation efficiencies, and four different values of excess noise in Tables \ref{table:max_KR} - \ref{table:keyRate_omit}. For $\beta = 0.90$ and $\xi = 0.04$ the secure key rates for QPSK modulation are zero, therefore the corresponding table entries are empty.

\subsection{Postselection strategies in the trusted detector scenario}\label{sec:trusted}
It is a well-known fact  (and has been confirmed in the previous subsection; see, for example, Figs. \ref{fig:p_pass_examination_8PSK_rps_95} and \ref{fig:8PSK_QPSK_no_PS}) that high noise levels negatively influence the secure key rate, as well as the maximum distance. If noise sources are assumed not to be under Eve's control ('trusted noise') we expect higher key rates for systems than for systems with the same overall noise level, where we cannot trust any noise. In this section we examine our three different postselection strategies in the trusted noise scenario with nonideal detectors for QPSK protocols. We expect a similar relation between QPSK and 8PSK results to that already seen in the previous section, hence we only investigate QPSK to avoid redundancies. To reduce the number of parameters in the presentation and to simplify the analytical results, we chose $\eta_q = \eta_p =: \eta_d = 0.72$ and $\nu_q = \nu_p =: \nu_{el}=0.04$ in our detector model (see also Footnote \footnote{Although it cannot be expected to have two identical detectors in real world applications, $\eta_d$ can be chosen to be the minimum efficiency of both detectors used and $\nu_{el}$ to be the maximum of the measured electronic noises. Then, the obtained secure key rates are still a valid lower bound in the trusted noise scenario for that system.}) which correspond to early-stage data of an experimental CV-QKD system of the Austrian Institute of Technology.

The parameter $|\alpha|$ is optimized via coarse-grained search in steps of $0.05$ and it turned out that the optimal value of the coherent state amplitude $|\alpha|$ in the untrusted detector scenario (cf. Figure \ref{fig:alpha_th_vs_num}) and in the trusted detector scenario is identical. 
In what follows, we fix the reconciliation efficiency $\beta = 0.95$ and examine two different levels of excess noise $\xi \in \{0.01, 0.02\}$. 

\begin{figure}[htbp!]
\subfloat[\label{fig:trusted_keyrates_vs_L_xi_001} $\xi = 0.01$, $\nu_{el} = 0.04$.]{
\includegraphics[width=0.48\textwidth]{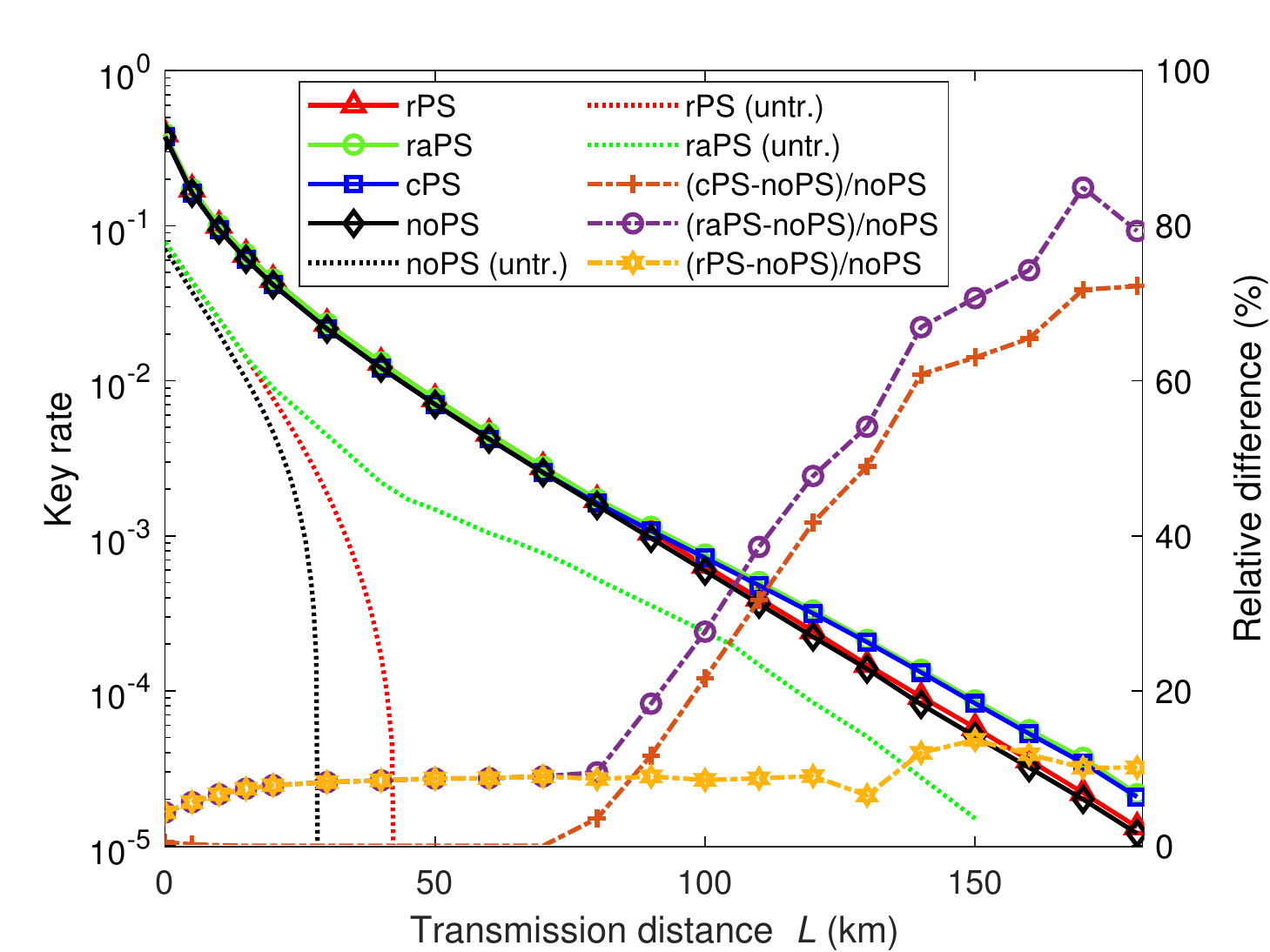}}\\
\subfloat[\label{fig:trusted_keyrates_vs_L_xi_002} $\xi = 0.02$, $\nu_{el} = 0.04$. ]{
\includegraphics[width=0.48\textwidth]{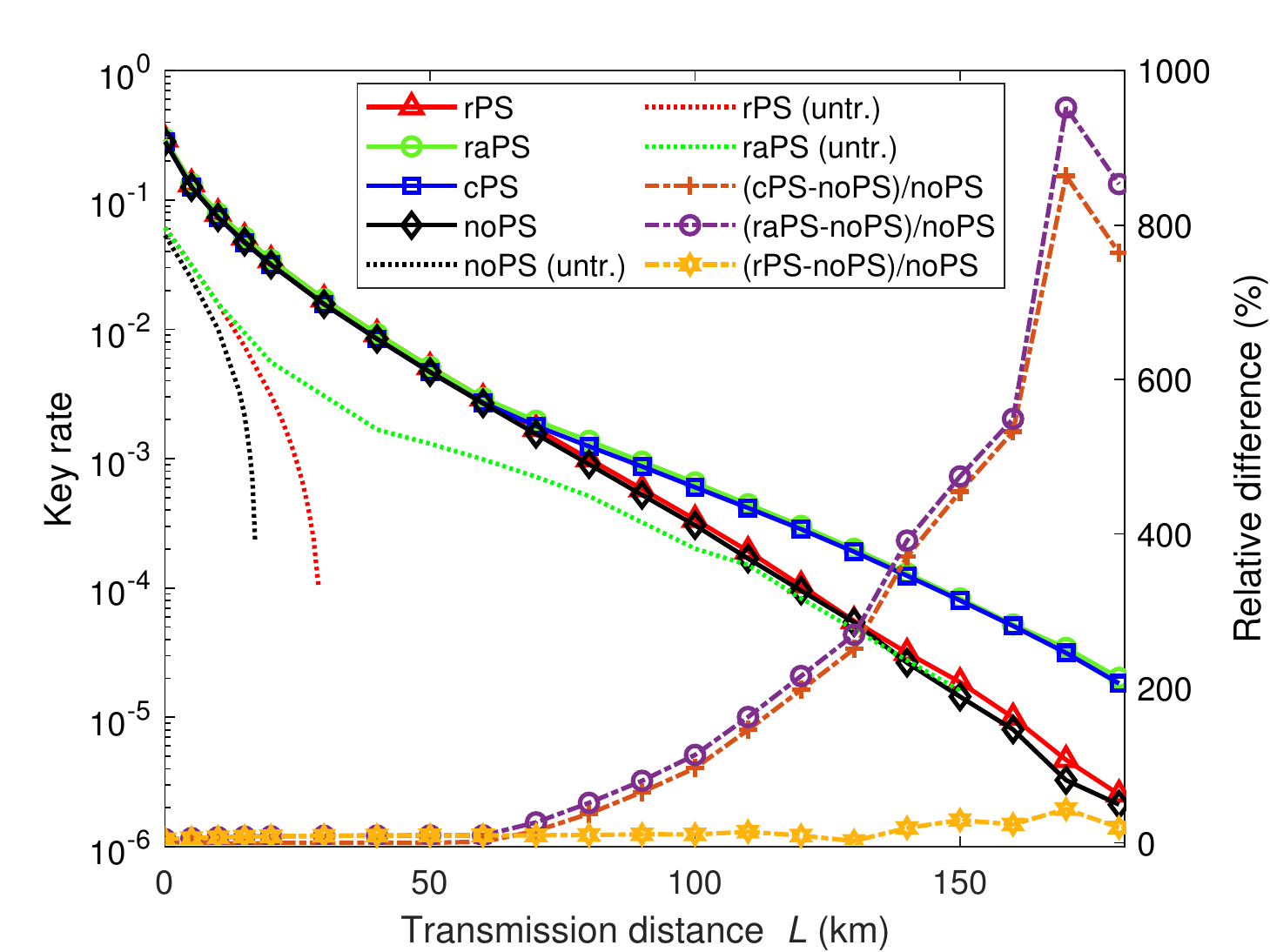}}
\makeatletter\long\def\@ifdim#1#2#3{#2}\makeatother
\caption{Secure key rate versus transmission length $L$ for different postselection schemes and with detector parameters $\eta_d = 0.72$ and $\nu_{el} = 0.04$ and reconciliation efficiency $\beta=0.95$. Dotted lines represent the key rates without trusting any noise, obtained with the same overall noise level (trusted plus untrusted noise) and the same total loss (for the same transmission distance) as the solid curves. We plotted relative (dot-dashed lines) differences between the trusted key rates obtained with radial postselection, cross-shaped postselection and radial and angular postselection, where the reference is the trusted secure key rate obtained without postselection, on the right $y$ axis.}\label{trusted_keyrate_vs_L}
\end{figure}

In Figure~\ref{fig:trusted_keyrates_vs_L_xi_001} ($\xi = 0.01$) and, Figure~\ref{fig:trusted_keyrates_vs_L_xi_002} ($\xi = 0.02$) we display the secure key rates obtained for all three postselection strategies and without postselection (noPS), as well as the relative differences to the key rates obtained without postselection (right $y$ axis). For reference, we additionally added curves representing the secure key rates for untrusted detectors with the same overall noise level (so, $\xi = 0.05$ for Figure~\ref{fig:trusted_keyrates_vs_L_xi_001} and $\xi = 0.06$ for Figure~\ref{fig:trusted_keyrates_vs_L_xi_002}) and detector efficiency $\eta_d = 0.72$.

 Similarly to our examinations for untrusted detectors, we observe a clear outperformance of the cross-shaped and radial and angular scheme over no postselection and the radial postselection strategy. For $\xi = 0.01$ (see Figure~\ref{fig:trusted_keyrates_vs_L_xi_001}) the radial postselection strategy performs about $10\%$ better than no postselection, while the cross-shaped and radial and angular scheme perform clearly better for distances greater than $80$km, peaking at relative improvements of $72\%$ and $79\%$ respectively for a transmission distance of $180$km. These advantages intensify for $\xi = 0.02$ (see Figure~\ref{fig:trusted_keyrates_vs_L_xi_002}), where cross-shaped and radial and angular postselection improve the secure key rates by factors of up to $7$-$9$.

For both levels of excess noise, we observe a clear improvement in the secure key rate for trusted detectors over those for untrusted detectors with the same overall noise level, in particular, compared with the no postselection and radial postselection curves for untrusted detectors, where the key rate curves drop steeply already at short transmission distances. In contrast, the secure key rates for untrusted detectors with radial and angular postselection remain nonzero up to $150$km. We observe that in the untrusted scenario, the optimal angular postselection parameter is nonzero for distances greater than $20$km for $\xi = 0.05$ and greater than $15$km for $\xi = 0.06$, compared with $80$km (for $\xi = 0.01$ and $\nu_{el} = 0.04)$ and $70$km (for trusted detectors with $\xi = 0.02$ and $\nu_{el} = 0.04)$. This can be seen in Figure~\ref{trusted_keyrate_vs_L}, where the difference between the dotted blue line and the solid blue line decreases slightly between $20$km and $80$km (Figure~\ref{fig:trusted_keyrates_vs_L_xi_001}) and between $15$km and $70$km (Figure~\ref{fig:trusted_keyrates_vs_L_xi_002}). Higher untrusted noise requires postselection already at shorter transmission distances, as expected. It is remarkable that for an overall noise level of $\xi = 0.06$ the curve for the untrusted detector scenario with radial and angular postselection performs comparably to the trusted curve with radial postselection for transmission distances of $120$-$150$km. Therefore, we conclude that in some scenarios radial and angular postselection for untrusted detectors yields key rates comparable with that for trusted detectors without or with radial postselection. Note that if we do not trust any noise, the security statement obtained is stronger. We expect similar results for cross-shaped postselection.

Although we conducted our analysis in the asymptotic limit, we are confident that our ideas apply as well to the finite-size regime. There are several techniques known to establish security against general attacks (providing security against collective attacks has been proven) \cite{Cirac_2009, Christandl_2009, Metger_2022}. These methods have the potential to lift our analysis. However, since finite-size analysis against general attacks is not the focus of the present paper, more detailed investigations are left for future work.
\section{\label{sec:Discussion} Conclusions}
In this work, we have adapted and  improved the numerical security analysis of \cite{Lin_2019} and analyzed and optimized continuous-variable quantum key distribution (CV-QKD) protocols with phase-shift keying (PSK) modulation with and without postselection. We have shown that having more than eight signal states does not lead to a significant improvement in the secure key rate, and thus have concentrated on QPSK and 8PSK modulation.

Our examinations for untrusted ideal detectors (Section~\ref{sec:Main}) have shown that for protocols with four signal states, both radial and angular postselection and cross-shaped postselection increase the secure key rate considerably compared with not performing postselection and perform clearly better than radial postselection. For low noise levels ($\xi = 0.01$), using cross-shaped or radial and angular postselection, the secure key rates can be improved by up to $70$-$80\%$, while for medium noise ($\xi = 0.02$) the secure key rates can be enhanced by up to $800\%$.

The key rates for the 8PSK protocols without or with radial postselection, are always superior to the key rates for the QPSK protocol with the same postselection strategy. The improvement is about $80\%$ for transmission distances up to $200$km and low noise and up to $300\%$ for medium to high noise levels, in particular, for longer transmission distances, where the QPSK key rates drop. However, radial and angular and cross-shaped postselection for QPSK perform comparably to 8PSK with radial postselection for long transmission distances, in particular for medium to high noise. 
This highlights that for certain scenarios, proper postselection, which is easy to implement, can have the same effect on the secure key rate as increasing the number of signal states, which is more challenging from an experimental point of view.

For trusted, nonideal detectors (Section~\ref{sec:trusted}) effects of postselection similar to those for untrusted detectors could be observed. Radial and angular postselection and cross-shaped postselection increase both the key rate and the maximal achievable transmission distance. 
We have compared the secure key rates in the trusted and untrusted detector noise scenarios with the same overall noise level. Secure key rates of protocols using radial and angular postselection and untrusted detectors are comparable to those without postselection but trusting the detectors. Therefore, postselection can achieve comparable secure key rates with weaker security assumptions. 

Postselection can reduce the computational bottleneck in error correction for CV-QKD (Section~\ref{Sec:p_pass_examinations}) because it can be used to reduce the length of the raw key (i.e., the data that has to be error corrected). 
Within this context, we showed that cross-shaped postselection is superior to radial postselection and pointed out how postselection can be applied to practical problems. 

Finally, we wish to highlight that postselection can be implemented easily both in new and existing QKD systems since it does not require additional hardware. Therefore, the aforementioned advantages can be utilized in any QKD system with PSK modulation.

\begin{acknowledgements}
We want to thank Martin Suda for beneficial discussions about various topics related to quantum optics and for proofreading early versions of this paper. Furthermore, we thank Michael Hentschel for discussing and providing realistic parameters for experimental implementations.

This work has received funding from the EU Horizon-2020 research and innovation programme under grant agreement No 857156 (OpenQKD) and 820466 (CiViQ).\\
\includegraphics[width=0.25\textwidth]{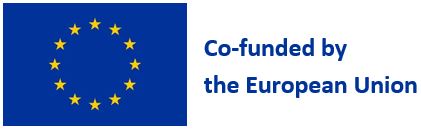} 

\end{acknowledgements}


\appendix
\begin{widetext}

\section{\label{sec:Th_Calc}Analytical validation for the noiseless channel}
In this section we briefly summarise our analytical calculations for loss-only channels. We use these results in Section~\ref{sec:number_signal_states} to argue why it is sufficient to consider only up to eight signal states. Furthermore, in what follows, we use these analytical results to validate our numerical method.

\subsection{Calculation of the secure key rate for loss-only channels}
We generalized the analytical approach from \cite{Heid_2006} to the general $N_{\textrm{St}}$ signal states for the protocols introduced in Section~\ref{sec:protocols} (where we do not perform postselection).  It is shown there that in the absence of channel-noise, it is sufficient to consider the generalized beamsplitter attack, i.e., Bob receives a coherent state whose amplitude is lowered by a factor of $\sqrt{\eta}$, while Eve obtains another coherent state whose amplitude is reduced by a factor of $\sqrt{1-\eta}$, $|\alpha_x\rangle_A \mapsto |\sqrt{\eta} \alpha_x \rangle_B \otimes |\sqrt{1-\eta} \alpha_x \rangle_E$. The secure key rate is given by the Devetak-Winter formula \cite{Devetak_Winter_2006},
\begin{equation*}
    R^{\infty} = \beta I(A:B)-\chi(B:E),
\end{equation*}
where $I(A:B)$ is the mutual information between Alice and Bob and $\chi$ is the Holevo information, an upper bound for Eve's information about Bob's signal. Therefore, the secure key rate can be obtained by calculating these quantities for the given situation.

Let us denote the states held by Eve by $|\epsilon_x\rangle := |\sqrt{1-\eta} \alpha_x \rangle$ for $x \in \{0,...,N_{\textrm{St}-1} \}$. Our goal is to describe Eve's system by an orthonormal system $\{ |e_0\rangle, ..., |e_{N_{\textrm{St}-1}}\rangle \}$. The idea is now to divide $\mathbb{N}_0$ into congruence classes of $0, ..., N_{\textrm{St}}-1$ (mod $N_{\textrm{St}}$) and to find basis vectors using only number states being in the same congruence class,
\begin{equation}
    |\tilde{e}_k\rangle = \sum_{n=0}^{\infty} \frac{\left(\sqrt{1-\eta} |\alpha| \right)^{N_{\textrm{St}}n+k}}{\sqrt{(N_{\textrm{St}}n+k)!}} (-1)^n |N_{\textrm{St}} n+k\rangle
\end{equation}
for $k \in \{0,...,N_{\textrm{St}-1} \}$. These vectors are pairwise orthogonal per construction. After defining
\begin{equation*}
    |e_k\rangle := \frac{1}{\sqrt{\langle\tilde{e}_k | \tilde{e}_k \rangle}} |\tilde{e}_k\rangle,
\end{equation*}
we obtain an orthonormal basis $B_{\textrm{ON}} := \{ |e_0\rangle, ..., |e_{N_{\textrm{St}}-1}\rangle \}$. Note that the normalisation constants can be expressed in terms of trigonometric and hyperbolic functions, hence can be calculated conveniently.

The Holevo information is given by $\chi(B:E) :=  H(\rho_E) - \sum_{j=0}^{N_{\textrm{St}}-1} P(z=j) H(\rho_{E,j})$, where $\rho_{E,j}$ is Eve's conditional state given that Bob measured the symbol labeled with $j$, defined as
\begin{equation*}
\rho_{E,j} :=    \sum_{i=0}^{N_{\textrm{St}}-1} \frac{P(x=i, z=j)}{P(z=j)} |\epsilon_i\rangle\langle \epsilon_i|
\end{equation*}
and $\rho_E$ is Eve's mixed state $\rho_E = \sum_{j=0}^{N_{\textrm{St}}-1} P(z=j) \rho_{E,j}$. After expressing $|\epsilon_i\rangle$ in terms of the basis vectors in $B_{\textrm{ON}}$, we obtain $(N_{\textrm{St}}-1) \times (N_{\textrm{St}}-1)$ matrices and can calculate the Holevo Information. The mutual information 
\begin{equation*}
    I(A:B) := H(\rho_A) + H(\rho_B) - H(\rho_A,\rho_B)
\end{equation*}
can be calculated directly. Inserting these values in the Devetak-Winter formula yields the secure key rate for loss-only channels. The maximal achievable secure key rate is then obtained by optimizing $R^{\infty}$ over $\alpha$ while holding the transmission distance constant.

\subsection{Analytical vs. numerical secure key rates}
We compared the theoretical model for a loss-only channel ($\xi = 0$) presented in the previous section to our numerical implementation for very low noise ($\xi = 10^{-5}$). We had to choose a low but non-zero $\xi$ to guarantee numerical stability. We fixed the photon-cutoff number to be $N_c = 12$ for the QPSK protocol and $N_c = 14$ for the 8PSK protocol. While data points with low coherent state amplitude $|\alpha|$ could have been calculated with a lower cutoff number, states with high coherent state amplitudes demand higher cutoff numbers to obtain reliable and tight results for the secure key rate. This is plausible since states with high coherent state amplitude have a high average photon number, hence it requires more photon states to represent those states properly. A brief discussion about the choice of the photon-cutoff can be found in Section~\ref{APDX:choice_cutoff}.

\begin{figure}
\includegraphics[width=0.48\textwidth]{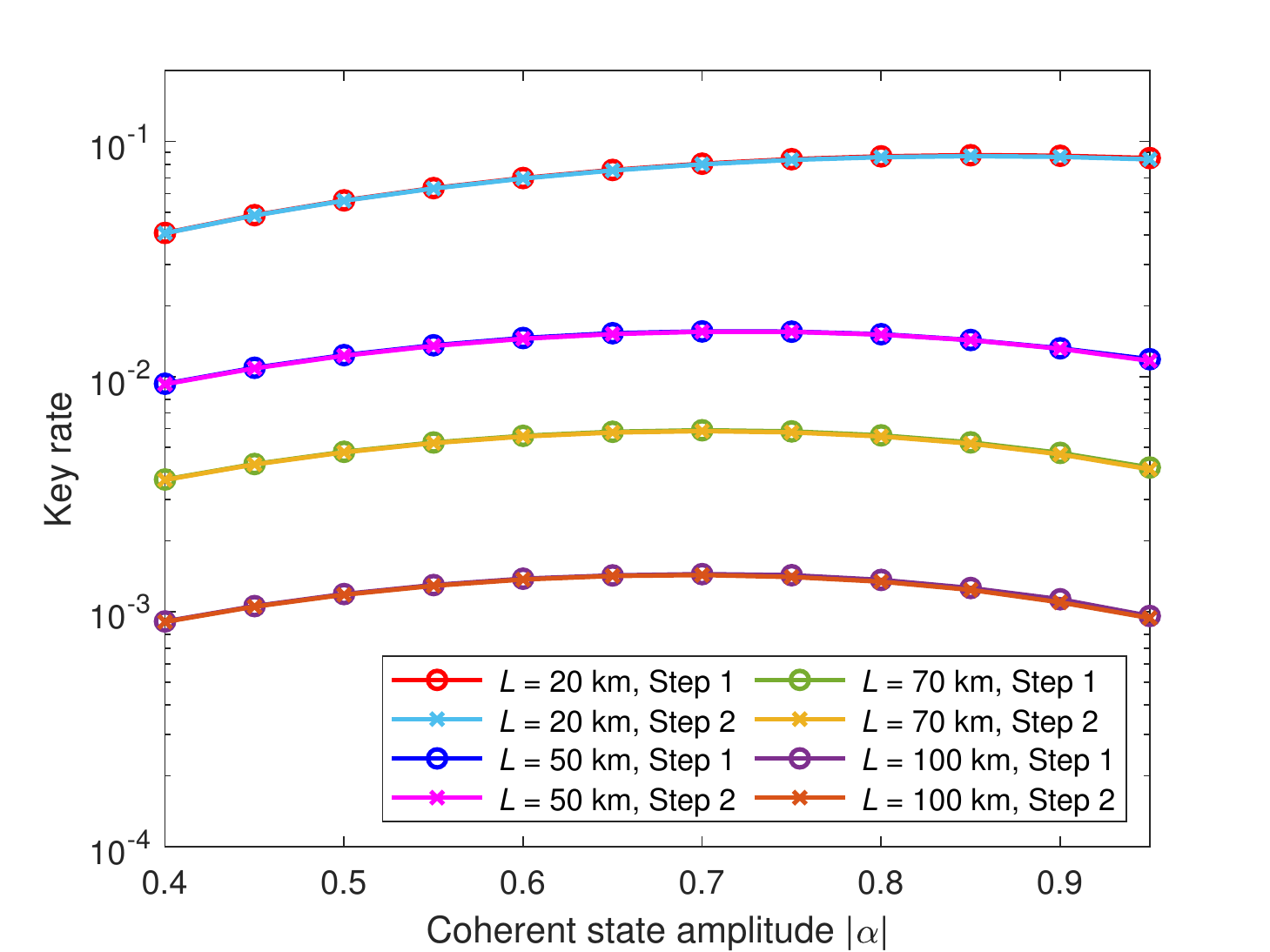}
\makeatletter\long\def\@ifdim#1#2#3{#2}\makeatother
\caption{Secure key rate vs. coherent state amplitude $|\alpha|$ for a noiseless channel (to achieve numerical stability we used $\xi = 10^{-5}$) for $L=20$km, $L=50$km , $L=70$km  and $L=100$km (from top to bottom). }\label{fig:keyrate_vs_alpha_ideal}
\end{figure}
We examined the achievable secure key rate for transmission distances between $0$ and $180$km by varying the coherent state amplitude in steps of $\Delta_{|\alpha|} = 0.05$. In Figure~\ref{fig:keyrate_vs_alpha_ideal}, we plot the obtained upper (step 1) and lower bounds (step 2) on the  secure key rates for four different transmission distances. It can be observed that the gaps between the upper and lower bounds are very small, indicating tight key rates. Therefore, we may omit the curves for step 1 in the present paper and plot only step 2, which is the relevant (lower) bound, to improve lucidity. The maxima of the curves in Figure~\ref{fig:keyrate_vs_alpha_ideal} represent the maximal achievable secure key rates and match perfectly with the predictions of the aforementioned analytical model. Note that the maximal secure key rate for $L=50$km is about ten times higher than the maximal secure key rate for $L=100$km, which meets the expectation for channel losses of $0.2$dB/km.

In Figure~\ref{fig:alpha_th_vs_num_th}, we compare the optimal coherent state amplitude obtained by analytical calculations with those obtained by our numerical implementation. The analytical coherent state amplitude was found by fine-grained search over different $|\alpha|$ for fixed transmission distance $L$ in steps of $0.005$. We observe an excellent agreement for both four- and eight-state protocols. In Figure~\ref{fig:th_calc}, we compare the theoretical prediction for the secure key rate with our numerical lower bound. Again, we observe an excellent accordance between our results and the analytical prediction with only minor deviations of less than $1\%$ (QPSK) and of less than $0.5\%$ (8PSK) for all data points except those for the lowest and highest transmission distance displayed, where the deviations are slightly higher. This can be explained by small numerical instabilities for very low and very high transmission distances at low values of excess noise. We did not observe such effects for practical values of excess noise. Summing up, our numerical results are very satisfying and meet the predictions by the analytical model excellently.

\begin{figure}
\subfloat[Optimal analytical $|\alpha|$ for different transmission distances $L$ and for $\xi = 0$. \label{fig:alpha_th_vs_num_th}]{
    \includegraphics[width=0.48\textwidth]{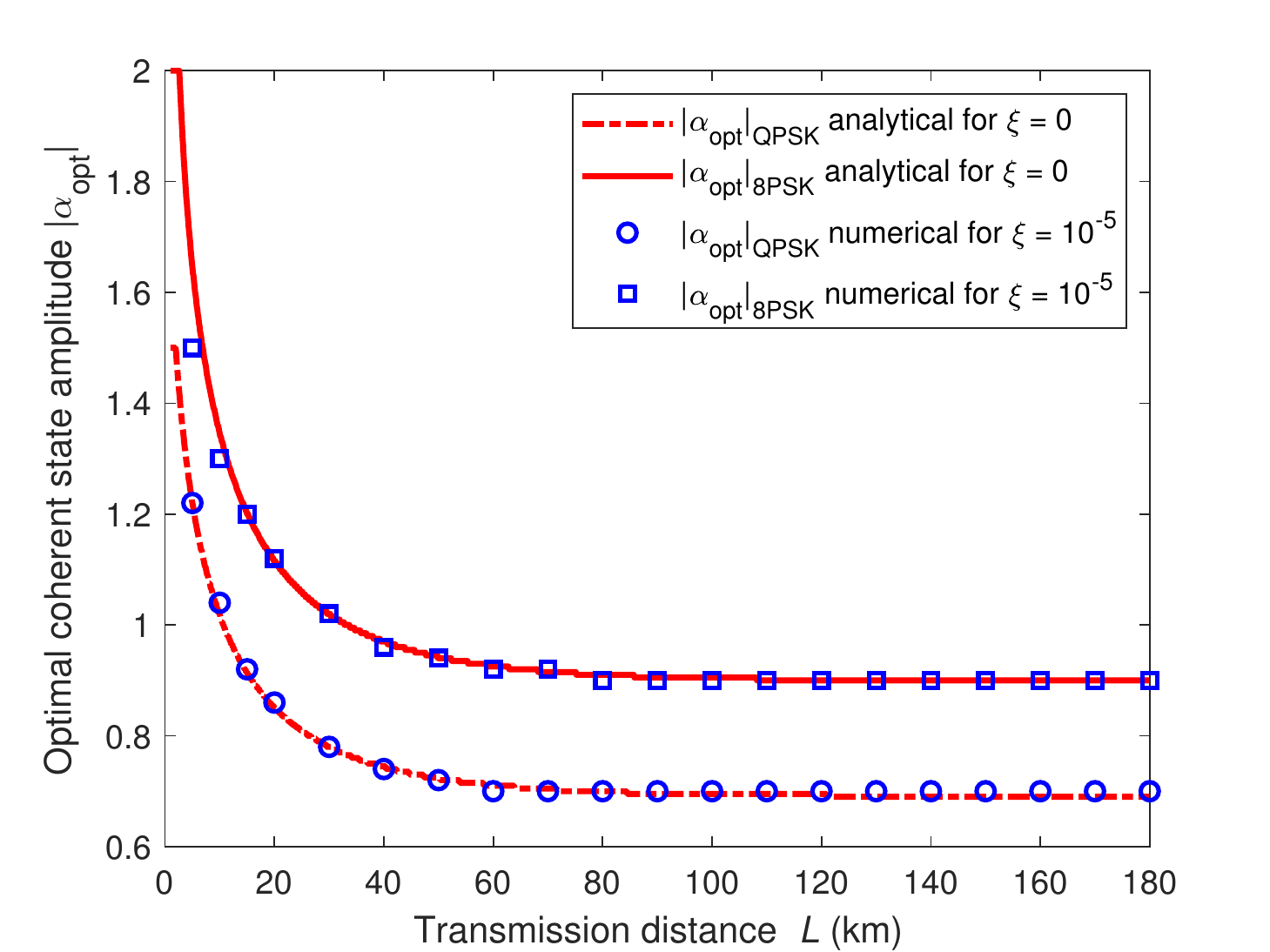}}
\subfloat[\label{fig:th_calc} Optimal analytical secure key rates for various modulation schemes and numerical lower bounds on the secure key rates for QPSK and 8PSK for transmission distances up to $180$km. The displayed lower bounds on the secure key rates (step 2) were obtained by optimizing over $|\alpha|$. The relative differences refer to the differences between numerical and analytical results for the same modulation scheme.]{
    \includegraphics[width=0.48\textwidth]{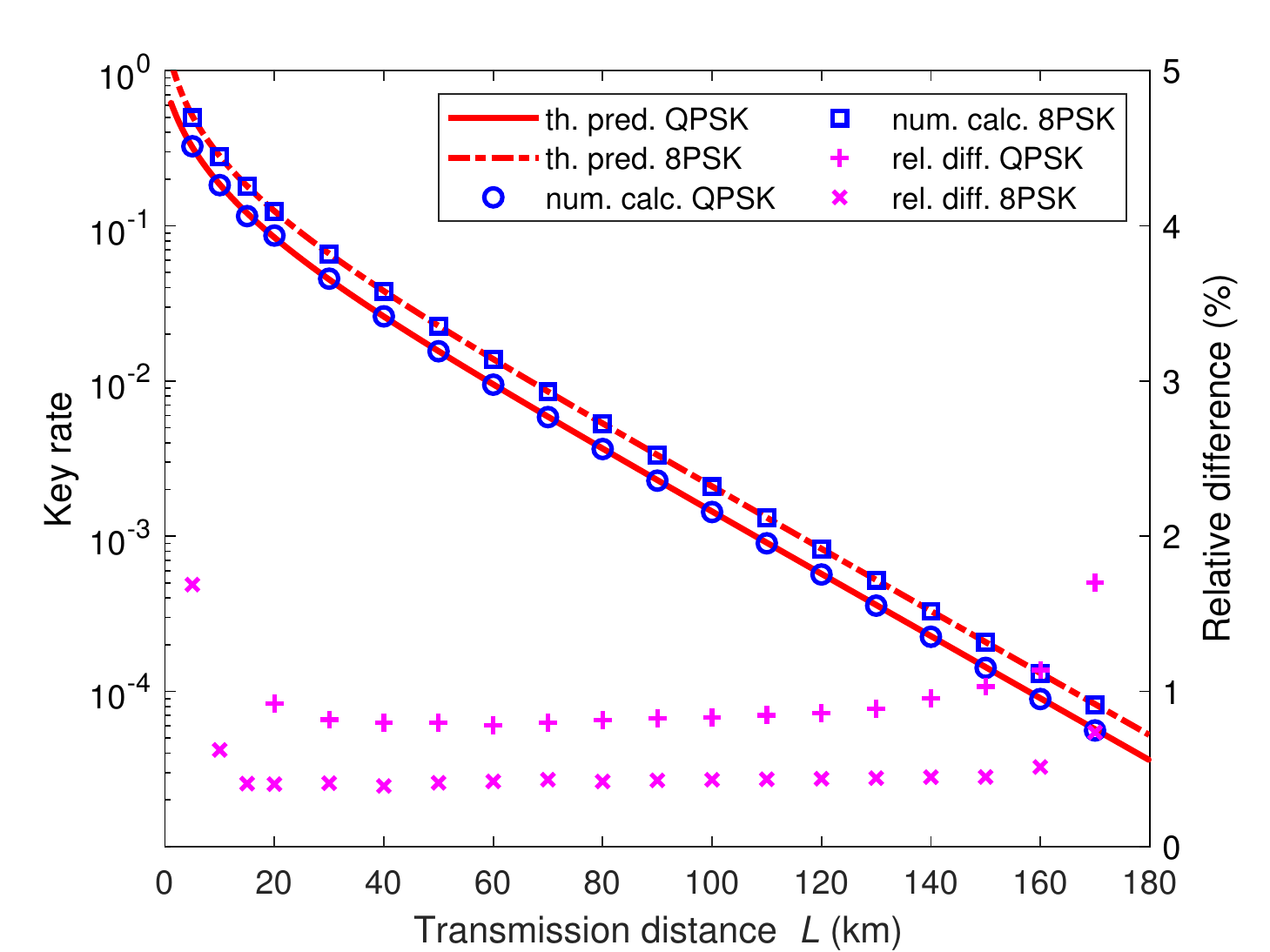}}
    \makeatletter\long\def\@ifdim#1#2#3{#2}\makeatother
\caption{Comparison between analytical prediction and numerical results for the secure key rate (almost) without noise and without performing postselection. The analytical curves were obtained by performing a fine-grained search in steps of $\Delta_{|\alpha|} = 0.005$ (red line) and the numerical results were obtained by numerical calculations and fine-grained search in steps of $\Delta_{|\alpha|} = 0.02$ (blue dots). The investigation shows that optimal coherent state amplitudes obtained by numerical calculation match perfectly with the analytical prediction and that the obtained secure key rates coincide with high accuracy.}
\end{figure}

\section{Discussion of the chosen postselection regions}\label{APDX:Discussion_PS}
In Section~\ref{sec:Motivation_PS} we motivated our choice for the postselection regions via the bit error rate for four-state protocols. These considerations naturally generalize to eight-state protocols by considering the corresponding sectors instead of quadrants. Bit-flips occur if signals that were prepared in one quadrant are measured in another quadrant. The probability density function for measuring a coherent state $|\alpha_k\rangle$ at $\gamma \in \mathbb{C}$ is given by $P(Y = \gamma|X = \alpha_k) = \frac{1}{\pi}e^{-|\alpha_k-\gamma|^2}$, in accordance with the (untrusted) POVM given in Equation~(\ref{eq:POVM_Ey}). By Bayes' theorem, the probability that the state $|\alpha_k\rangle$ was sent conditioned on Bob measured $\gamma$ is given by
\begin{equation}
    P(X = \alpha_k|Y = \gamma) = \frac{P(Y = \gamma|X = \alpha_k) P(X = \alpha_k)}{P(Y = \gamma)}.
\end{equation}
Then $P(Y = \gamma)$ is given by 
\begin{equation}
    P(Y = \gamma) = \sum_{k=1}^{4} P(Y = \gamma|X = \alpha_k) P(X = \alpha_k),  
\end{equation}
and $P(X = \alpha_k) = \frac{1}{4}$ since Alice prepares her state following the uniform distribution. 

To obtain the bit error probability for the first quadrant, we additionally need to take into account that errors from arising from different signals contribute differently, depending on the encoding of our QPSK scheme. Note that the ideas presented in what follows generalize easily to the other quadrants as well. We consider a symmetric scheme, where 'horizontal' and 'vertical' bit-flips contribute equally. This constellation is sketched in Figure~\ref{fig:phase_space}. Measuring $|\alpha_3\rangle$ in the first quadrant causes a bit-flip both in the MSB and the LSB, while bit-flips from the second and forth quadrant cause only single bit-flips (in the LSB for $|\alpha_2\rangle$ and in the MSB for $|\alpha_3\rangle$). Therefore, we build the weighted sum of the probabilities $P(X = \alpha_k|Y = \gamma)$ with weights $w_1 = 0,~w_2 = 1,~w_3 = 2,~w_4 = 1$ and divide the result by $2$ since we are interested in the probability of bit errors in any bit. Note that $w_1=0$ because measuring the first signal in the first quadrant does not contribute to the bit error rate. We obtain for the (expectation value of) bit error rate (BER)
\begin{equation}
    \langle\textrm{BER} \rangle = \frac{1}{2} \frac{ \sum_{k=1}^{4} w_k e^{-|\alpha_k-\gamma|^2}}{ \sum_{k=1}^{4} e^{-|\alpha_k-\gamma|^2}}.
\end{equation}


As outlined in the main text, the postselection regions can be chosen along the contour-line (which contour-line is subject to an optimization) corresponding to the bit error probability. However, there are two issues. First, a-priory, the exact shape of such a contour line is not analytically known and it is not immediately clear that the shape remains the same once we consider noisy channels. Second, even if the shape was (approximately) known we cannot expect to describe it by an easy and elementary function, but require more than two postselection parameters to be parametrised. Besides this, computing secure key rates for postselection strategies with complicated shapes and a high number of postselection parameters is computationally very costly for two reasons: (i) the occurring operators for the exact postselection regions need to be calculated numerically (instead of using analytical results, as derived in Appendix \ref{apdx:untrusted} and \ref{apdx:trusted}) and, (ii) the search for the optimal set of parameters would involve a much larger parameter space. Furthermore, we observed that region operators, obtained by numerical integration (in particular, if the integrations range over regions with 'exotic' boundaries) lead to a less stable algorithm at all. 

The reason behind is the following. Small numerical deviations/errors in the matrix elements of the region operators tend to intensify over many iterations of the present algorithm. This often causes a couple of slightly negative eigenvalues. Since density matrices are required to be positive semidefinite, negative eigenvalues have to be removed by a transformation. Due to this transformation, the obtained density matrix is altered slightly, causing small constraint violations. In order to obtain secure lower bounds, step 2 of the present algorithm transforms the obtained upper bound from step 1 into a reliable lower bound. Part of this process is to take the influence of constraint violations into account, using a relaxation theorem. Therefore, the whole security proof critically demands low constraint violations, as otherwise the secure key rate drops significantly.

\section{Explanation of the two-step process of the security proof\label{APP:Two-step-process}}
For completeness, in this chapter, we give details about the two-step process that was introduced in \cite{Winick_2018} to calculate the secure key rate and specify the optimization problem related to the protocols examined in the present work.
\subsection{Finding an almost optimal attack}
In the first step of the used method, we have to find an attack that is close to optimal. The objective-function $f(\rho) = D(\mathcal{G}(\rho) ||\mathcal{Z}(\mathcal{G}(\rho)))$ is  non-linear. Therefore, we tackle the problem iteratively by approximating $f$ to first order solving a linear semidefinite program. Since we face a constrained optimization problem, we require an iterative algorithm that is guaranteed to stay in the feasible set, like the Frank-Wolfe algorithm \cite{Frank_Wolfe_1956} (alternatively, we may apply an algorithm that leaves the feasible set, combined with a projection that brings us back). The classical Frank-Wolfe algorithm can be expedited, if a line search towards the found optimal direction is added. The following algorithm is suggested in  \cite{Winick_2018}.\\
\begin{algorithm}[H]
  \caption{Modified Frank-Wolfe for step 1}
  \label{ALG:FW}
   \begin{algorithmic}[1]
   \State Choose $\epsilon_{\textrm{FW}} > 0$, $\rho_0 \in \mathcal{S}$ and set $k=0$
   \State Find $\Delta \rho :=\textrm{arg min}_{\Delta \rho}\textrm{Tr}\left[\left(\Delta \rho\right)^{\top} \nabla f(\rho_k)\right]$ subject to $\rho_k+ \Delta \rho \in \mathcal{S}$
   \State STOP if $\textrm{Tr}\left[\left(\Delta \rho\right)^{\top} \nabla f(\rho_k)\right] < \epsilon_{\textrm{FW}}$
   \State Find $\lambda \in (0,1) $ that minimizes $f(\rho_k + \lambda \Delta \rho)$
   \State $\rho_{k+1} := \rho_{k} + \lambda \Delta \rho$, $k \leftarrow k+1$, proceed with 2.
   \end{algorithmic}
\end{algorithm}
The optimization problem can be brought into a more advantageous form. Following \cite{Winick_2018}, we orthonormalize the (cutoff representation of) observables $\Gamma_i$ and obtain an orthonormal set $\{\overline{\Gamma}_i~:~ i \in I\}$, which we extend to an orthonormal basis $\{\overline{\Gamma}_i~:~ i \in I\} \cup \{\Omega_j ~ :~ j \in J\}$ of $\mathcal{D}(\mathcal{H}_{AB}^{N_c})$ with respect to the Hilbert-Schmidt norm, where $\mathcal{H}_{AB}^{N_c} :=\mathcal{H}_A \otimes \mathcal{H}_B^{N_c}$ and $J$ is another finite set. The expectation values of the $i$-th orthonormalised operator is denoted by $\overline{\gamma}_i$. Then, the feasible set can be reformulated as
\begin{equation}
    \mathcal{S} = \left\{\left. \sum_{i\in I} \overline{\gamma}_i \overline{\Gamma}_i + \sum_{j \in J} \omega_j \Omega_j~ \right|~ \vec{\omega} \in \mathbb{R}^{|J|}  \right\},
\end{equation}
where the first part represents the subspace fixed by the constraints, and the second part represents the free subspace.
Finally, the present minimization problem (with objective function $f$) in every Frank-Wolfe step reads \cite{Winick_2018} 
\begin{align}
    \vec{\omega'} =\textrm{arg min}_{\vec{\omega}} \sum_{ j \in J} \omega_j\textrm{Tr}\left[ \Omega_j^{\top} \nabla f(\rho_i) \right] \\
   \textrm{subject to } \sum_{j \in J} \omega_j \Omega_j + \rho_i \in \mathcal{D}(\mathcal{H}_{AB}^{N_c}),
\end{align}
and the next iterate can be obtained by 
\begin{equation}
    \rho_{i+1} = \rho_i + \lambda \sum_{j \in J} \omega'_j \Omega_j,
\end{equation}
where $\lambda \in [0,1]$ can be found by a line-search to speed-up the algorithm (or set to $1$ otherwise). For $\lambda = \frac{2}{k+2}$, it is known \cite{Jaggi_2013} that the Frank-Wolfe algorithm with target function $f$ satisfies $f\left(\rho_k\right) - f\left(\rho^*\right) \leq \mathcal{O}\left( \frac{1}{k} \right)$, where $\rho^*$ is the optimal solution and $\rho_k$ is the $k$-th iterate. As performing a line-search includes the case of $\lambda = \frac{2}{k+2}$, the Frank-Wolfe algorithm combined with a line-search is known to converge at least at that rate. For our application, we observed that the Frank-Wolfe algorithm with additional line-search converges considerably faster. As mentioned earlier, for our implementation we used the bisection method.
\subsection{Obtaining a tight lower bound on the key rate}
The second step aims to convert the upper bound, obtained in step 1, into a lower bound. We only state the basic idea, following \cite{Winick_2018}, where step 2 is justified by a sequence of three theorems.

The basic idea of step 2 is to convert every upper bound on the secure key rate for a feasible $\rho$, obtained from step 1, into a lower bound on the secure key rate by linearisation and solving the dual problem of the occurring semidefinite program. Therefore, both step 1 and step 2 involve the evaluation of the gradient of $f$, which might not exist for every $\rho$ (e.g., if $\mathcal{G}$ has not full rank). To address this issue, a perturbed map 
\begin{equation}
    \mathcal{G}_{\tilde{\epsilon}}(\rho) := \mathcal{D}_{\tilde{\epsilon}}(\mathcal{G}(\rho))
\end{equation}
is introduced, where $0 < \tilde{\epsilon} < 1$ and 
\begin{equation}
\mathcal{D}_{\tilde{\epsilon} }(\rho) := (1-\tilde{\epsilon} ) \rho + \tilde{\epsilon}  \frac{1}{\textrm{dim}(\mathcal{G}(\rho))}\mathbbm{1}_{N_c}.
\end{equation}
Computational evaluations and differences between the exact constraints and their representation due to finite precision can introduce little numerical errors in the secure key rate calculations and therefore have to be taken into account for a reliable security proof. Let us denote the computer representation of variables with tildes, for example, $\tilde{\Gamma}_i$ is the representation of $\Gamma_i$. It is shown in \cite{Winick_2018} that if the constraints are satisfied up to some small number $\epsilon' \in \mathbb{R}$, $\forall i \in I: ~\left|\textrm{Tr}\left[ \tilde{\Gamma}_i \rho - \tilde{\gamma}_i \right] \right| \leq \epsilon'$, the following statement holds.\\
\textbf{Theorem: } Let $\rho \in \left\{ \rho \in \mathcal{D}(\mathcal{H}_A \otimes \mathcal{H}_B^{N_c}) ~ : ~ \left|\textrm{Tr}\left[ \tilde{\Gamma}_i \rho - \tilde{\gamma}_i \right] \right| \leq \epsilon' \right\}$ where $\epsilon' > 0$ and $0 < \epsilon \leq \frac{1}{e(\textrm{dim}(\mathcal{G}(\rho))-1)}$. Then
\begin{equation}
    \min_{\rho \in \mathcal{S}}f(\rho) \geq \beta_{\epsilon \epsilon'}(\rho) - \zeta_{\epsilon}
\end{equation}
where $\zeta_{\epsilon} := 2\epsilon(\textrm{dim}(\mathcal{G}(\rho))-1) \log\left( \frac{\textrm{dim}(\mathcal{G}(\rho))}{\epsilon(\textrm{dim}(\mathcal{G}(\rho))-1)} \right)$ and 
\begin{equation}
\begin{aligned}
    \beta_{\epsilon, \epsilon'}(\sigma) := f_{\epsilon}(\sigma) -\textrm{Tr}\left[ \sigma^{\top} \nabla f_{\epsilon}(\sigma) \right] + \max_{(\vec{y},\vec{z}) \in \tilde{\mathcal{S}}_{\epsilon}^*(\rho)} \left( \vec{\tilde{\gamma}} \cdot \vec{y} - \epsilon' \sum_{i=1}^{|I|} z_i \right).
\end{aligned}
\end{equation}
The set $\tilde{\mathcal{S}}_{\epsilon}^*(\sigma)$ is given by
\begin{equation}
\begin{aligned}
    &\tilde{\mathcal{S}}_{\epsilon}^*(\rho) := \left\{ (\vec{y}, \vec{z}) \in (\mathbb{R}^{|I|}, \mathbb{R}^{|I|}) ~|~ -\vec{z}  \leq \vec{y} \leq \vec{z}, ~ \sum_{i=1}^{|I|} y_i \tilde{\Gamma}_i^{\top} \leq \nabla f_{\epsilon}(\sigma) \right\}.
\end{aligned}
\end{equation}

\section{Details about the implementation}\label{APDX:Details_Implementation}
\subsection{Specifying the optimization problem}\label{sec:postprocessing}
To formulate the relevant optimization problem, we need to specify the quantities appearing in (\ref{eq:SDP}), i.e., we require $\Gamma_i$ as well as the right-hand sides $\gamma_i$ to formulate the constraints, and the postprocessing map $\mathcal{G}$ as well as the pinching channel $\mathcal{Z}$ to fully characterize the objective function. 

We start with briefly explaining the constraints in the optimization problem (\ref{eq:SDP}). The constraints in (\ref{eq:SDP}) have two different origins: (i) some of the constraints arise from Bob's measurement results, depending on the state Alice has sent. If Eve manipulates the quantum channel (which is assumed being under her control), she has to do it in a way such that the communicating parties do not recognise her actions. Therefore, the density matrix is constrained by $4 N_{\textrm{St}}$ equations from Bob's measurements. (ii) Since we assume that Eve cannot access Alice's lab, her share of the state (in the entanglement-based picture) is fixed. This gives a matrix-valued constraint, which is transformed by quantum state-tomography (see, e.g., \cite{Altepeter_2005}) into $N_{\textrm{St}}^2$ scalar-valued constraints. Thereby, the basis of Alice's system is chosen in a way such that the set of constraints ($4N_{\textrm{St}}$ from the measurements, $N_{\textrm{St}}^2$ from the state-tomography) is sufficient to linear-combine the trace-equal-to-one condition with sufficient numerical precision. Hence, in contrast to \cite{Lin_2019}, we do not add the constraint $\textrm{Tr}\left[ \rho_{AB}\right] = 1$ since the resulting density matrix already has trace equal to one with satisfying numerical accuracy without requiring this condition explicitly. For numerical reasons, it is beneficial to avoid (almost) linearly dependent conditions in our set of constraints.

Next, we discuss the postprocessing maps. In this work, we follow the (postprocessing-)framework of \cite{Lin_2019} and the indices $A$, $B$, and $R$ label Alice's and Bob's system and a classical register, respectively.  Therefore, the occurring postprocessing map $\mathcal{G}(\sigma) := K \sigma K^{\dagger}$ is defined by the Kraus operator
\begin{equation}\label{eq:Kraus_operator}
    K := \sum_{z=0}^{N_{\textrm{St}}-1} |z\rangle_R \otimes \mathbbm{1}_A \otimes \left(\sqrt{R_z} \right)_B,
\end{equation}
where $\left(R_z\right)_{z \in\{0,..., N_{\textrm{St}}-1\}}$ are the so-called region operators, whose form depends on the actual key map (as specified in Section~\ref{sec:protocols}). If $E_y$ denotes the POVM of Bob's measurements, they are given by 
\begin{equation}\label{eq:RegionOpUntrusted}
    R_z := \int_{\mathcal{A}_z} E_y \,d^2y,
\end{equation}
where $\mathcal{A}_z$ is the set corresponding to the symbol $z$ in the key map. For ideal homodyne measurements the corresponding POVM \cite{Sanders_2004} is given by 
\begin{equation}\label{eq:POVM_Ey}
    E_y = \frac{1}{\pi} |y\rangle \langle y|.
\end{equation}

According to the definition of the postselection maps (raPS, cPS, 8raPS) in Section~\ref{sec:protocols}, we define the following subsets of the phase-space ($\mathbb{C})$:
\begin{equation}\label{eq:Ara}
\begin{aligned}
 A_k^{\textrm{ra}} &:= \left\{ \zeta \in \mathbb{C}:  \arg(\zeta) \in \left[\frac{k\pi}{2} + \Delta_a, \frac{(k+1) \pi}{2}-\Delta_a\right) \land |\zeta| \geq \Delta_r\right\},
\end{aligned}
\end{equation}
for $k\in\{0,1,2,3\}$ and
\begin{equation}\label{eq:Ac}
\begin{aligned}
A_0^{\textrm{c}} &:= \left\{ \zeta \in \mathbb{C}: \Re(\zeta) \geq \Delta_c  \land \Im(\zeta) \geq \Delta_c  \right\},\\
A_1^{\textrm{c}} &:= \left\{ \zeta \in \mathbb{C}: \Re(\zeta) \leq -\Delta_c \land \Im(\zeta) \geq \Delta_c  \right\},\\
A_2^{\textrm{c}} &:= \left\{ \zeta \in \mathbb{C}: \Re(\zeta) \leq -\Delta_c \land \Im(\zeta) \leq -\Delta_c \right\},\\
A_3^{\textrm{c}} &:= \left\{ \zeta \in \mathbb{C}: \Re(\zeta) \geq \Delta_c  \land \Im(\zeta) \leq -\Delta_c \right\}
\end{aligned}
\end{equation}
for four-state protocols and
\begin{equation}\label{eq:Ara8}
\begin{aligned}
 A_k^{\textrm{8ra}} &:= \left\{ \zeta \in \mathbb{C}:  \arg(\zeta) \in \left[\frac{(2k-1)\pi}{8} + \Delta_a, \frac{(2k+1) \pi}{8}-\Delta_a\right) \land |\zeta| \geq \Delta_r\right\},
\end{aligned}
\end{equation}
for $k\in \{0,...,7\}$ in the case of eight signal states.
Here, the superscript labels the chosen postselection strategy, and the subscript labels the symbol we associate with the defined set. Note that the radial postselection-scenario is included in the radial and angular case by setting $\Delta_a = 0$ and the no postselection scenario is included in both schemes by setting all postselection parameters to zero. Therefore, we do not need to define separate sets for the radial scheme and for no postselection.

For trusted detectors, we cannot use the POVM of ideal detectors any more. Hence, we have to replace $E_y$ in Equation~(\ref{eq:RegionOpUntrusted}) by a POVM corresponding to the aforementioned detector model. In \cite{Lin_2020} the following POVM elements
\begin{equation}\label{eq:POVM_Gy}
    G_y = \frac{1}{\eta_d \pi} \hat{D}\left(\frac{y}{\sqrt{\eta_d}} \right) \rho_{th}\left( \frac{1-\eta_d+\nu_{el}}{\eta_d} \right) \hat{D}^{\dagger}\left(\frac{y}{\sqrt{\eta_d}} \right)
\end{equation}
have been derived, where $\hat{D}$ is the displacement operator and $\rho_{th}(\overline{n})$ denotes a thermal state with mean photon number $\overline{n}$.

Finally, the pinching channel $\mathcal{Z}$ is given by
\begin{equation}
\mathcal{Z}(\sigma) := \sum_{j=0}^{N_{\textrm{St}}-1} \left(|j\rangle\langle j|_R \otimes \mathbbm{1}_{AB} \right) \sigma \left(|j\rangle\langle j|_R \otimes \mathbbm{1}_{AB} \right).
\end{equation}
 Hence, we have completely specified the objective function $f(\rho)$. We refer to Section \ref{sec:RO_untrusted} for our explicit analytical expressions for the region operators for untrusted ideal detectors, as well as to Section \ref{sec:RO_trusted} for our explicit analytical expressions for the region operators for trusted nonideal detectors for each of the proposed postselection strategies. The detailed derivations are given in Appendix \ref{apdx:Region_op} (untrusted) and Appendix \ref{apdx:trusted} (trusted). Using these analytical expressions instead of calculating the matrix elements for the region operators numerically increases the accuracy of our results, speeds up the whole algorithm considerably and eliminates the (so-far not considered) influence of inaccuracies of numerical integration on the secure key rate.

\subsection{Channel model}
It remains to specify the right-hand sides of the constraints of the present optimization problem. Therefore, we simulate the quantum channel connecting Alice and Bob as a phase-invariant Gaussian channel with transmittance $\eta$ and excess noise $\xi$, which is a common model for optical fibres. The right-hand sides of the constraints can be found similarly to \cite{Lin_2019}, where the expectation values for QPSK states on the axes are calculated using Husimi-Q-functions. The rotation in the phase space that transforms the arrangement of the states on the axes to our 'QPSK-like' constellation does not lead to significant changes in that approach. Furthermore, this idea can be generalized easily to $N_{\textrm{St}}$ signal states. Recall that we measure the excess noise in multiples of the shot noise. The expectation values read
\begin{align}
    \langle \hat{q} \rangle_x &= \sqrt{2\eta}~ \Re(\alpha_x), \\
    \langle \hat{p} \rangle_x &= \sqrt{2\eta}~ \Im(\alpha_x), \\
    \langle \hat{n} \rangle_x &= \eta |\alpha_x|^2 + \frac{\eta \xi}{2},\\
    \langle \hat{d} \rangle_x &= \eta \left(\alpha_x^2 + (\alpha_x^*)^2 \right)
\end{align}
for $x \in \{0,..., N_{\textrm{St}}-1\}$, where $\alpha_x$ is a complex number associated with the coherent state Alice prepares. Note that $\hat{n}$ and $\hat{d}$ are related to the second-moment observables $\hat{q}^2$ and $\hat{p}^2$. Therefore, the constraints for $\langle \hat{n}\rangle_x$ and $\langle \hat{d}\rangle_x$ can be replaced by expressions for $\langle \hat{q}^2\rangle_x$ and $\langle \hat{p}^2\rangle_x$.

For trusted detectors, the POVM, given in Equation~(\ref{eq:POVM_Gy}), is used to define first- and second-moment observables \cite{Lin_2020}
\begin{align} 
    \hat{F}_Q &= \int \frac{y^*+y}{\sqrt{2}} G_y \,d^2y, \label{eq:observable_FQ}\\
    \hat{F}_P &= \int i \frac{y^*-y}{\sqrt{2}} G_y\,d^2y, \label{eq:observable_FP}\\
    \hat{S}_Q &= \int \left(\frac{y^*+y}{\sqrt{2}}\right)^2 G_y \,d^2y, \label{eq:observable_SQ}\\
     \hat{S}_P &= \int \left(i \frac{y^*-y}{\sqrt{2}}\right)^2 G_y\,d^2y \label{eq:observable_SP}
\end{align}
with expectation values
\begin{align}
    \langle \hat{F}_Q \rangle_x &= \sqrt{2 \eta_d \eta}~ \Re(\alpha_x),\\
    \langle \hat{F}_P \rangle_x &= \sqrt{2 \eta_d \eta}~ \Im(\alpha_x),\\
    \langle \hat{S}_Q \rangle_x &= 2\eta_d \eta \left(\Re(\alpha_x)\right)^2 + 1 + \frac{1}{2}\eta_d \eta \xi + \nu_{el}, \\
    \langle \hat{S}_P \rangle_x &= 2\eta_d \eta \left(\Im(\alpha_x)\right)^2 + 1 + \frac{1}{2}\eta_d \eta \xi + \nu_{el}.
\end{align}

Hence, for the trusted detector model, we face a slightly modified problem, where changes occur in the constraints due to measurements and the objective function (as the map $\mathcal{G}$ depends on the region operators, hence on the POVM). For details regarding the trusted detector approach, we redirect the reader to \cite{Lin_2020}.

\subsection{Information-leakage during reconciliation}
Finally, it remains to calculate the information leakage in the error correction phase and the probability to pass the postselection. By construction, we find the probability that Bob obtains the symbol $z=k$ conditioned that Alice has prepared the state $x=l$ by 
\begin{equation}\label{eq:pCond}
P(z=k |x=l ) =\textrm{Tr}\left[\rho_B^l R_k\right],
\end{equation}
where 
\begin{equation}
    \rho_B^l = \frac{1}{p_l}\textrm{Tr}_A\left[ \rho_{AB} \left(|l\rangle\langle l|_A \otimes \mathbbm{1}_B \right) \right]
\end{equation}
and $R_k$ denotes one of the region operators introduced and specified in the previous sections. By knowing this and taking the error correction efficiency into account, we can calculate the information leakage during reconciliation per signal $\delta_{EC}$ and the probability that a signal passes the postselection phase $p_{\textrm{pass}}$. 
If we were able to perform information reconcilation at the Slepian-Wolf limit \cite{Slepian_Wolf_1973} $\delta_{EC} = H(\mathbf{Z} |\mathbf{X}) = H(\mathbf{Z}) - I(\mathbf{X}:\mathbf{Z})$ would hold, where $H(\mathbf{Z})$ is the von-Neumann entropy of the string $\mathbf{Z}$, $H(\mathbf{Z}|\mathbf{X})$ is the conditioned von-Neumann entropy, and $I(\mathbf{X}:\mathbf{Z})$ denotes the mutual information between the strings $\mathbf{Z}$ and $\mathbf{X}$. As we assume to perform error correction with efficiency $\beta<1$ (depending on the error correction procedure), we replace the mutual information between $\mathbf{Z}$ and $\mathbf{X}$ by $\beta I(\mathbf{X}:\mathbf{Z})$ and then rewrite the expression in terms of entropies again. Therefore, we obtain
\begin{equation}
\begin{aligned}
\delta_{EC} &= H(\mathbf{Z}) - \beta \left(H(\mathbf{Z}) - H(\mathbf{Z}|\mathbf{X})\right)\\
&= (1-\beta) H(\mathbf{Z}) + \beta H(\mathbf{Z}|\mathbf{X}).
\end{aligned}
\end{equation}
The entropies can be calculated using the probabilities given in Equation~(\ref{eq:pCond}) and the law of total probability.

Furthermore, we obtain the probability that a signal passes the postselection phase by
\begin{equation}
    p_{\textrm{pass}} = \sum_{l=0}^{N_{\textrm{St}}-1} \sum_{k=0}^{N_{\textrm{St}}-1} p_l P(z=k|x=l), 
\end{equation}
where $p_l$ denotes the probability that Alice prepares the state corresponding to the symbol $l$ (which is given in the protocol description; for symmetry reasons, we choose $\forall l \in\{0, ..., N_{\textrm{St}}-1\}:~p_l = \frac{1}{N_{\textrm{St}}}$).

\section{\label{apdx:Region_op}Fock state representation of region operators for the untrusted noise scenario} \label{apdx:untrusted}
Here, we present the explicit calculations leading to to the matrix representations of the region operators with respect to the Fock basis, as stated in Section~\ref{sec:postprocessing}. Both for the calculation of the radial and angular and the cross-shaped postselection strategy, the projection of a coherent state with amplitude $|\alpha|$ and phase $\theta$ on a number state 
\begin{equation}
\left\langle |\alpha| e^{i \theta} \right|\left. n\right\rangle = e^{-\frac{|\alpha|^2}{2}} \frac{|\alpha|^n e^{-i n \theta}}{\sqrt{n!}},   
\end{equation}
or, in Cartesian coordinates, $|\alpha| e^{i\theta} = x+iy$,
\begin{equation}
    \langle x+iy | n\rangle = e^{-\frac{x^2+y^2}{2}} \frac{(x-iy)^n}{\sqrt{n!}}
\end{equation}
will be useful. This relation is obtained readily by expressing the coherent state in the number basis and applying the inner product with $|n\rangle$.

Before we start with the calculation, we derive an integral that occurs multiple times in the following derivations. For $p>0$ and $k > 0$ we have
\begin{equation}\label{eq:commonInt}
    \int_{\Delta}^{\infty} \gamma^p e^{-k \gamma^2}\,d\gamma = \frac{1}{2 k^{\frac{p+1}{2}}} \Gamma\left( \frac{p+1}{2}, k\Delta^2 \right).
\end{equation}
This can be seen using the substitution $z := k\gamma^2 $
\begin{align*}
   \int_{\Delta}^{\infty} \gamma^p e^{-k \gamma^2}\,d\gamma = \frac{1}{2k^{\frac{p+1}{2}}}\int_{k \Delta^2}^{\infty} z^{\frac{p-1}{2}} e^{-z} \,dz = \frac{1}{2 k^{\frac{p+1}{2}}}\int_{k \Delta^2}^{\infty} z^{\frac{p+1}{2} -1} e^{-z} \,dz.
\end{align*}
According to the definition of the incomplete gamma function, the integral in the last line is equal to $\Gamma\left(\frac{p+1}{2}, k \Delta^2 \right)$.

\subsection{Radial\&angular postselection}

We start with the expression for the region operators given in Equation~(\ref{eq:Rra}) and insert the definition of the sets $A_0^{\textrm{ra}}, A_1^{\textrm{ra}}, A_2^{\textrm{ra}}$ and $A_3^{\textrm{ra}}$ from (\ref{eq:Ara}),
\begin{equation}
R_z^{\textrm{ra}} = \frac{1}{\pi} \int_{\Delta_r}^{\infty} \int_{\frac{\pi}{2}z+ \Delta_a}^{\frac{\pi}{2}(z+1)-\Delta_a} \gamma | \gamma e^{i \theta}\rangle \langle \gamma e^{i \theta}|~\mathrm{d}\theta  ~\mathrm{d}\gamma.
\end{equation}
Note that we transformed the integral to polar coordinates, which explains the additional $\gamma$ coming from the Jacobi-determinant. By using the completeness relation, $\mathbbm{1} = \sum_n |n\rangle\langle n|$, twice, we obtain 

\begin{align*}
R_z^{\textrm{ra}} &= \frac{1}{\pi} \int_{\Delta_r}^{\infty} \int_{\frac{z \pi}{2} + \Delta_a}^{\frac{(z+1) \pi}{2} - \Delta_a} \sum_{n,m}|n\rangle \langle m |  \gamma \langle n|\gamma e^{i \theta}\rangle \langle \gamma e^{i \theta}|m\rangle ~\mathrm{d}\theta  ~\mathrm{d}\gamma \\
&= \frac{1}{\pi} \sum_{n,m}  |n\rangle\langle m| \int_{\Delta_r}^{\infty} \frac{ \gamma^{n+m+1} e^{-\gamma^2}}{\sqrt{m!~ n!}}~ d\gamma ~ \int_{\frac{z \pi}{2} + \Delta_a}^{\frac{(z+1) \pi}{2} - \Delta_a} e^{i \theta (n-m)} ~d\theta.
\end{align*}

The radial integral can be expressed by the incomplete gamma function $\Gamma(x,a)$ using the integral given in Equation~(\ref{eq:commonInt}).
\begin{align*}
\int_{\Delta_r}^{\infty} \frac{ \gamma^{n+m+1} e^{-\gamma^2}}{\sqrt{m!~ n!}}~ d\gamma  = \frac{1}{2 \sqrt{m!~ n!}}  \Gamma\left(\frac{m+n}{2}+1, \Delta_r^2 \right).
\end{align*}

If $m=n$, the angular integral simplifies to $\frac{\pi}{2}-2 \Delta_a$. For the case $m \neq n$ we obtain  $\frac{2}{m-n} e^{-i(m-n)\left(z+\frac{1}{2} \right)\frac{\pi}{2}} \sin\left[\left(\frac{\pi}{4}-\Delta_a \right)(m-n) \right]$.

Summing up, we have
\begin{equation}
R_z^{\textrm{ra}} :=\frac{1}{2\pi} \sum_{n} \sum_{m} \frac{\Gamma(\frac{m+n}{2}+1, \Delta_r^2)}{ \sqrt{m!~ n!}}   |n\rangle\langle m | \\
\cdot \left\{
\begin{array}{ll}
\frac{\pi}{2}-2\Delta_a \,& m = n \\
\frac{2}{m-n} e^{-i(m-n)\left(z+\frac{1}{2} \right)\frac{\pi}{2}} \sin\left[\left(\frac{\pi}{4}-\Delta_a \right)(m-n) \right] & \, n \neq m \\
\end{array}
\right. .
\end{equation}
Similarly, the corresponding expression for the 8PSK radial and angular protocol can be obtained, where only the angular integral has to be adapted accordingly.
\subsection{Cross-shaped postselection} \label{sec:calculations_cPS_untrusted}
We start by using the definition of the region operators in Equation~(\ref{eq:Rc}) and the sets $A_0^{\textrm{c}}, A_1^{\textrm{c}}, A_2^{\textrm{c}}$ and $A_3^{\textrm{c}}$ from Equation~(\ref{eq:Ac}), 
\begin{align*}
    R_0^{\textrm{c}} &= \frac{1}{\pi} \int_{\Delta_c}^{\infty} \int_{\Delta_c}^{\infty} |x+iy\rangle \langle x+iy| ~\mathrm{d}y ~\mathrm{d}x,\\
    R_1^{\textrm{c}} &= \frac{1}{\pi} \int_{-\infty}^{-\Delta_c} \int_{\Delta_c}^{\infty} |x+iy\rangle \langle x+iy| ~\mathrm{d}y ~\mathrm{d}x,\\
    R_2^{\textrm{c}} &= \frac{1}{\pi} \int_{-\infty}^{-\Delta_c} \int_{-\infty}^{-\Delta_c} |x+iy\rangle \langle x+iy| ~\mathrm{d}y ~\mathrm{d}x,\\
    R_3^{\textrm{c}} &= \frac{1}{\pi} \int_{\Delta_c}^{\infty} \int_{-\infty}^{-\Delta_c} |x+iy\rangle \langle x+iy| ~\mathrm{d}y ~\mathrm{d}x.
\end{align*}
All integrals have the same form and differ only by the boundaries of the occurring integrals. Hence, we derive only the expression for $R_0^{\textrm{c}}$ and argue to obtain the remaining integrals. We start by using the completeness relation, $\mathbbm{1} = \sum_{n}|n\rangle \langle n|$, twice and obtain

\begin{align*}
R_0^{\textrm{c}} &= \frac{1}{\pi} \sum_{n,m} |n\rangle\langle m| \int_{\Delta_c}^{\infty} \int_{\Delta_c}^{\infty} \langle n | x+iy \rangle \langle x+iy | m\rangle  ~\mathrm{d}y ~\mathrm{d}x =  \frac{1}{\pi} \sum_{n,m} \frac{|n\rangle\langle m|}{\sqrt{n!}\sqrt{m!}} \int_{\Delta_c}^{\infty} \int_{\Delta_c}^{\infty} e^{-(x^2+y^2)} (x+iy)^n(x-iy)^m  ~\mathrm{d}y~ \mathrm{d}x. 
\end{align*}

For $m=n$ we find
\begin{align*}
\int_{\Delta_c}^{\infty} \int_{\Delta_c}^{\infty} e^{-(x^2+y^2)} (x^2+y^2)^n  ~\mathrm{d}y~ \mathrm{d}x &=\sum_{k=0}^{n} \binom{n}{k} \int_{\Delta_c}^{\infty}e^{-x^2} x^{2k}~ \mathrm{d}x \int_{\Delta_c}^{\infty} e^{-y^2} y^{2(n-k)}~\mathrm{d}y\\
&= \frac{1}{4}\sum_{k=0}^{n} \binom{n}{k} \Gamma\left(k+\frac{1}{2}, \Delta_c^2 \right) \Gamma\left(n-k+\frac{1}{2}, \Delta_c^2\right).
\end{align*}
Where we used Equation~(\ref{eq:commonInt}) to express the occurring integrals by the incomplete gamma function.

For $m \neq n$ we deduce
\begin{align*}
&\int_{\Delta_c}^{\infty} \int_{\Delta_c}^{\infty} e^{-(x^2+y^2)} (x+iy)^n(x-iy)^m  ~\mathrm{d}y~ \mathrm{d}x\\
&= \sum_{j=0}^{n} \sum_{k=0}^{m} \binom{n}{j}\binom{m}{k} \int_{\Delta_c}^{\infty} e^{-x^2} x^{j+k}~ \mathrm{d}x  \int_{\Delta_c}^{\infty} e^{-y^2} y^{n+m-j-k} (-1)^{m-k} i^{n+m-j-k}  ~ \mathrm{d}y\\
&= \frac{1}{4} \sum_{j=0}^{n}\sum_{k=0}^{m} \binom{n}{j} \binom{m}{k} (-1)^{m-k} i^{n+m-j-k} \Gamma\left(\frac{j+k+1}{2}, \Delta_c^2 \right) \Gamma\left(\frac{n+m-j-k+1}{2}, \Delta_c^2 \right),
\end{align*}
where we used again Equation~(\ref{eq:commonInt}). Note that $(-1)^{m-k} i^{n+m-j-k} = i^{n+3m-j-3k} = i^{n-m+k-j}$. Including this in the expression for the region operator, we obtain
\begin{align}
R_0^{\textrm{c}}  &= \sum_{n,m}\frac{|n\rangle\langle m|}{4\pi \sqrt{n!} \sqrt{m!}} \cdot \left\{
\begin{array}{ll}
\sum_{j=0}^{n} \binom{n}{j} \Gamma\left(j+\frac{1}{2}, \Delta_c^2 \right) \Gamma\left(n-j+\frac{1}{2}, \Delta_c^2\right) & n = m \\
\sum_{j=0}^{n}\sum_{k=0}^{m} \binom{n}{j}\binom{m}{k} \Gamma\left(\frac{j+k+1}{2}, \Delta_c^2 \right) \Gamma\left(\frac{n+m-j-k+1}{2}, \Delta_c^2 \right) i^{n-m+k-j}
 & \, n \neq m. \\
\end{array}
\right.    
\end{align}
We observe that the integral for the case $m=n$ consists only of squares of $x$ and $y$, hence this part is not sensitive to sign-changes and therefore equal for all four operators $R_z^c, ~ z = 0,1,2,3$.

For $m \neq n$, when we calculate  $R_1^{\textrm{c}}$, we face an integral of the same form as we do for $R_0^{\textrm{c}}$ once we substitute $x \mapsto - \tilde{x}$. This leads to
\begin{align*}
\int_{-\infty}^{-\Delta_c} e^{-x^2} x^{j+k} ~ \mathrm{d}x = (-1)^{j+k} \int_{\Delta_c}^{\infty} e^{-\tilde{x}^2} \tilde{x}^{j+k} ~ \mathrm{d}\tilde{x}.
\end{align*}
So, this substitution introduces a factor of $(-1)^{j+k}$ leaving the remaining expression unchanged. If we substitute $y \mapsto - \tilde{y}$, as required for the calculation of $R_3^{\textrm{c}}$, we find
\begin{align*}
    \int_{-\infty}^{-\Delta_c} e^{-y^2} y^{n+m-j-k} ~ \mathrm{d}y = (-1)^{n+m-j-k} \int_{ \Delta_c}^{\infty} e^{-\tilde{y}^2} \tilde{y}^{n+m-j-k} ~ \mathrm{d}\tilde{y}.
\end{align*}
Here, we obtain a factor of $(-1)^{n+m-j-k}$. Finally, the calculation for $R_2^{\textrm{c}}$ requires two substitutions, namely $x\mapsto -\tilde{x}$ and $y\rightarrow -\tilde{y}$, which introduces a factor of $(-1)^{j+k}(-1)^{m+n-j-k}$. Let us denote the power of $-1$ that occurs in the expression for the region operator $z$ by $D_{j,k,m,n}^{(z)}$. According to the consideration above, we find
\begin{align*}
\tilde{D}_{j,k,m,n}^{(0)} & = 1,\\
\tilde{D}_{j,k,m,n}^{(1)} & = (-1)^{j+k} = (-1)^{k-j},\\
\tilde{D}_{j,k,m,n}^{(2)} & = (-1)^{j+k} (-1)^{m+n-j-k} = (-1)^{m+n} = (-1)^{n-m}, \\
\tilde{D}_{j,k,m,n}^{(3)} & = (-1)^{m+n-j-k} =(-1)^{n-m+k-j}.
\end{align*}

To include the power of $i$ in this factor, we define $D_{j,k,m,n}^{(z)} := \tilde{D}_{j,k,m,n}^{(z)} ~ i^{n+m-j-k}$. Therefore, we finally arrive at
\begin{align}
R_z^{\textrm{c}}  &= \sum_{n,m}  \frac{|n\rangle\langle m|}{4\pi \sqrt{n!} \sqrt{m!}} \cdot \left\{
\begin{array}{ll}
\sum_{j=0}^{n} \binom{n}{j} \Gamma\left(j+\frac{1}{2}, \Delta_c^2 \right) \Gamma\left(n-j+\frac{1}{2}, \Delta_c^2\right) & n = m \\
\sum_{j=0}^{n}\sum_{k=0}^{m} \binom{n}{j}\binom{m}{k} \Gamma\left(\frac{j+k+1}{2}, \Delta_c^2 \right) \Gamma\left(\frac{n+m-j-k+1}{2}, \Delta_c^2 \right)  D_{j,k,m,n}^{(z)}
 & \, n \neq m. \\
\end{array}
\right.    
\end{align}

\section{\label{apdx:trusted}Fock state representation of region operators for the trusted detector scenario}
First, we express the POVM Element, given in Equation~(\ref{eq:POVM_Gy}) in the number basis, where we use Equations (6.13) and (6.14) in \cite{Glauber_1967}. After defining $C_{n,m} := \frac{1}{\pi \eta_d{\frac{m-n}{2}+1}} \sqrt{\frac{n!}{m!}} \frac{\overline{n}_d^n}{(1+\overline{n}_d)^{m+1}}$, $a := \frac{1}{\eta_d(1+\overline{n}_d)}$ and $b:= \eta_d \overline{n}_d(1+\overline{n}_d)$, we obtain for $n\leq m$
\begin{equation}\label{eq:Gy_numberbasis}
    \langle n |G_y | m \rangle = C_{n,m} e^{-a |y|^2} (y^*)^{m-n} L_n^{(m-n)}\left( - \frac{|y|^2}{b} \right),
\end{equation}
where 
\begin{equation}\label{eq:laguerre}
  L_k^{\alpha}(x) = \sum_{j=0}^{k} (-1)^j \binom{k+\alpha}{k-j} \frac{x^j}{j!}   
\end{equation}
is the generalized Laguerre polynomial of degree $k$ and with parameter $\alpha$ \cite{Oldham_2008}.

\subsection{Radial\&angular postselection}

We start with the expression for the region operators given in Equation~(\ref{eq:Rra}), where we replaced the POVM for the ideal homodyne detector by that for the nonideal, trusted detector, and inserted the definition of the sets $A_0^{\textrm{ra}}, A_1^{\textrm{ra}}, A_2^{\textrm{ra}}$ and $A_3^{\textrm{ra}}$ from Equation~(\ref{eq:Ara}),
\begin{align*}
R_z^{\textrm{ra, tr}} = \int_{\Delta_r}^{\infty} \int_{\frac{\pi}{2}z+ \Delta_a}^{\frac{\pi}{2}(z+1)-\Delta_a} \gamma~ G_{\gamma e^{i \theta}}~\mathrm{d}\theta  ~\mathrm{d}\gamma = \sum_{n=0}^{N_c} \sum_{m=0}^{N_c}|n\rangle\langle m| \int_{\Delta_r}^{\infty} \int_{\frac{\pi}{2}z+ \Delta_a}^{\frac{\pi}{2}(z+1)-\Delta_a} \gamma~ \langle n|G_{\gamma e^{i \theta}} |m\rangle  \,d\theta \, d\gamma.
\end{align*}

Inserting the expression for $G_y$ from Equation~(\ref{eq:Gy_numberbasis}) yields
\begin{equation*}
    R_z^{\textrm{ra, tr}} =\sum_{n=0}^{N_c} \sum_{m=0}^{N_c} C_{n,m} |n\rangle\langle m| \int_{\Delta_r}^{\infty}  e^{-a \gamma^2} \gamma^{m-n+1}  L_n^{(m-n)}\left( - \frac{\gamma^2}{b} \right) \, d\gamma \int_{\frac{\pi}{2}z+ \Delta_a}^{\frac{\pi}{2}(z+1)-\Delta_a}  e^{-i \theta(m-n)} \,d\theta.
\end{equation*}

For $n=m$ the angular integral simplifies to $\frac{\pi}{2} - 2\Delta_a$ and the radial integral can be expressed as
\begin{align*}
  \int_{\Delta_r}^{\infty}  e^{-a \gamma^2} \gamma  L_n^{(0)}\left( - \frac{\gamma^2}{b} \right) \, d\gamma
  &=\sum_{j=0}^{n} \binom{n}{n-j} \frac{1}{b^j j!} \int_{\Delta_r}^{\infty} \gamma^{2j+1} e^{-a\gamma^2} \,d\gamma,
\end{align*}
where we used the sum-representation (\ref{eq:laguerre}) of the generalized Laguerre polynomials. Using Equation~(\ref{eq:commonInt}), we obtain
\begin{align*}
    \langle n | R_z^{\textrm{ra, tr}} |n\rangle = \frac{C_{n,n}}{2} \left( \frac{\pi}{2} - 2\Delta_a\right)\sum_{j=0}^{n} \binom{n}{n-j} \frac{1}{a^{j+1} b^j j!} \Gamma\left( j+1, a\Delta_r^2 \right).
\end{align*}
For $n\neq m$, we obtain for the angular integral $\frac{2}{ (m-n)} e^{-i(m-n)\left(z+\frac{1}{2} \right) \frac{\pi}{2}} \sin\left[(m-n) \left(\frac{\pi}{4}- \Delta_a \right) \right]$ and derive for the radial integral
\begin{align*}
\int_{\Delta_r}^{\infty}  e^{-a \gamma^2} \gamma^{m-n+1}  L_n^{(m-n)}\left( - \frac{\gamma^2}{b} \right) \, d\gamma
  &=\sum_{j=0}^{n}\binom{m}{n-j} \frac{1}{b^j j!} \int_{\Delta_r}^{\infty} \gamma^{2j+m-n+1} e^{-a\gamma^2}\,d\gamma \\
  &= \frac{1}{2} \sum_{j=0}^{n}\binom{m}{n-j} \frac{1}{a^{j+1+\frac{m-n}{2}} b^j j!} \Gamma\left(j+1+\frac{ m - n }{2}, a\Delta_r^2 \right).
\end{align*}
We do not need to calculate the matrix element for $m<n$ separately, as the region operator has to be Hermitian. 

In conclusion, we found for $R_z^{\textrm{ra, tr}}=\sum_{n=0}^{\infty}\sum_{m=0}^{\infty} \langle n |R_z^{\textrm{ra, tr}} |m\rangle |n\rangle \langle m|$  the matrix elements
\begin{equation}
 \langle n |R_z^{\textrm{ra, tr}} |m\rangle = \left\{
\begin{array}{ll}
\frac{C_{n,n}}{2}\left[\frac{\pi}{2}-2\Delta_a \right] \sum\limits_{j=0}^{n} \binom{n}{n-j} \frac{\Gamma(j+1, a\Delta_r^2)}{ a^{j+1} b^j  j!} & n = m \\
\frac{C_{n,m}}{m-n} e^{-i(m-n)\left(z+\frac{1}{2}\right)\frac{\pi}{2}} \sin\left[(m-n)\left(\frac{\pi}{4}-\Delta_a \right) \right] \sum\limits_{j=0}^{n} \binom{m}{n-j} \frac{\Gamma\left(j+1+\frac{m-n}{2},a\Delta_p^2\right)}{  a^{j+1+\frac{m-n}{2}} b^j j!} & \, n < m \\
\overline{\langle m |R_z^{\textrm{ra, tr}} |n\rangle} &\, n>m
\end{array}
\right.   .
\end{equation}
Similarly, we obtain the corresponding expression for the 8PSK radial and angular protocol, where only the angular integral has to be adapted accordingly.
\subsection{Cross-shaped postselection} 
Similarly to the calculations for the untrusted scenario, we start by using the definition of the region operators in Equation~(\ref{eq:Rc}) and the sets $A_0^c, A_1^c, A_2^c$ and $A_3^c$ from Equation~(\ref{eq:Ac}), 
\begin{align*}
    R_0^{\textrm{c, tr}} &= \int_{\Delta_c}^{\infty} \int_{\Delta_c}^{\infty} G_{x+iy} ~\mathrm{d}y ~\mathrm{d}x,\\
    R_1^{\textrm{c, tr}} &= \int_{-\infty}^{-\Delta_c} \int_{\Delta_c}^{\infty} G_{x+iy} ~\mathrm{d}y ~\mathrm{d}x,\\
    R_2^{\textrm{c, tr}} &= \int_{-\infty}^{-\Delta_c} \int_{-\infty}^{-\Delta_c} G_{x+iy} ~\mathrm{d}y ~\mathrm{d}x,\\
    R_3^{\textrm{c, tr}} &= \int_{\Delta_c}^{\infty} \int_{-\infty}^{-\Delta_c} G_{x+iy} ~\mathrm{d}y ~\mathrm{d}x.
\end{align*}
Again, all integrals have the same form and differ only by the boundaries of the occurring integrals. Hence, we derive only the expression for $R_0^{\textrm{c, tr}}$ and reason the changes to obtain the remaining integrals.

For $n\leq m$ we obtain
\begin{align*}
R_0^{\textrm{c, tr}} &= \sum_{n,m} |n\rangle\langle m| \int_{\Delta_c}^{\infty}\int_{\Delta_c}^{\infty} \langle n|G_{x+iy}|m\rangle  \,dy\,dx\\
&= \sum_{n,m} |n\rangle\langle m| C_{n,m} \int_{\Delta_c}^{\infty}\int_{\Delta_c}^{\infty} e^{-a(x^2+y^2)} (x-iy)^{m-n} L_n^{(m-n)}\left(- \frac{x^2+y^2}{b} \right)\,dy\,dx,
\end{align*}
where we inserted the expression for $G_y$ from Equation~(\ref{eq:Gy_numberbasis}) in the last line.

First, we treat the case $m=n$, where we have
\begin{align*}
    \langle n | R_0^{\textrm{c, tr}} |m\rangle &= C_{n,n} \int_{\Delta_c}^{\infty}\int_{\Delta_c}^{\infty} e^{-a(x^2+y^2)} L_n^{(0)}\left(- \frac{x^2+y^2}{b} \right)\,dy\,dx\\ 
    &= C_{n,n} \sum_{j=0}^{n}\binom{n}{n-j}\frac{(-1)^j}{j!} \int_{\Delta_c}^{\infty}\int_{\Delta_c}^{\infty} e^{-a(x^2+y^2)} \frac{(x^2+y^2)^j}{b^j} (-1)^j\,dy\,dx\\
    &=C_{n,n} \sum_{j=0}^{n}\binom{n}{n-j}\frac{1}{b^j j!}  \sum_{k=0}^{j} \binom{j}{k} \int_{\Delta_c}^{\infty} e^{-a x^2} x^{2k} \,dx \int_{\Delta_c}^{\infty} e^{-a y^2} y^{2(j-k)} \,dy.
\end{align*}
For the second equality we inserted the sum representation of the Laguerre polynomials (\ref{eq:laguerre}) and for the third equality we used the binomial theorem to express $(x^2+y^2)^j$ as sum. Both the integrals over $x$ and $y$ are of the same form as discussed in Equation~(\ref{eq:commonInt}), therefore we obtain
\begin{equation}
\langle n|R_0^{\textrm{c, tr}} |n\rangle = C_{n,n} \sum_{j=0}^{n} \binom{n}{n-j} \frac{1}{b^j j!} \sum_{k=0}^{j} \binom{j}{k} \frac{1}{a^{j+1}} \Gamma\left(k+\frac{1}{2}, a\Delta_c^2 \right)\Gamma\left(j-k+\frac{1}{2}, a\Delta_c^2 \right).
\end{equation}

Second, we deal with  $n<m$. We have
\begin{align*}
 \langle m |R_0^{\textrm{c, tr}} | n \rangle &= C_{n,m}  \int_{\Delta_c}^{\infty}\int_{\Delta_c}^{\infty} e^{-a(x^2+y^2)} L_n^{(m-n)}\left(- \frac{x^2+y^2}{b} \right)\,dy\,dx\\ 
 &=C_{n,m}\sum_{j=0}^{n} \binom{m}{n-j} \frac{1}{b^j j!}  \int_{\Delta_c}^{\infty}\int_{\Delta_c}^{\infty}e^{-a(x^2+y^2)} (x-iy)^{m-n} (x^2+y^2)^j  \,dy\,dx\\
 &= C_{n,m}\sum_{j=0}^{n} \binom{m}{n-j} \frac{1}{b^j j!} \sum_{l=0}^{j}\binom{j}{l} \sum_{k=0}^{m-n} \binom{m-n}{k} (-i)^{m-n-k} \int_{\Delta_c}^{\infty} e^{-a x^2} x^{k+2l} \,dx\int_{\Delta_c}^{\infty}e^{-a y^2} y^{2j-2l+m-n-k}  \,dy.
\end{align*}
For the second equality we inserted the sum representation of the Laguerre polynomials (\ref{eq:laguerre}) and for the third equality we used the binomial theorem twice; first, to express $(x^2+y^2)^j$ as sum and second, to write $(x-iy)^{m-n}$ as a sum too. Again, the occurring integrals are of the form given in Equation~(\ref{eq:commonInt}). Therefore, we obtain
\begin{equation}
\begin{aligned}
 &\langle m |R_0^{\textrm{c, tr}} | n \rangle \\
 &= \frac{C_{n,m}}{4} \sum_{j=0}^{n} \binom{m}{n-j} \frac{1}{b^j j!} \sum_{l=0}^{j}\binom{j}{l} \sum_{k=0}^{m-n} \binom{m-n}{k} \frac{i^{m-n-k} (-1)^{m-n-k}}{a^{j+1+\frac{m-n}{2}}} \Gamma\left(l+\frac{k+1}{2}, a\Delta_c^2 \right) \Gamma\left(j-l+\frac{m-n-k+1}{2}, a\Delta_c^2 \right).
\end{aligned}
\end{equation}
As the region operators have to be Hermitian, we do not need to calculate the matrix elements for $n>m$ separately. Summing up, we found for $R_0^{\textrm{c, tr}} = \sum_{n,m} |n\rangle\langle m| \langle n |R_0^{\textrm{c, tr}} |m\rangle$ the matrix elements
\begin{equation}
\begin{aligned}
 \langle n &|R_0^{\textrm{c, tr}} |m\rangle = \\
& \left\{
\begin{array}{ll}
 C_{n,n} \sum_{j=0}^{n} \binom{n}{n-j} \frac{1}{b^j j!} \sum_{k=0}^{j} \binom{j}{k} \frac{1}{a^{j+1}} \Gamma\left(k+\frac{1}{2}, a\Delta_c^2 \right)\Gamma\left(j-k+\frac{1}{2}, a\Delta_c^2 \right) & n = m \\
\frac{C_{n,m}}{4} \sum_{j=0}^{n} \binom{m}{n-j} \frac{1}{b^j j!} \sum_{l=0}^{j}\binom{j}{l} \sum_{k=0}^{m-n} \binom{m-n}{k} \frac{i^{m-n-k} (-1)^{m-n-k}}{a^{j+1+\frac{m-n}{2}}} \Gamma\left(l+\frac{k+1}{2}, a\Delta_c^2 \right) \Gamma\left(j-l+\frac{m-n-k+1}{2}, a\Delta_c^2 \right) & \, n < m \\
\overline{\langle m |R_z^{\textrm{c, tr}} |n\rangle} &\, n>m
\end{array}
\right. .    
\end{aligned}
\end{equation}
Similarly to the cross-shaped postselection in the untrusted scenario, we observe that the integral for $m=n$ contains only even powers of $x$ and $y$. Hence, this part is not affected by sign-changes in the boundaries of the occurring integrals. In contrast for $n<m$ we have odd powers of $x$ and $y$, so we expect additional powers of $-1$ in the expressions for $\langle n | R_z^{\textrm{c, tr}} | m \rangle$, $z\in{1,2,3}$ compared with $\langle n | R_0^{\textrm{c, tr}} | m \rangle$. By similar considerations as carried out in Section~\ref{sec:calculations_cPS_untrusted}, we obtain
\begin{align*}
\tilde{D}_{k,m,n}^{(0)} &= (-1)^{m-n-k},\\ 
\tilde{D}_{k,m,n}^{(1)} &= (-1)^{m-n}\\
\tilde{D}_{k,m,n}^{(2)} &= (-1)^{k}, \\ 
\tilde{D}_{k,m,n}^{(3)} &= 1,\\
\end{align*}
where $\tilde{D}_{m,n,k}^{(z)}$ denotes the power of $-1$ that occurs in the expression for the region operator $z$. Note that we have already included the factor $(-1)^{m-n-k}$, which occurs in the expression for all $z$. We define $D_{m,n,k}^{(z)}:= \tilde{D}_{m,n,k}^{(z)}~ i^{m-n-k}$ and obtain for $R_z^{\textrm{c, tr}} = \sum_{n,m} |n\rangle\langle m| \langle n |R_z^{\textrm{c, tr}} |m\rangle$ the representation with respect to the number basis
\begin{equation}
\begin{aligned}
\langle n |& R_z^{\textrm{c, tr}} |m\rangle = \\
& \left\{
\begin{array}{ll}
 C_{n,n} \sum_{j=0}^{n} \binom{n}{n-j} \frac{1}{a^{j+1} b^j j!} \sum_{k=0}^{j} \binom{j}{k} \Gamma\left(k+\frac{1}{2}, a\Delta_c^2 \right)\Gamma\left(j-k+\frac{1}{2}, a\Delta_c^2 \right) & n = m \\
\frac{C_{n,m}}{4 a^{\frac{m-n}{2}}} \sum_{j=0}^{n} \binom{m}{n-j} \frac{1}{a^{j+1} b^j j!} \sum_{l=0}^{j}\binom{j}{l} \sum_{k=0}^{m-n} \binom{m-n}{k} D_{k,m,n}^{(z) }   \Gamma\left(l+\frac{k+1}{2}, a\Delta_c^2 \right) \Gamma\left(j-l+\frac{m-n-k+1}{2}, a\Delta_c^2 \right) &  n < m \\
\overline{\langle m |R_z^{\textrm{c, tr}} |n\rangle} & n>m
\end{array}
\right. .    
\end{aligned}
\end{equation}
\subsection{First- and second-moment observables}\label{sec:observables}
For the sake of completeness, we give explicit number-basis representations of the first- and second-moment observables, defined in equations (\ref{eq:observable_FQ}-\ref{eq:observable_SP}). We note that \cite{Lin_2020} gives explicit representations in the appendix, too, which again depend on the coefficients of some Taylor expansion. In contrast to, we give explicit expressions and solve the integrals similar to our calculations for the region operators in the preceding sections. In what follows, we give only expressions for $n \leq m$, as all operators need to be Hermitian, hence $\langle k|\hat{O} | l \rangle = \overline{\langle l | \hat{O} |k \rangle}$ gives the missing matrix elements.

We start with $\hat{F}_Q$, whose matrix elements with respect to the number basis are given by
\begin{equation*}
    \langle n|\hat{F}_Q|m\rangle = \frac{1}{\sqrt{2}} \int (y + y^*) \langle n | G_{y} |m\rangle
\end{equation*}
Choosing polar coordinates and inserting the expression for $G_{\gamma e^{i\theta}}$ from Equation~(\ref{eq:Gy_numberbasis}) leads to
\begin{align*}
    \langle n|\hat{F}_Q|m\rangle &= \frac{C_{n,m}}{\sqrt{2}} \int_{0}^{2\pi}  \left(e^{i \theta}+e^{-i\theta} \right)e^{-i\theta(m-n)} \,d\theta \int_{0}^{\infty} \gamma^{m-n+2} e^{-a\gamma^2} L_{n}^{(m-n)}\left(- \frac{\gamma^2}{b}\right) \,d\gamma \\
    &= \frac{2\pi C_{n,m}}{\sqrt{2}} \delta_{m,n\pm 1}\int_{0}^{\infty} \gamma^{m-n+2} e^{-a\gamma^2} L_{n}^{(m-n)}\left(- \frac{\gamma^2}{b}\right) \,d\gamma\\
    &= \frac{2\pi C_{n,m}}{\sqrt{2}} \delta_{m,n\pm 1} \sum\limits_{j=0}^{n} \binom{m}{n-j} \frac{(-1)^j}{b^j j!} \int_{0}^{\infty} \gamma^{m-n+j+2} e^{-a\gamma^2} \,d\gamma.
\end{align*}
The remaining integral can be solved using (\ref{eq:commonInt}) for the special case where $\Delta = 0$. Therefore, we obtain
\begin{equation}
    \langle n | \hat{F}_Q | n+1 \rangle = \frac{\pi C_{n,n+1}}{\sqrt{2}} \sum_{j=0}^{n} \binom{n+1}{n-j} \frac{1}{a^{j+2}b^j j!} \Gamma(j+2) = \frac{\pi C_{n,n+1}}{\sqrt{2}} \sum_{j=0}^{n} \binom{n+1}{n-j} \frac{j+1}{a^{j+2}b^j}
\end{equation}
and $\langle n|\hat{F}_Q|m\rangle = 0$ for $m \neq n\pm 1$, where we used the definition of the gamma function. Similarly, starting from Equation~(\ref{eq:observable_FP}),  we derive
\begin{equation}
    \langle n | \hat{F}_P | n+1 \rangle =i \frac{\pi C_{n,n+1}}{\sqrt{2}} \sum_{j=0}^{n} \binom{n+1}{n-j} \frac{j+1}{a^{j+2}b^j} = i \langle n | \hat{F}_Q | n+1 \rangle
\end{equation}
and $\langle n | \hat{F}_P | m \rangle = 0$ if $m \neq n\pm 1$.

The matrix elements of the second-moment observables read
\begin{align}
    \langle n | \hat{S}_Q | n \rangle = -\langle n | \hat{S}_P | n \rangle&= \pi C_{n,n} \sum_{j=0}^{n} \binom{n}{n-j} \frac{j+1}{a^{j+2}b^j},\\
     \langle n | \hat{S}_Q | n+2 \rangle = -\langle n | \hat{S}_P | n+2 \rangle &= \pi C_{n,n+2} \sum_{j=0}^{n} \binom{n+2}{n-j} \frac{(j+2)(j+1)}{a^{j+3}b^j}
\end{align}
and $\langle n | \hat{S}_Q | m \rangle  = \langle n | \hat{S}_P | m \rangle = 0$ otherwise.

\section{Choice of the cutoff number $N_c$}\label{APDX:choice_cutoff}
In this section, we briefly discuss our choice of the photon cutoff number $N_c$. Following \cite{Lin_2019}, we use the first $N_c$ Fock states to approximate Bob's infinite-dimensional Hilbert space with sufficient accuracy. Choosing $N_c$ too small leads to inaccurate results, while choosing $N_c$ too large increases the runtime unnecessarily (the problem size increases quadratically with increasing $N_c$). Therefore, we examined the change in the secure key rate when increasing $N_c$ for different transmission distances and different postselection parameters for fixed parameters $\beta = 0.95$ and $\xi = 0.01$. The examinations in this section are carried out for QPSK protocols exemplarily, but a similar behaviour can be observed for eight-state protocols as well.  The result is visualised in Figure~\ref{fig:num_valid}, where it can be seen that the secure key rate remains (almost) constant for $N_c \geq 12$ for all three curves. The relative changes between neighbouring  data points for $N_c \geq 12$ are lower than $0.5\%$ and for $N_c \geq 14$ lower than $0.2\%$. This fluctuations are mainly caused by the gap between step 1 and step 2, which is more sensitive to numerical errors. The relative differences between neighbouring data points of the results for the first step are smaller than $0.01\%$ for $N_c \geq 12$. This motivates our choice of $N_c = 12$ for the QPSK protocols in the present paper, being a good compromise between accuracy and computational feasibility. Similarly, we chose $N_c = 14$ as a sound compromise for 8PSK. The reason for the higher cutoff compared with protocols with four signal states is the slightly higher optimal coherent state amplitude for 8PSK protocols, hence higher average photon numbers. In general, higher coherent state amplitudes lead to higher non-negligible Fock-number states which increases the required cutoff.

\begin{figure}
\includegraphics[width=0.45\textwidth]{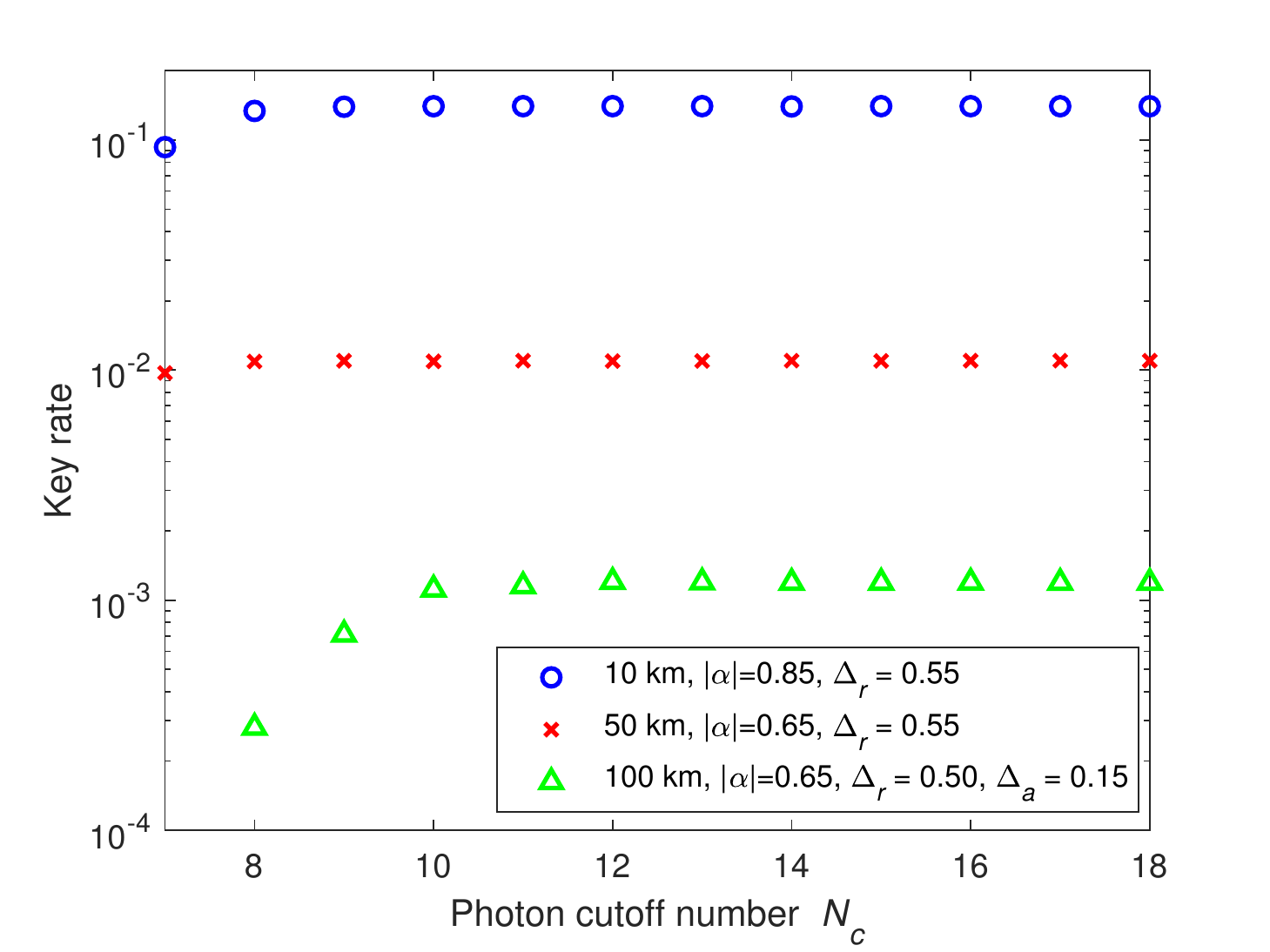}
\makeatletter\long\def\@ifdim#1#2#3{#2}\makeatother
\caption{\label{fig:num_valid} Secure key rate versus chosen cutoff-number $N_c$ for QPSK protocols and three different distances and choices of postselection parameter. For all three curves, we set $\beta = 0.95$ and $\xi = 0.01$.}
\end{figure}
\end{widetext}

\bibliography{Bibliography}

\end{document}